\documentclass[english]{article}
\usepackage[T1]{fontenc}
\usepackage[utf8]{inputenc}
\usepackage{geometry}
\usepackage{color}
\usepackage{babel}
\usepackage{array}
\usepackage{float}
\usepackage{multirow}
\usepackage{amsmath}
\usepackage{amssymb}
\usepackage{graphicx}
\usepackage{setspace}
\usepackage{wasysym}
\usepackage[unicode=true,pdfusetitle,
 bookmarks=true,bookmarksnumbered=false,bookmarksopen=false,
 breaklinks=false,pdfborder={0 0 0},pdfborderstyle={},backref=false,colorlinks=true]
 {hyperref}
\hypersetup{
 linkcolor=blue,citecolor=blue,urlcolor=blue}

\makeatletter

\providecommand{\tabularnewline}{\\}

\newcommand{\lyxaddress}[1]{
	\par {\raggedright #1
	\vspace{1.4em}
	\noindent\par}
}

\makeatother

\begin{document}
\title{Boundary states, overlaps, nesting and bootstrapping AdS/dCFT}
\author{Tamas Gombor, Zoltan Bajnok }
\maketitle

\lyxaddress{\begin{center}
\emph{Wigner Research Centre for Physics}\\
\emph{Konkoly-Thege Miklós u. 29-33, 1121 Budapest , Hungary}\\
\par\end{center}}
\begin{abstract}
Integrable boundary states can be built up from pair annihilation
amplitudes called $K$-matrices. These amplitudes are related to mirror
reflections and they both satisfy Yang Baxter equations, which can
be twisted or untwisted. We relate these two notions to each other
and show how they are fixed by the unbroken symmetries, which, together
with the full symmetry, must form symmetric pairs. We show that the
twisted nature of the $K$-matrix implies specific selection rules
for the overlaps. If the Bethe roots of the same type are paired the
overlap is called chiral, otherwise it is achiral and they correspond
to untwisted and twisted $K$-matrices, respectively. We use these
findings to develop a nesting procedure for $K$-matrices, which provides
the factorizing overlaps for higher rank algebras automatically. We
apply these methods for the calculation of the simplest asymptotic
all-loop 1-point functions in AdS/dCFT. In doing so we classify the
solutions of the YBE for the $K$-matrices with centrally extended
$\mathfrak{su}(2\vert2)_{c}$ symmetry and calculate the generic overlaps
in terms of Bethe roots and ratio of Gaudin determinants.
\end{abstract}

\section{Introduction}

Recently there have been renewed interest and relevant progress in
calculating overlaps between periodic multiparticle states and integrable
boundary states. They appear in quite distinct parts of theoretical
physics including statistical physics and the gauge/string duality.

In statistical physics there are significant activities in analyzing
the behavior of both integrable and non-integrable systems after a
quantum quench \cite{Caux:2013ra}. In a typical situation a parameter
of the Hamiltonian is suddenly changed implying that the ground-state
of the pre-quenched Hamiltonian is no longer an eigenstate of the
post-quench Hamiltonian. The after quench time evolution can be fully
described once the overlaps of this initial state with the eigenstates
of the post-quench Hamiltonian are known. Integrable quenches are
those when the initial state is an integrable boundary state \cite{Piroli:2017sei,Piroli:2018don,Piroli:2018ksf}.

In the AdS/CFT correspondence there are at least two places where
overlaps appeared so far. Recently a new class of three-point functions
were investigated in the $AdS_{5}/CFT_{4}$ correspondence involving
a local gauge invariant single trace operator and two determinant
operators dual to maximal giant gravitons \cite{Jiang:2019xdz,Jiang:2019zig}.
The authors showed that the three-point function can be calculated
as an overlap between the finite volume multiparticle state and a
finite volume integrable boundary state.

In the other, much more investigated, application a codimension one
defect is introduced in the gauge theory, which breaks part of the
gauge symmetry of the model. As a result some scalar fields develop
vacuum expectation values and one-point functions of local gauge invariant
operators can be non zero. Their space-time dependence is fixed by
the unbroken conformal symmetry up to an operator dependent normalization
constant. As the operators are normalized by their two point functions
far away from the defect this coefficient is a physical quantity.
Since in the integrable description of the AdS/CFT correspondence
local gauge-invariant operators are related to finite volume multiparticle
states their one-point functions can be interpreted as finite volume
overlaps with a boundary state created by the defect. Much progress
has been achieved so far which included D3-D5 defects \cite{deLeeuw:2015hxa,Buhl-Mortensen:2015gfd,deLeeuw:2018mkd}
and D3-D7 defects \cite{deLeeuw:2016ofj,deLeeuw:2019ebw}. Most of
the analysis is restricted however for the leading order results in
the coupling and only subsectors for the whole theory, although some
partial one-loop results are also available in a diagonal subsector
\cite{Buhl-Mortensen:2016pxs,Grau:2018keb}. In \cite{Buhl-Mortensen:2017ind}
the authors proposed an all loop formula for the one point functions
in the $\mathfrak{su}(2)$ sector in the asymptotic domain, i.e. neglecting
wrapping corrections. One of the aim of our paper is to go beyond
these results and try to bootstrap the overlaps to be valid for any
couplings and any sectors in the asymptotic domain.

Motivated by the above mentioned applications there were already significant
activities and progress in calculating the matrix elements of integrable
boundary states and eigenstates of the transfer matrix for integrable
spin chains. The calculations of these on-shell overlaps have been
already performed for various setups. In \cite{Brockmann_2014_1,Brockmann_2014_2}
the authors investigated the Neél state in the XXZ spin chain and
derived that the on-shell overlaps are non-vanishing only when the
Bethe roots are paired and they can be written as the product of one
particle overlap functions and a ratio of Gaudin-like determinants.
This finding turned out to be true for several other models when the
boundary states were integrable. The integrability condition in the
XXZ spin chain was proposed in \cite{Piroli:2017sei}. In \cite{Pozsgay:2018ixm}
the author proposed a same type of formula for arbitrary states, built
from solutions of the boundary Yang Baxter equation (two-site states),
and proved its validity numerically. For matrix product states it
was proposed that the on-shell overlap formulas cannot be written
in the previous product form, rather a sum of those and this claim
was verified numerically for the XXX spin chain \cite{Buhl-Mortensen:2015gfd}.
The integrability condition can be generalized to nested systems and
there are some known overlap formulas for integrable states in these
cases. In \cite{deLeeuw:2016umh,Piroli:2018ksf,Piroli:2018don} there
are numerically verified formulas for two-site and matrix product
states in the $\mathfrak{su}(3)$ spin chain. The authors found that
the non-vanishing overlaps require pair structures for the two types
of Bethe roots. In \cite{deLeeuw:2018mkd,deLeeuw:2019ebw} there are
numerically verified overlap formulas for $\mathfrak{so}(6)$ spin
chains and each type of Bethe roots have their pair structure for
non-vanishing overlaps. In \cite{Jiang:2019xdz} the authors investigated
a different boundary state of the $\mathfrak{so}(6)$ spin chain and
found a novel pair structure which was different from what had been
found so far. All of these results are in specific low rank models
and are not exactly derived\footnote{In a recent paper \cite{Jiang:2020sdw} derived some overlap formulas
for non-nested system but it is not obvious whether this can be generalized
to the nested ones.}. Another aim of our paper is to perform a systematic study of higher
rank spin chains, derive the pair structure of their Bethe roots and
develop a nesting procedure which could provide the factorizing overlaps.
In doing so we found that the unbroken symmetries play a crucial role.
This residual symmetries together with the original symmetry must
form a symmetric pair and the nature of these maximal subalgebras
are intimately related to the nature of the pair structures whose
knowledge is essential in formulating the right nesting.

Keeping both the spin chain and AdS/dCFT applications in mind we try
to be as general as required. In order to make the presentations lighter
new notions and methods are demonstrated in various examples.

The rest of the paper is organized as follows: In the next section
we introduce integrable boundary states and their relations to boundary
reflections emphasizing that for non-crossing invariant theories they
are not equivalent. We then turn to analyze overlaps between integrable
boundary states and periodic states. Periodic states are the eigenstates
of the transfer matrices, which can be constructed from scattering-
or spin chain $R$-matrices. We first recall the eigenvalues and Bethe
root structures for $\mathfrak{su}(N)$, $\mathfrak{so}(4)$, $\mathfrak{so}(6)$
and $\mathfrak{su}(2\vert2)_{c}$ chains, as they will be relevant
in what follows. We then analyze the consequences of the integrability
requirement for the root structure and conclude that for algebras
with a nontrivial Dynkin diagram symmetry roots can be paired in a
chiral and an achiral way. In the next section we investigate integrable
two sites boundary states related to solutions of the KYBE, the YBE
for boundary states. We observe that there are two types of solutions
of the KYBE with quite distinct symmetries, which we relate to the
chirality/achirality of the overlaps. Keeping also the AdS/dCFT applications
in mind we derive the most general bosonic solution of the KYBE for
the centrally extended $\mathfrak{su}(2\vert2)_{c}$ algebra, and
identify its symmetries. In the next section we invent a version of
the nesting, which enables us to calculate the overlaps in various
spin chains including the $\mathfrak{su}(2\vert2)_{c}$ symmetric
ones, relevant for AdS/dCFT. Symmetry argumentations combined with
nesting and the selection rules for roots can be used to investigate
the possible K-matrices for the AdS/dCFT correspondence. We close
this section with an all-loop asymptotic proposal for the simplest
D3-D5 1-point functions. Finally we conclude and provide a list of
open problems. Technical details are relegated to Appendices.

\section{Boundary states, K- and boundary Yang-Baxter equations}

There are two ways to place an integrable boundary in a two dimensional
system: it can be placed either in space or in time. We formulate
these two cases at such a level of generality which can cover both
the AdS/CFT scattering matrix and all rational spin chains.

\subsection{Boundary states and KYBE}

\begin{figure}[h]
\begin{centering}
\includegraphics{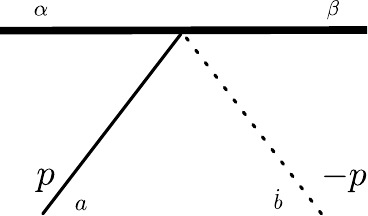}~~~~~~~~~~~~~~~~\includegraphics{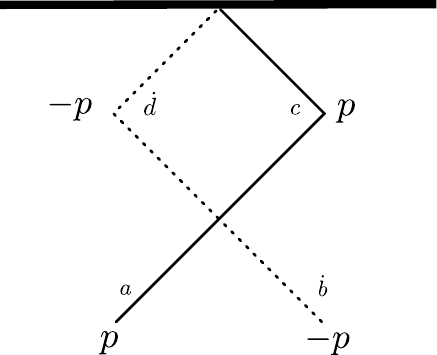}
\par\end{centering}
\caption{Graphical representation of the K-matrix and its crossing property.
The boundary state might have an inner degree of freedom, which is
labeled by Greek letters. It might also annihilate particles from
different representations, which is indicated by straight and dashed
lines and by a dot on the indices of the dashed particles. One typical
example is when it annihilates a pair consisting of a particle and
an anti-particle, which transform wrt. a representation and its contragradient
representation, respectively. Such a case can appear for instance
in $\mathfrak{su}(N)$ spin chains.}

\label{KandUni}
\end{figure}

If the boundary is placed in time it serves as an initial or finite
state. The final state annihilates pairs of particles, while the initial
state creates those. In a QFT an integrable boundary state is annihilated
by the infinitely many parity odd charges of the theory \cite{Ghoshal:1993tm}.
This in particular implies that both the initial and the final boundary
states can be described by the two particle K-matrix and one is related
to the other by conjugation. We will focus on a final state, which
can be depicted on the left of Figure \ref{KandUni}. 

The K-matrix $K_{a\dot{b}}^{\alpha\beta}(p)$ is the amplitude of
annihilating a pair of particles with labels $a,\dot{b}$ and momenta
$p,-p$ and having boundary degrees of freedom $\alpha$ and $\beta$
on the two ends. To keep the discussion on a general level we allowed
that particles with momentum $p$ and $-p$ transform wrt. different
representations (of the same dimension) which we differentiated by
a dot on the index. 

In integrable theories particle-trajectories can be shifted without
altering the amplitudes, see Figure \ref{KandUni}. As a consequence,
the K-matrix satisfies the crossing equation 

\begin{equation}
K_{a\dot{b}}^{\alpha\beta}(p)=S_{a\dot{b}}^{c\dot{d}}(p,-p)K_{\dot{d}c}^{\alpha\beta}(-p)\quad;\qquad K_{1\dot{2}}(p)=S_{1\dot{2}}(p,-p)K_{\dot{2}1}(-p)
\end{equation}
where we also introduced a compact notation by indicating only in
which place and representation the scattering $S$- and $K$-matrices
act, i.e $(p\to1,-p\to2)$ and suppressed to write out explicitly
the boundary degrees of freedom. 

\begin{figure}[h]
\begin{centering}
\includegraphics[height=3cm]{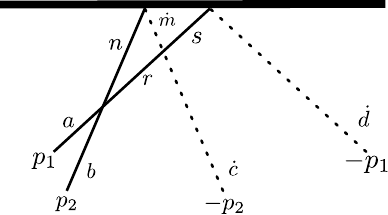}~~~~~~~~~~~~~~\includegraphics[height=3cm]{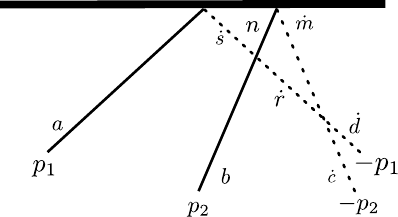}
\par\end{centering}
\caption{K-matrix YBE from shifting particle lines. Dashed lines and dotted
indices might transform wrt. different representations. If they are
different the KYBE is called twisted, otherwise it is called untwisted.}

\label{KYBE}
\end{figure}

Two pairs of particles can be annihilated in two different ways, see
Figure \ref{KYBE} leading to the Yang-Baxter equation for the K-matrix
(KYBE):
\begin{equation}
S_{ab}^{rn}(p_{1},p_{2})K_{n\dot{m}}^{\alpha\beta}(p_{2})S_{r\dot{c}}^{s\dot{m}}(p_{1},-p_{2})K_{s\dot{d}}^{\beta\gamma}(p_{1})=K_{a\dot{s}}^{\alpha\beta}(p_{1})S_{b\dot{r}}^{n\dot{s}}(p_{2},-p_{1})K_{n\dot{m}}^{\beta\gamma}(p_{2})S_{\dot{c}\dot{d}}^{\dot{m}\dot{r}}(-p_{2},-p_{1})\label{eq:KYBEmat}
\end{equation}
or alternatively 

\begin{equation}
K_{3\dot{4}}(p_{2})K_{1\dot{2}}(p_{1})S_{1\dot{4}}(p_{1},-p_{2})S_{13}(p_{1},p_{2})=K_{1\dot{2}}(p_{1})K_{3\dot{4}}(p_{2})S_{3\dot{2}}(p_{2},-p_{1})S_{\dot{4}\dot{2}}(-p_{2},-p_{1})\label{eq:KYBE}
\end{equation}
where again for each particle we associated a vector space as $(p_{1}\to1,-p_{1}\to2,p_{2}\to3,-p_{2}\to4)$.
If the dotted representation is really different from the undotted
one the KYBE is called \emph{twisted}. If the two representations
are the same, the dots can be neglected and the KYBE is called \emph{untwisted}. 

Thanks to integrability the general multiparticle annihilation process
can be written in terms of the two particle annihilation amplitudes
and the two particle scattering matrices. In particular, in terms
of the ZF operators, the boundary state has an exponential form
\begin{equation}
\langle B\vert=\langle0\vert\exp\left\{ \int_{-\infty}^{\infty}\frac{dp}{4\pi}K_{AB}(p)Z^{A}(-p)Z^{B}(p)\right\} 
\end{equation}
where $Z^{A}(p)$ is the operator, which annihilates a particle of
type $A$ and momentum $p$, and these operators form the ZF algebra
\begin{equation}
Z^{A}(p_{1})Z^{B}(p_{2})=S_{CD}^{AB}(p_{1},p_{2})Z^{D}(p_{2})Z^{C}(p_{1})
\end{equation}
The boundary state contains the contributions of all possible particles,
which we indicated by summing over indices of all types. Typically
either $K_{ab}^{\alpha\beta}=0=K_{\dot{a}\dot{b}}^{\alpha\beta}$
or $K_{a\dot{b}}^{\alpha\beta}=0=K_{\dot{a}b}^{\alpha\beta}$, which
is related to the fact that $S_{a\dot{b}}^{\dot{c}d}=0$. Consistency
of the boundary state, i.e. invariance for $p\to-p$ and uniqueness
of the exponentials, implies the crossing property and the KYBE.

\subsection{Reflection matrices and BYBE}

\begin{figure}
\begin{centering}
\includegraphics[height=2.5cm]{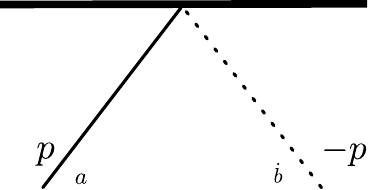}~~~~~~~~~~~~~~\includegraphics[width=4cm]{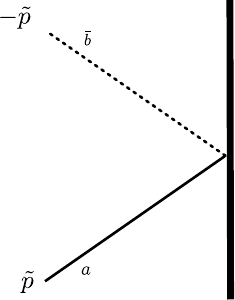}
\par\end{centering}
\caption{Reflection matrix in the mirror theory as obtained by a mirror (rotation
by $\pi/2)$ and a parity transformation: $z\to\frac{\omega}{2}-\tilde{z}.$
The $K$-matrix connects the a representation and a dotted one, while
the mirror reflection connects the same representation and the contragradient
of the dotted one. }

\label{R}
\end{figure}

Alternatively, we can place the boundary in space and characterize
it by specifying how particles scatter off it. Integrable boundaries
have infinitely many parity even conserved charges. As a consequence,
multiparticle reflections factorize into one-particle reflections
and pairwise scatterings, thus it is enough to determine the one-particle
reflections. This reflection amplitude is not independent from the
K-matrix above. Indeed, one can perform a rotation in exchanging the
role of space and time, which gives rise to the mirror theory. In
non-relativistic theories the mirror dispersion relation $\tilde{E}(\tilde{p})$,
obtained by analytical continuation, can be different from the original
one $E(p)$. Here and from now on we indicate mirror quantities by
tilde. Typically we parametrize the dispersion relation with a generalized
rapidity parameter $p(z),E(z)$, and then the scattering matrix depends
on these rapidities $S(z_{1},z_{2})$. For a properly chosen rapidity
parameter the mirror rapidity, $\tilde{z}$ is simply the shifted
version of the original one $z=\tilde{z}+\frac{\omega}{2}$. The notation
indicates that $z\to z+\omega$ is the crossing transformation which
maps $E\to-E$ and $p\to-p$ and replaces particles with antiparticles\footnote{In relativistic theories $z$ is the rapidity and $\omega=i\pi$,
while for the $AdS_{5}/CFT_{4}$ integrable model $z$ is a torus
variable and $\omega=\omega_{2}$ \cite{Arutyunov:2007tc}. For physical
processes both in the theory and its mirror version the rapidity variables
are real. Since $\omega$ is imaginary the $z\to\tilde{z}+\frac{\omega}{2}$
transformation involves an analytical continuation. }. Thus going from the original theory to the mirror theory is half
of a crossing transformation. The mirror dispersion relation is $\tilde{E}(\tilde{z})=-ip(\tilde{z}+\frac{\omega}{2})$,
$\tilde{p}(\tilde{z})=-iE(\tilde{z}+\frac{\omega}{2})$, while the
mirror scattering matrix is $\tilde{S}(\tilde{z}_{1},\tilde{z}_{2})=S(\tilde{z}_{1}+\frac{\omega}{2},\tilde{z}_{2}+\frac{\omega}{2}).$
In the following we show that the KYBE is equivalent to the boundary
YBE (BYBE) for the reflection matrix in this mirror theory. In doing
so we suppress to write out the spectator boundary degrees of freedom.
The index form of the KYBE, (\ref{eq:KYBEmat}), can be rewritten
by introducing the charge conjugation matrix $C$ and its inverse
$C^{n\bar{m}}C_{\bar{m}k}=\delta_{k}^{n}$ which intertwine between
the particle and antiparticle representations, as
\begin{align}
S_{ab}^{rn}(z_{1},z_{2})C^{\bar{m}\dot{m}}K_{n\dot{m}}(z_{2})S_{r\dot{c}}^{s\dot{m}}(z_{1},-z_{2})C_{\dot{m}\bar{m}}K_{s\dot{d}}(z_{1})C^{\bar{c}\dot{c}}C^{\bar{d}\dot{d}} & =\\
C^{\bar{c}\dot{c}}C^{\bar{d}\dot{d}}K_{a\dot{s}}(z_{1})C^{\dot{s}\bar{s}}C_{\bar{s}\dot{s}}S_{b\dot{r}}^{n\dot{s}}(z_{2},-z_{1})K_{n\dot{m}}(z_{2})C^{\dot{m}\bar{m}}C_{\bar{m}\dot{m}}S_{\dot{c}\dot{d}}^{\dot{m}\dot{r}}(-z_{2},-z_{1})\nonumber 
\end{align}
Let us introduce the matrix $K(z)$ with indices: $C^{\bar{d}\dot{d}}K_{s\dot{d}}(z)=K_{s}^{\bar{d}}(z)$.
It connects particles to twisted antiparticles labeled by bar, i.e.
to the contragradient representation of the dotted representation.
In terms of these quantities the KYBE takes a form 
\begin{equation}
S_{12}(z_{1},z_{2})K_{2}(z_{2})C_{2}S_{1\dot{2}}^{t_{2}}(z_{1},-z_{2})C_{2}^{-1}K_{1}(z_{1})=K_{1}(z_{1})C_{1}S_{2\dot{1}}^{t_{1}}(z_{2},-z_{1})C_{1}K_{2}(z_{2})S_{\bar{2}\bar{1}}(-z_{2},-z_{1})
\end{equation}
Using the crossing symmetry of the $S$-matrix $C_{1}^{-1}S_{1\dot{2}}^{t_{1}}(z_{1},z_{2})C_{1}=S_{\bar{2}1}(z_{2},z_{1}+\omega)$
and parity invariance $S_{21}(-z_{2},-z_{1})=S_{12}(z_{1},z_{2})$
the equation takes the form: 
\begin{equation}
S_{12}(z_{1},z_{2})K_{2}(z_{2})S_{\bar{2}1}(-z_{2}+\omega,z_{1})K_{1}(z_{1})=K_{1}(z_{1})S_{\bar{1}2}(-z_{1}+\omega,z_{2})K_{2}(z_{2})S_{\bar{1}\bar{2}}(z_{1},z_{2})
\end{equation}
We now use analytical continuation to go to the mirror theory, together
with relabeling and a parity transformation $z_{1}\to\frac{\omega}{2}-\tilde{z}_{2}$
and $z_{2}\to\frac{\omega}{2}-\tilde{z}_{1}$, to reach
\begin{equation}
\tilde{S}_{12}(\tilde{z}_{1},\tilde{z}_{2})\tilde{R}_{1}(\tilde{z}_{1})\tilde{S}_{2\bar{1}}(\tilde{z}_{2},-\tilde{z}_{1})\tilde{R}_{2}(\tilde{z}_{2})=\tilde{R}_{2}(\tilde{z}_{2})\tilde{S}_{1\bar{2}}(\tilde{z}_{1},-\tilde{z}_{2})\tilde{R}_{1}(\tilde{z}_{1})\tilde{S}_{\bar{2}\bar{1}}(-\tilde{z}_{2},-\tilde{z}_{1})
\end{equation}
where the mirror S-matrix was used (which is also parity invariant)
and we introduced a mirror reflection matrix $\tilde{R}(\tilde{z})=K(\frac{\omega}{2}-\tilde{z})$,
see also Figure \ref{R}: 
\begin{equation}
\tilde{R}_{a}^{\bar{b}}(\tilde{z})=C^{\bar{b}\dot{b}}K_{a\dot{b}}(\frac{\omega}{2}-\tilde{z})\label{eq:RK}
\end{equation}
Thus we can conclude that if the $K$-matrix satisfies the KYBE, (\ref{eq:KYBE}),
then the mirror reflection factor defined by (\ref{eq:RK}) satisfies
the BYBE. If the representation with the bar is different from the
one without the bar, the BYBE is called \emph{twisted}, otherwise
it is called \emph{untwisted}. Let us point out that the mirror BYBE
is not always equivalent to the BYBE in the original theory. It can
differ in two ways. For AdS/CFT, particles and antiparticles are in
the same representation but the dispersion relation and the scattering
matrix is not relativistic invariant, thus $\tilde{S}$ and $S$ are
different. For rational spin chains and corresponding quantum field
theories particles and antiparticles can transform wrt. different
representations, thus an untwisted KYBE can results in twisted BYBE
and vica versa. We elaborate on the possible cases in section 5.

\begin{figure}
\begin{centering}
\includegraphics[height=5cm]{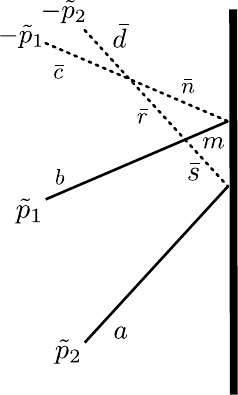}~~~~~~~~~~~~~~~~~~~~~~~~~~~~\includegraphics[height=5cm]{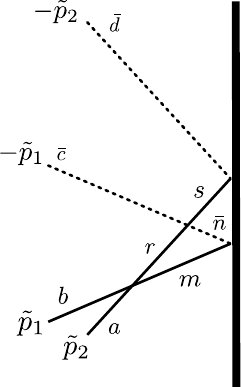}
\par\end{centering}
\caption{Boundary Yang Baxter equation in the mirror model. Dashed line indicates
that the representation of the reflected particle can be different
from the original one. If it is different, the BYBE is called twisted,
otherwise it is called untwisted. }

\label{BYBE}
\end{figure}

We obtained the mirror BYBE from the KYBE by a mirror and a parity
transformation: $z\to\frac{\omega}{2}-\tilde{z}$. Graphically the
resulting equation can be depicted as on Figure \ref{BYBE}. 

The mirror reflection matrix satisfies the unitarity relation 
\begin{equation}
\tilde{R}_{i}^{j}(\tilde{z})\tilde{R}_{j}^{k}(-\tilde{z})=\delta_{i}^{k}\quad;\qquad\tilde{R}_{1}(\tilde{z})\tilde{R}_{1}(-\tilde{z})=\mathbb{I}
\end{equation}
. 

In summarizing, in quantum field theories the boundary state in the
physical theory can be represented by the K-matrix, which satisfies
the crossing equation and the KYBE. It is related to a reflection
matrix of the mirror theory as (\ref{eq:RK}), which satisfies unitarity. 

Let us finally note that we have exactly the same equations for integrable
spin chains, where the $R$-matrix plays the role of the scattering
matrix and the $K$-matrix, solution of the boundary YBE is the reflection
amplitude. 

So far our considerations were in the infinite volume setting. In
practical applications, however the boundary states are in finite
volume and we are interested in the overlap of the finite volume boundary
state and the finite volume multiparticle states. In the following
we recall the finite volume periodic spectrum in various models we
need. 

\section{Spin chains and asymptotic spectrum }

Integrable spin chains are interesting in their own rights, but they
also have a direct connection to integrable QFTs. For any integrable
QFT with inner degrees of freedom the scattering matrix has a scalar
factor $S_{0}$ and a matrix part, $R$: 
\begin{equation}
S(z_{1},z_{2})=S_{0}(z_{1},z_{2})R(z_{1},z_{2})
\end{equation}
 The matrix part satisfies the YBE and can be considered as an R-matrix
of an integrable spin chain. The large volume (asymptotic) spectrum
of the QFT, neglecting exponentially small volume corrections, is
simply the infinite volume spectrum $E_{n}(L)=\sum_{i}E(z_{i})$ only
the momenta are quantized. This momentum quantization can be determined
from the eigenvalues of the transfer matrix of the spin chain $t(z,\{z_{i}\})$,
which is the trace of the monodromy matrix, built from the $R$-matrices
as 
\begin{equation}
t(z,\{z_{i}\})=\text{Tr}_{0}(T_{0}(z))\quad;\qquad T_{0}(z)=R_{0L}(z,z_{L})\dots R_{01}(z,z_{1})
\end{equation}
 Here $0$ labels an auxiliary particle with rapidity $z$, whose
representation space is traced over. This space can carry the same
representation as the physical particle or some different representations.
Transfer matrices for different representations and spectral parameters
commute with each other and can be diagonalized in a spectral parameter
independent basis.

In the following we recall the results of the nested Bethe ansatz
for the rational $\mathfrak{su}(N)$, $\mathfrak{so}(4)$ and $\mathfrak{so}(6)$
spin chains, together with the centrally extended $\mathfrak{su}(2\vert2)_{c}$,
which will be relevant for the later investigations. In spin chains
the spectral parameter is traditionally denoted by $u$, which is
not necessarily the rapidity, it might be some non-trivial function
of it $u(z)$.

In the $AdS_{5}/CFT_{4}$ correspondence spin chains appear also on
the Yang-Mills side. Indeed, the one-loop dilatation operator of the
scaling dimensions of local operators can be related to an $\mathfrak{su}(2,2\vert4)$
nearest neighbors spin chain \cite{Beisert:2010jr}. At higher loops
the interaction range in the spin chain gets extended and spoils the
$\mathfrak{su}(2,2\vert4)$ structure. What carries over is the nested
Bethe ansatz obtained by choosing a pseudo vacuum. This pseudo vacuum
breaks the $\mathfrak{su}(2,2\vert4)$ symmetry to $\mathfrak{su}(2\vert2)_{c}\oplus\mathfrak{su}(2\vert2)_{c}$,
which gets central extended at higher loop orders.

\subsection{Spectrum of the $\mathfrak{su}(N)$ spin chain}

In the $\mathfrak{su}(N)$ symmetric spin chains the R-matrix is a
function of the differences of the spectral parameters and has the
form 
\begin{equation}
R(u_{1},u_{2})\equiv R(u_{1}-u_{2})=\mathbf{1}+\frac{1}{u_{1}-u_{2}}\mathbf{P}\label{eq:Rsun}
\end{equation}
where $\mathbf{1}$ is the identity and $\mathbf{P}$ is the permutation
operator acting on $N$-dimensional spaces, carrying the fundamental
representations. The eigenvectors of the transfer matrix can be parametrized
by Bethe roots $\vert u_{1}^{(a)},\dots u_{n_{a}}^{(a)}\rangle\equiv\bigl|\mathbf{u}^{(a)}\bigr\rangle$,
with $a=1,\dots,N-1$ and the eigenvalues can be built up from the
elementary building blocks $z_{k}$:

\begin{equation}
t(u)\bigl|\mathbf{u}^{(a)}\bigr\rangle=\Lambda(u)\bigl|\mathbf{u}^{(a)}\bigr\rangle\quad;\qquad\Lambda(u)=\sum_{k=1}^{N}\frac{Q_{k-1}^{[k+1]}\left(u\right)}{Q_{k-1}^{[k-1]}\left(u\right)}\frac{Q_{k}^{[k-2]}\left(u\right)}{Q_{k}^{[k]}\left(u\right)}
\end{equation}
where $f^{[k]}(u)=f(u+\frac{k}{2})$ \cite{Arnaudon_2005}. The blocks,
$z_{k}$, contribute also to the eigenvalues of the transfer matrices,
where the auxiliary representation corresponds to a rectangular Young
diagram and can be written in terms of the $Q$-functions which encode
the Bethe roots
\begin{equation}
Q_{0}(u)=u^{L}\quad,\qquad Q_{k}(u)=\prod_{i=1}^{n_{k}}\left(u-u_{i}^{(k)}\right),\qquad k=1,\dots,N-1,\quad;\qquad Q_{N}(u)=1.
\end{equation}
Bethe roots can be obtained from the Bethe Ansatz equations, which
arise by demanding the regularity of the transfer matrix at $u=u_{i}^{(k)}$.

\subsection{Spectrum of the $\mathfrak{so}(4)$ spin chain}

The $R$-matrix in the $\mathfrak{so}(4)$ spin chain can be written
as 
\begin{equation}
R(u)=\mathbf{1}+\frac{1}{u}\mathbf{P}-\frac{1}{1+u}\mathbf{K}.\label{eq:Rso4-1}
\end{equation}
where $\mathbf{K}$ is the trace operator, $K_{ij}^{kl}=\delta_{ij}\delta^{kl}$.
As the $\mathfrak{so}(4)$ Lie algebra can be written as $\mathfrak{so}(4)\equiv\mathfrak{su}(2)\oplus\mathfrak{su}(2)$
the $R$-matrix has a factorized form 
\begin{equation}
R(u)\cong\frac{u}{u+1}\left(\mathbf{1}+\frac{1}{u}\mathbf{P}\right)\otimes\left(\mathbf{1}+\frac{1}{u}\mathbf{P}\right).
\end{equation}
which carries over to the transfer matrix and the Bethe roots 
\begin{align}
t(u) & =\left(\frac{u}{u+1}\right)^{L}t_{+}(u)\otimes t_{-}(u), & \bigl|\mathbf{u}_{+},\mathbf{u}_{-}\bigr\rangle & =\bigl|\mathbf{u}_{+}\bigr\rangle\otimes\bigl|\mathbf{u}_{-}\bigr\rangle,
\end{align}
such that the eigenvalues can be written in terms of the $\mathfrak{su}(2)$
$Q$-functions as 
\begin{equation}
\Lambda(u)=\left(\frac{u}{u+1}\right)^{L}\Lambda_{+}(u)\Lambda_{-}(u)\quad;\qquad\Lambda_{\pm}(u)=\left(\frac{u+1}{u}\right)^{L}\frac{Q_{\pm}^{[-1]}\left(u\right)}{Q_{\pm}^{[1]}\left(u\right)}+\frac{Q_{\pm}^{[3]}\left(u\right)}{Q_{\pm}^{[1]}(u)}
\end{equation}

\subsection{Spectrum of the $\mathfrak{so}(6)$ spin chain }

The $\mathfrak{so}(6)$ symmetric $R$-matrix can be written as

\begin{equation}
R(u)=\mathbf{1}+\frac{1}{u}\mathbf{P}-\frac{1}{2+u}\mathbf{K}.\label{eq:Rso6}
\end{equation}
The eigenvalue of the transfer matrix with the fundamental representations
of $\mathfrak{so}(6)$ can be written in terms of three types of Bethe
roots with $Q_{1},Q_{\pm}$ as follows \cite{deVega:1986xj}: 
\begin{equation}
\Lambda(u)=\frac{Q_{0}^{[2]}}{Q_{0}}\frac{Q_{1}^{[-1]}}{Q_{1}^{[1]}}+\frac{Q_{1}^{[3]}}{Q_{1}^{[1]}}\frac{Q_{+}}{Q_{+}^{[2]}}\frac{Q_{-}}{Q_{-}^{[2]}}+\frac{Q_{+}}{Q_{+}^{[2]}}\frac{Q_{-}^{[4]}}{Q_{-}^{[2]}}+\frac{Q_{0}^{[2]}}{Q_{0}^{[4]}}\frac{Q_{1}^{[5]}}{Q_{1}^{[3]}}+\frac{Q_{1}^{[1]}}{Q_{1}^{[3]}}\frac{Q_{+}^{[4]}}{Q_{+}^{[2]}}\frac{Q_{-}^{[4]}}{Q_{-}^{[2]}}+\frac{Q_{+}^{[4]}}{Q_{+}^{[2]}}\frac{Q_{-}}{Q_{-}^{[2]}}
\end{equation}
 where $Q_{0}=u^{L}$ and we did not write out explicitly their argument,
which is $u$.

\subsection{Spectrum of the $\mathfrak{su}(2\vert2)_{c}\oplus\mathfrak{su}(2\vert2)_{c}$
spin chain}

This is the spin chain which appears in the asymptotic limit of the
spectral problem in the $AdS_{5}/CFT_{4}$ correspondence. The $R$-matrix
is invariant under the centrally extended $\mathfrak{su}(2\vert2)_{c}$
algebra. Details about notations and conventions can be found in appendix
A. As this is a superalgebra the transfer matrix is the supertrace
of the product of graded $R$-matrices. The full $R$-matrix is the
tensor product of two copies of the $\mathfrak{su}(2\vert2)_{c}$$R$-matrices,
thus the transfer matrix has a factorized form 
\begin{equation}
t(u)=t_{+}(u)t_{-}(u)\label{eq:tAdS}
\end{equation}
Each transfer matrix is related to an $\mathfrak{su}(2\vert2)_{c}$
symmetry and their eigenvectors and eigenvalues can be written in
terms of $y$-roots $\{y_{k}\}_{k=1..N}$ and $w$-roots $\{w\}_{l=1..M}$
\begin{equation}
t_{\pm}(u)\vert\mathbf{y}_{\pm},\mathbf{w}_{\pm}\rangle=\Lambda_{\pm}(u)\vert\mathbf{y}_{\pm},\mathbf{w}_{\pm}\rangle\label{eq:rootsu22}
\end{equation}
These are usually referred to as the left and the right wings. Focusing
only on one of them the eigenvalue on a $2L$ long chain with inhomogeneities
$\{p_{i}\}_{i=1..2L}$ takes the form \cite{Martins:2007hb}
\begin{equation}
\Lambda(u)=e^{-i\frac{p(u)}{2}(N-2L)}\frac{\mathcal{R}^{(+)[1]}}{\mathcal{R}^{(+)[-1]}}\left\{ \frac{\mathcal{R}^{(-)[1]}\mathcal{R}_{y}^{[-1]}}{\mathcal{R}^{(+)[1]}\mathcal{R}_{y}^{[1]}}-\frac{\mathcal{R}_{y}^{[-1]}Q_{w}^{[2]}}{\mathcal{R}_{y}^{[1]}Q_{w}}-\frac{\mathcal{B}_{y}^{[1]}Q_{w}^{[-2]}}{\mathcal{B}_{y}^{[-1]}Q_{w}}+\frac{\mathcal{B}^{(+)[-1]}\mathcal{B}_{y}^{[1]}}{\mathcal{B}^{(-)[-1]}\mathcal{B}_{y}^{[-1]}}\right\} 
\end{equation}
where 
\begin{equation}
\mathcal{R}_{y}(u)=\prod_{j=1}^{N}(x(u)-y_{j})\quad;\quad;\quad Q_{w}(u)=\prod_{l=1}^{M}(u-w_{l})\quad;\qquad\mathcal{R}^{(\pm)}(u)=\prod_{i=1}^{2L}(x(u)-x^{\pm}(p_{i}))
\end{equation}
and the $\mathcal{B}$ quantities can be obtained from the $\mathcal{R}$-s
by replacing $x(u)$ with $1/x(u)$: 
\begin{equation}
\mathcal{B}_{y}(u)=\prod_{j=1}^{N}(1/x(u)-y_{j})\quad;\qquad\mathcal{B}^{(\pm)}(u)=\prod_{i=1}^{2L}(1/x(u)-x^{\pm}(p_{i}))
\end{equation}
Shifts are understood as $x^{[\pm1]}(u)\equiv x^{\pm}(u)=x(u\pm\frac{i}{2g})$
and we assumed that the total momentum vanishes: $\sum_{i}p_{i}=0$.
Bethe ansatz equations for the roots can be obtained from the regularity
of the transfer matrix at $x^{+}(u)=y_{j}$ and $u=w_{l}$, see (\ref{eq:su22BA}).

\section{Selection rules for integrable overlaps}

Boundary states can be analyzed by computing the overlaps with bulk
states. Nonzero overlaps require a pair structure and in the following
we elaborate on the possible structures. In particular, we analyze
the $\mathfrak{su}(N)$, $\mathfrak{so}(4)$, $\mathfrak{so}(6)$
and $\mathfrak{su}(2\vert2)_{c}\oplus\mathfrak{su}(2\vert2)_{c}$
spin chains. Similarly to the large volume QFT spectrum the large
volume overlaps are basically the overlaps in the corresponding spin
chains, i.e. the matrix element of a boundary state $\langle\Psi\vert$
with the eigenstates of the transfer matrix. By generalizing the notion
of integrable boundaries from QFT to spin chains the authors of \cite{Piroli:2017sei}
came up with the definition of an integrable boundary state. This
is a state which is annihilated by the odd conserved charges of the
theory. Since the transfer matrix $t(u)$ generates the conserved
charges the integrability requirement translates into 
\begin{equation}
\left\langle \Psi\right|t(u)=\left\langle \Psi\right|\Pi t(u)\Pi\label{eq:intcond}
\end{equation}
where $\Pi$ is the space reflection operator \cite{Doikou:1998cz}.

In the following we refine this definition and analyze its consequences
for the allowed pair structures appearing in the spin chains introduced
above. 

\subsection{Pair structure in the $\mathfrak{su}(N)$ spin chain\label{subsec:Pair-structure-insuN}}

The parity transformation $\Pi t(u)\Pi$ reverses the order in the
product of $R$-matrices in the definition of the transfer matrix:
\[
\Pi t(u)\Pi=\mathrm{Tr}_{0}R_{01}(u)\dots R_{0L}(u).
\]
Since the R-matrix with fundamental and with anti-fundamental representations
are related by crossing symmetry
\begin{equation}
\bar{R}_{01}(u)=R_{01}^{t_{0}}(-u-N/2)=R_{01}^{t_{1}}(-u-N/2)
\end{equation}
where $t_{0}$ and $t_{1}$ denote transposition in spaces $0$ and
$1$, respectively, the parity transformed transfer matrix can be
related to the transfer matrix where the auxiliary space is the anti-fundamental
representation 
\begin{equation}
\Pi t(u)\Pi=\mathrm{Tr}_{0}\bar{R}_{0L}(-u-N/2)\dots\bar{R}_{01}(-u-N/2)=\bar{t}(-u-N/2)\label{eq:tbar}
\end{equation}
The eigenvalues of the anti-fundamental transfer matrix has the same
structure as the fundamental one and can be written in terms of the
same $Q$-functions: 
\begin{equation}
\bar{t}(u)\bigl|\mathbf{u}^{(a)}\bigr\rangle=\bar{\Lambda}(u)\bigl|\mathbf{u}^{(a)}\bigr\rangle\quad;\qquad\bar{\Lambda}(u)=\sum_{k=1}^{N}\frac{Q_{k-1}^{[N-k-1]}\left(u\right)}{Q_{k-1}^{[N-k+1]}\left(u\right)}\frac{Q_{k}^{[N-k+2]}\left(u\right)}{Q_{k}^{[N-k]}\left(u\right)},
\end{equation}
We now investigate the overlap of an integrable boundary state and
the Bethe state. In doing so we insert the transfer matrices into
the overlap

\begin{equation}
\Lambda(u)\bigl\langle\Psi\bigl|\mathbf{u}^{(a)}\bigr\rangle=\bigl\langle\Psi\bigr|t(u)\bigl|\mathbf{u}^{(a)}\bigr\rangle=\bigl\langle\Psi\bigr|\bar{t}(-u-N/2)\bigl|\mathbf{u}^{(a)}\bigr\rangle=\bar{\Lambda}(-u-N/2)\bigl\langle\Psi\bigl|\mathbf{u}^{(a)}\bigr\rangle,
\end{equation}
Thus the non-vanishing overlap requires $\Lambda(u)=\bar{\Lambda}(-u-N/2)$.
This actually implies the same relation for the fused transfer matrices
and leads to similar relations to each building block, $z_{k}$. This
is equivalent to $Q_{k}(u)=Q_{k}(-u)$, which implies that each type
of root must have the following pair structure:
\begin{equation}
\mathbf{u}^{(a)}=\left\{ u_{1}^{(a)},-u_{1}^{(a)},\dots,u_{n_{a}/2}^{(a)},-u_{n_{a}/2}^{(a)}\right\} ,\qquad\text{for all }a=1,\dots,N-1.
\end{equation}
where here and from now on it is understood that for odd $n_{a}$
we have a zero rapidity. In the following we show that in the other
models the pair structure can be even richer. 

\subsection{Pair structure in the $\mathfrak{so}(4)$ spin chain\label{subsec:Pair-structure-inso4}}

The integrability condition involves the space reflected transfer
matrix, which can be related by the crossing symmetry to the original
transfer matrix as 
\begin{equation}
\Pi t(u)\Pi=t(-u-1).\label{eq:tbar-1}
\end{equation}
Inserting the transfer matrix into the matrix element $\langle\Psi\bigl|\mathbf{u}_{+},\mathbf{u}_{-}\bigr\rangle$
we can easily conclude that the non-vanishing overlap requires that
\begin{equation}
\Lambda(u)\equiv\left(\frac{u}{u+1}\right)^{L}\Lambda_{+}(u)\Lambda_{-}(u)=\Lambda(-u-1).
\end{equation}
Now this integrability requirement (\ref{eq:tbar-1}) can be satisfied
in \emph{two} \emph{different} ways:
\begin{equation}
Q_{\pm}\left(u\right)=Q_{\pm}\left(-u\right)\qquad\text{or}\qquad Q_{+}\left(u\right)=Q_{-}\left(-u\right)
\end{equation}
Accordingly, we can have two different pair structures, what we call
\emph{chiral} and \emph{achiral}: 
\begin{enumerate}
\item Chiral pair structure, where 
\begin{equation}
\mathbf{u}^{(\pm)}=\left\{ u_{1}^{(\pm)},-u_{1}^{(\pm)},\dots,u_{n_{\pm}/2}^{(\pm)},-u_{n_{\pm}/2}^{(\pm)}\right\} .
\end{equation}
\item Achiral pair structure, where ($n_{+}=n_{-}=n$)
\begin{equation}
\mathbf{u}^{(+)}=\left\{ +u_{1},+u_{2},\dots,+u_{n-1},+u_{n}\right\} =-\mathbf{u}^{(-)}=-\left\{ -u_{1},-u_{2},\dots,-u_{n-1},-u_{n}\right\} .
\end{equation}
\end{enumerate}
Thus the naive generalization of the integrability condition is too
''weak'' since it does not fix the pair structure uniquely. The reason
is that we did not use the ''elementary'' transfer matrix, rather
the product of two elementary ones. Based on the ''elementary'' transfer
matrices $t_{\pm}(u)$, we can define two types of integrable states:

\begin{align}
\text{chiral integrable state: }\left\langle \Psi\right|t_{\pm}(u) & =\left\langle \Psi\right|\Pi t_{\pm}(u)\Pi.\label{eq:chiralSO4}\\
\text{achiral integrable state: }\left\langle \Psi\right|t_{+}(u) & =\left\langle \Psi\right|\Pi t_{-}(u)\Pi.\label{achiralSO4}
\end{align}

After a simple calculation one can check that the non-vanishing overlaps
of chiral and achiral integrable states require chiral and achiral
pair structures, respectively.

\subsection{Pair structure in the $\mathfrak{so}(6)$ spin chain\label{subsec:Pair-structure-inso6}}

From the naive integrability condition (\ref{eq:intcond}), one can
derive the following requirement for the eigenvalues of the Bethe
states with non-vanishing overlaps
\begin{equation}
\Lambda(u)=\Lambda(-u-2).\label{eq:naivcondSO6}
\end{equation}
Similarly to the $\mathfrak{so}(4)$ model the integrability condition
(\ref{eq:naivcondSO6}) can be satisfied in two alternative ways:
\begin{enumerate}
\item Chiral pair structure, where 
\begin{equation}
\mathbf{u}^{(1)}=\left\{ u_{1}^{(1)},-u_{1}^{(1)},\dots,u_{n_{1}/2}^{(1)},-u_{n_{1}/2}^{(1)}\right\} \quad;\qquad\mathbf{u}^{(\pm)}=\left\{ u_{1}^{(\pm)},-u_{1}^{(\pm)},\dots,u_{n_{\pm}/2}^{(\pm)},-u_{n_{\pm}/2}^{(\pm)}\right\} ,
\end{equation}
\item Achiral pair structure, where ($n_{+}=n_{-}=n$)
\begin{equation}
\mathbf{u}^{(1)}=\left\{ u_{1}^{(1)},-u_{1}^{(1)},\dots,u_{n_{1}/2}^{(1)},-u_{n_{1}/2}^{(1)}\right\} \quad;\qquad\mathbf{u}^{(+)}=\left\{ +u_{1},+u_{2},\dots,+u_{n-1},+u_{n}\right\} =-\mathbf{u}^{(-)},
\end{equation}
\end{enumerate}
The reason why the integrability condition did not fix completely
the pair structure is similar to the $\mathfrak{so}(4)$ case. Namely,
we did not use the elementary transfer matrices to relate the transfer
matrix and the parity transformed one. The ''elementary'' transfer
matrices $t^{(\pm)}(u)$ corresponds to auxiliary spaces carrying
the spinor representations of the $SO(6)$ group. Let us define these
transfer matrices as
\begin{align}
t^{(\pm)}(u)=\mathrm{Tr}_{0}R_{0L}^{(\pm)}(u)\dots R_{01}^{(\pm)}(u)\quad;\qquad R^{(+)}(u) & =\mathbf{1}+\frac{1}{u+1-\frac{1}{2}}e_{ij}\otimes E_{ji}\\
R^{(-)}(u) & =\mathbf{1}-\frac{1}{u+1+\frac{1}{2}}e_{ij}\otimes E_{ij}
\end{align}
where $e_{ij}$ and $E_{ij}$ are the defining and the six dimensional
representations of $\mathfrak{gl}(4)$. The two representations, and
such a way the two $R$-matrices, are connected by crossing symmetry
\begin{equation}
R^{(+)t}(-u-2)=R^{(-)}(u)
\end{equation}
which implies the connection between the space reflected transfer
matrices
\begin{equation}
\Pi t^{(\pm)}(u)\Pi=\mathrm{Tr}_{0}R_{01}^{(\pm)t}(u)\dots R_{0L}^{(\pm)t}(u)=t^{(\mp)}(-u-2)
\end{equation}
The eigenvalues of $t^{(\pm)}(u)$ can be written in terms of the
$Q$-functions as
\begin{align}
\Lambda^{(+)} & =\frac{Q_{0}^{[3]}}{Q_{0}^{[1]}}\frac{Q_{+}^{[-1]}}{Q_{+}^{[1]}}+\frac{Q_{0}^{[3]}}{Q_{0}^{[1]}}\frac{Q_{1}}{Q_{1}^{[2]}}\frac{Q_{+}^{[3]}}{Q_{+}^{[1]}}+\frac{Q_{-}^{[5]}}{Q_{-}^{[3]}}+\frac{Q_{1}^{[4]}}{Q_{1}^{[2]}}\frac{Q_{-}^{[1]}}{Q_{-}^{[3]}}\quad;\quad\Lambda^{(-)}=\frac{Q_{0}^{[1]}}{Q_{0}^{[3]}}\frac{Q_{+}^{[5]}}{Q_{+}^{[3]}}+\frac{Q_{0}^{[1]}}{Q_{0}^{[3]}}\frac{Q_{1}^{[4]}}{Q_{1}^{[2]}}\frac{Q_{+}^{[1]}}{Q_{+}^{[3]}}+\frac{Q_{-}^{[-1]}}{Q_{-}^{[1]}}+\frac{Q_{1}}{Q_{1}^{[2]}}\frac{Q_{-}^{[3]}}{Q_{-}^{[1]}}
\end{align}
Using these formulas, together with the chiral/achiral pair structures,
we can define two types of integrable initial states\footnote{The particular prefactor is related to our normalization of the $R^{(\pm)}$-matrices.
Clearly by renormalizing them the prefactor can be eliminated.}:
\begin{align}
\text{chiral integrable state: }\left\langle \Psi\right|t^{(\pm)}(u) & =\left\langle \Psi\right|\Pi t^{(\pm)}(u)\Pi.\\
\text{achiral integrable state: }\left\langle \Psi\right|t^{(+)}(u) & =\left(\frac{u+\frac{3}{2}}{u+\frac{1}{2}}\right)^{L}\left\langle \Psi\right|\Pi t^{(-)}(u)\Pi.
\end{align}

Let us finally note that the $\mathfrak{su}(4)$ and $\mathfrak{so}(6)$
spin chains are equivalent. The only difference between the model
of this paragraph and the above investigated $\mathfrak{su}(4)$ model
is that the quantum spaces are in different representations.

\subsection{Pair structure in the $\mathfrak{su}(2\vert2)_{c}\oplus\mathfrak{su}(2\vert2)_{c}$
spin chain}

Similarly to the previous spin chains we can define two types of integrable
states based on how we relate the transfer matrices of the two wings
$t_{\pm}(u)$ to each other: 
\begin{align}
\text{chiral integrable state: }\left\langle \Psi\right|t_{\pm}(u) & =\left\langle \Psi\right|\Pi t_{\pm}(u)\Pi.\\
\text{achiral integrable state: }\left\langle \Psi\right|t_{+}(u) & =\left\langle \Psi\right|\Pi t_{-}(u)\Pi.
\end{align}
For the achiral integrable state roots of the left wing are opposite
to the roots of the right wing: 
\begin{equation}
\mathbf{y}_{+}=-\mathbf{y}_{-}\quad;\qquad\mathbf{w}_{+}=-\mathbf{w}_{-}
\end{equation}
while for chiral integrable states we have within each wing the following
root structure: 
\begin{equation}
\mathbf{y}=\{y_{1},-y_{1},\dots,y_{N/2},-y_{N/2}\}\quad;\qquad\mathbf{w}=\{w_{1},-w_{1},\dots,w_{M/2},-w_{M/2}\}
\end{equation}

\subsection{General structures}

Let us summarize our observations in the previous cases. We have seen
that there exist two types of pair structures, chiral and achiral
and for both integrability conditions can be defined. Based on these
examples it is a natural assumption that the achiral structure is
related to an outer automorphism of the symmetry algebra, i.e. to
a symmetry of the Dynkin diagram. Non-trivial Dynkin diagram symmetries
exist only for the the $\mathfrak{sl}(N)$ and $\mathfrak{so}(2N)$
algebras, and for algebras being the direct sum of two identical copies.
Therefore we expect that achiral pair structures can exist only for
these cases. In the following we show how the chirality of the overlap
can be read off from the symmetries of the $K$-matrices. A more detailed
explanation in the case of rational spin chains can be found in Appendix
\ref{sec:KYBE,-symmetries-and}.

\section{Integrable states from K-matrices and their symmetries}

In the previous section we defined chiral and achiral integrable states,
but it is not clear if they are realized at all. In \cite{Pozsgay:2018dzs}
it was shown that a large class of integrable boundary states can
be obtained from the solution of the KYBE as two-site and matrix product
states
\begin{align}
\left\langle \Psi\right| & =\left\langle \psi_{1}\right|\otimes\left\langle \psi_{2}\right|\otimes\dots\otimes\left\langle \psi_{L/2}\right|,\quad\left\langle \mathrm{MPS}\right|=\mathrm{Tr}\left[\omega_{i_{1}}\omega_{i_{2}}\dots\omega_{i_{L}}\right]\sum_{i_{1},\dots,i_{L}}\left\langle i_{1}\right|\otimes\left\langle i_{2}\right|\otimes\dots\otimes\left\langle i_{L}\right|,
\end{align}
where $\left\langle \psi_{k}\right|\in\mathcal{H}\otimes\mathcal{H}$
and $\left\langle i\right|\in\mathcal{H}$, $\omega_{i}\in\mathrm{End}(V)$,
$\mathcal{H}$ being a one site Hilbert space and $V$ is the boundary
vector space. For two site states we can take
\begin{equation}
\left\langle \psi_{i}\right|=\left\langle a\right|\otimes\left\langle b\right|K_{ab}(u_{i})
\end{equation}
where $K(u)$ is the solution of the KYBE. Indeed, the integrability
condition is satisfied since from the KYBE equation $\left\langle \Psi\right|K_{0}(u)T_{0}^{t_{0}}(u)=\left\langle \Psi\right|\Pi T_{0}(u)\Pi K_{0}(u)$
follows, which after inverting $K_{0}$ and tracing over the auxiliary
space provides the required equation (\ref{eq:intcond}). Matrix product
states can be obtained from specific solutions of the KYBE with inner
degrees of freedom. Indeed if $K_{ab}^{\alpha\beta}$ factorizes as
$K_{ab}^{\alpha\beta}=\omega_{a}^{\alpha\gamma}\omega_{b}^{\gamma\beta}$
then the MPS is integrable due to a similar argument. 

In the following we classify the solutions of the KYBE or BYBE and
reveal how the unbroken symmetry can be related to the chirality of
the overlap. In rational spin chains when the boundary breaks the
original symmetry algebra $\mathfrak{g}$ to $\mathfrak{h}$ then
integrability requires that $(\mathfrak{g},\mathfrak{h})$ has to
be a symmetric pair \cite{Gombor:2019bun}. Thus the residual symmetry
algebra has to form an invariant sub-algebra for a Lie algebra involution
$\alpha$ i.e. $\mathfrak{h}:=\left\{ X\in\mathbb{\mathfrak{g}}|\alpha(X)=X\right\} $.
The residual symmetry $\mathfrak{h}$ is called \emph{regular} if
$\alpha$ is an inner involution, while \emph{special} if it is an
outer one. They are also related to an isometry of the Dynkin diagram.
For inner involutions the symmetry is the identity, while for outer
it acts non-trivially.

In the analysis of rational spin chains it was found that for regular
residual symmetries the BYBEs are always untwisted. For special residual
symmetries one can always find particles which reflect back into a
different representation, such that their BYBE is twisted \cite{Gombor:2019bun}.
In this sense boundaries with regular residual symmetries can be called
untwisted, while those with special symmetries as twisted. This so
far concerned the BYBE. The chirality of the overlap, however is determined
by the nature of the $K$-matrix and the twistedness of the KYBE.
In order to understand this we analyze some examples. 

\subsection{K-matrices in the $\mathfrak{so}(4)$ spin chain}

For the $\mathfrak{so}(4)\cong\mathfrak{su}(2)\oplus\mathfrak{su}(2)$
spin chain, there are two types of solutions of the KYBE, which by
their nature can be called factorizing and non-factorizing. For the
factorizing solutions, the $K$-matrix is factorized in the spinor
basis as
\begin{equation}
K_{ab,a'b'}(u)=\sigma_{aa'}^{i}\sigma_{bb'}^{j}K_{ij}(u)=K_{ab}^{(+)}(u)K_{a'b'}^{(-)}(u)
\end{equation}
where $K^{(\pm)}(u)$ are solutions of the $\mathfrak{su}(2)$ KYBE
see in (\ref{eq:XXXK}) and $\sigma^{i}$-s are the Pauli matrices
for $i=1,2,3$ and $\sigma^{4}=iI$. The two-site state built from
this $K$-matrix satisfy the chiral integrability condition (\ref{eq:chiralSO4})
thus factorizing $K$-matrices correspond to chiral overlaps. The
residual symmetries of these $K$-matrices are factorized $H_{+}\times H_{-}$
where $H_{\pm}$ can be independently either $\mathfrak{u}(1)$ or
$\mathfrak{su}(2)$. The corresponding residual symmetry algebras
are all regular.

The non-factorizing $K$-matrix reads as
\begin{equation}
K_{ab,a'b'}(u)=K_{ad'}^{0}K_{bc'}^{0}R_{a'b'}^{c'd'}(2u)
\end{equation}
where $K^{0}\in\mathrm{End}(\mathbb{C}^{2})$. The two-site state
built from this $K$-matrix satisfies the achiral integrability condition
and the residual symmetry of $K$ is the diagonal $\mathfrak{su}(2)_{D}\cong\mathfrak{so}(3)$,
which is a special sub-algebra. 

For the $\mathfrak{so}(4)$ algebra representations and conjugate
representations are equivalent, thus the twisted nature of the BYBE
and KYBE equations are equivalent, too. Here we found that regular
residual symmetry algebras correspond to chiral, while special ones
to achiral overlaps.

\subsection{K-matrices in the $\mathfrak{so}(6)$ spin chain}

For the $\mathfrak{so}(6)$ spin chain, there are five types of solutions
of the KYBE \cite{Aniceto:2017jor}. They can be classified according
to their symmetries. Since the unbroken symmetry together with $\mathfrak{s0}(6)$
has to form a symmetric pair we have the following possibilities for
the residual symmetry: $\mathfrak{so}(6)$, $\mathfrak{so}(5)$, $\mathfrak{so}(2)\oplus\mathfrak{so}(4)$,
$\mathfrak{so}(3)\oplus\mathfrak{so}(3)$ or $\mathfrak{u}(3)$. The
regular subalgebras are $\mathfrak{so}(6),\mathfrak{so}(2)\oplus\mathfrak{so}(4)$,
$\mathfrak{u}(3)$ while the special ones are $\mathfrak{so}(5)$
and $\mathfrak{so}(3)\oplus\mathfrak{so}(3)$. The full symmetry is
preserved by the identity, while the rest of the explicit $K/R$-matrices
can be found in Table (\ref{O6R}).

\begin{table}[H]
\begin{centering}
\begin{tabular}{|c|c|c|c|}
\hline 
$\mathfrak{so}(5)$ & $\mathfrak{so}(3)\oplus\mathfrak{so}(3)$ & $\mathfrak{so}(2)\oplus\mathfrak{so}(4)$ & $\mathfrak{u}(3)$\tabularnewline
\hline 
special & special & regular & regular\tabularnewline
\hline 
{\footnotesize{}$\left(\begin{array}{cccccc}
a & 0 & 0 & 0 & 0 & 0\\
0 & 1 & 0 & 0 & 0 & 0\\
0 & 0 & 1 & 0 & 0 & 0\\
0 & 0 & 0 & 1 & 0 & 0\\
0 & 0 & 0 & 0 & 1 & 0\\
0 & 0 & 0 & 0 & 0 & 1
\end{array}\right)$} & {\footnotesize{}$\left(\begin{array}{cccccc}
-1 & 0 & 0 & 0 & 0 & 0\\
0 & -1 & 0 & 0 & 0 & 0\\
0 & 0 & -1 & 0 & 0 & 0\\
0 & 0 & 0 & 1 & 0 & 0\\
0 & 0 & 0 & 0 & 1 & 0\\
0 & 0 & 0 & 0 & 0 & 1
\end{array}\right)$} & {\footnotesize{}$\left(\begin{array}{cccccc}
b & c & 0 & 0 & 0 & 0\\
-c & b & 0 & 0 & 0 & 0\\
0 & 0 & 1 & 0 & 0 & 0\\
0 & 0 & 0 & 1 & 0 & 0\\
0 & 0 & 0 & 0 & 1 & 0\\
0 & 0 & 0 & 0 & 0 & 1
\end{array}\right)$} & {\footnotesize{}$\left(\begin{array}{cccccc}
d & u & 0 & 0 & 0 & 0\\
-u & d & 0 & 0 & 0 & 0\\
0 & 0 & d & u & 0 & 0\\
0 & 0 & -u & d & 0 & 0\\
0 & 0 & 0 & 0 & d & u\\
0 & 0 & 0 & 0 & -u & d
\end{array}\right)$}\tabularnewline
$\mathfrak{sp}(4)$ & $\mathfrak{so}(4)$ & $\mathfrak{su}(2)\oplus\mathfrak{su}(2)\oplus\mathfrak{u}(1)$ & $\mathfrak{su}(3)\oplus\mathfrak{u}(1)$\tabularnewline
chiral & chiral & achiral & achiral\tabularnewline
\hline 
\end{tabular}
\par\end{centering}
\caption{Nontrivial solutions of the KYBE for the $\mathfrak{so}(6)$ spin
chain. The first line is the unbroken part of $\mathfrak{so}(6)$,
together with the nature of their remaining symmetries. The parameters
in the solutions are $a=\frac{u-1}{u+1}$, $b=\frac{\frac{1}{2}+d^{2}-u^{2}}{d^{2}+\left(\frac{1}{2}+u\right)^{2}}$,
$c=\frac{du}{d^{2}+\left(\frac{1}{2}+u\right)^{2}}$ and $d$ is a
constant. In the third line the symmetries in the $\mathfrak{su}(4)$
language is shown, while the forth line reflects the type of the overlap. }

\label{O6R}
\end{table}

In order to decide whether these K-matrices define chiral or achiral
integrable states, we have to switch to the $\mathfrak{su}(4)$ description
as we know that chiral boundary states connects the $\mathfrak{su}(4)$
fundamental representations to themselves, but achiral boundary states
involves a conjugation. Since for $\mathfrak{su}(4)$ models particles
and anti-particles transform in different representations there are
two types of BYBE or KYBE, depending on how particles reflect back
off the boundary or annihilated by the boundary state. If the particle
reflects back as an antiparticle the boundary BYBE is a twisted one,
with a special residual symmetry, but the KYBE is an untwisted one
with a chiral overlap. If however the particle reflects back as a
particle then the KYBE is a twisted one, it involves a conjugation
and leads to achiral overlaps. 

\subsection{General Lie algebras }

We have seen that the twisted nature of the residual symmetry, which
is the same for the reflection matrix and the $K$-matrix is purely
determined by the type of the BYBE, while the chiral nature of the
overlap by the type of the KYBE. For algebras such as $\mathfrak{so}(2n+1)$
and $\mathfrak{sp}(2n)$ the Dynkin diagram has no nontrivial symmetry,
all residual symmetries are regular. Both the BYBEs and the KYBEs
are untwisted and the overlaps are chiral. For algebras $\mathfrak{su}(n)$
and $\mathfrak{so}(2n)$ the BYBE has the same nature as the symmetry.
When switching to the KYBE we have to introduce a charge conjugation.
This charge conjugation correspond also to a symmetry of the Dynkin
diagram. For $\mathfrak{su}(n)$ and $\mathfrak{so}(4n+2)$ this is
the same symmetry which defined the twisted BYBE. As a consequence
the nature of the KYBE is just the opposite to that of the BYBE. This
happens when the dotted representation is the contragradient in the
boundary state. In case of $\mathfrak{so}(4n)$, however the charge
conjugation is trivial and the nature of the BYBE and KYBE are the
same. This is summarized in the Table \ref{tab:class}. We elaborate
further on this in Appendix B.

\begin{table}
\begin{onehalfspace}
\begin{centering}
\begin{tabular}{|c|c|c|c|c|}
\hline 
Symmetry algebra & \multirow{1}{*}{Type of sub-algebra} & Explicit subalgebras & conjugation & Pair structure\tabularnewline
\hline 
\hline 
\multirow{2}{*}{$\mathfrak{gl}(N|M)$} & untwisted & $\mathfrak{gl}(n|m)\oplus\mathfrak{gl}(N-n|M-m)$ & non-trivial & achiral\tabularnewline
\cline{2-5} \cline{3-5} \cline{4-5} \cline{5-5} 
 & twisted & $\mathfrak{osp}(N|M)$ & non-trivial & chiral\tabularnewline
\hline 
\hline 
\multirow{2}{*}{$\mathfrak{so}(4k)$} & untwisted & $\mathfrak{so}(2l)\oplus\mathfrak{so}(4k-2l)$, $\mathfrak{u}(2k)$ & trivial & chiral\tabularnewline
\cline{2-5} \cline{3-5} \cline{4-5} \cline{5-5} 
 & twisted & $\mathfrak{so}(2l+1)\oplus\mathfrak{so}(4k-2l-1)$ & trivial & achiral\tabularnewline
\hline 
\hline 
\multirow{2}{*}{$\mathfrak{so}(4k+2)$} & untwisted & $\mathfrak{so}(2l)\oplus\mathfrak{so}(4k+2-2l)$, $\mathfrak{u}(2k+1)$ & non-trivial & achiral\tabularnewline
\cline{2-5} \cline{3-5} \cline{4-5} \cline{5-5} 
 & twisted & $\mathfrak{so}(2l+1)\oplus\mathfrak{so}(4k+1-2l)$ & non-trivial & chiral\tabularnewline
\hline 
\end{tabular}
\par\end{centering}
\end{onehalfspace}
\caption{Regular/special and chiral/achiral classification together with the
conjugation properties.}

\label{tab:class}
\end{table}

\subsection{K-matrices in the $\mathfrak{su}(2\vert2)_{c}\oplus\mathfrak{su}(2\vert2)_{c}$
spin chain}

Similarly to the $\mathfrak{so}(4)$ case the symmetry algebra is
the direct sum of two identical algebras. As a consequence there are
two types of solutions of the KYBE. 

The general non-factorizing $K$-matrix was found in \cite{Jiang:2019xdz,Jiang:2019zig}
and involves the scattering matrix:
\begin{equation}
K_{ab,\dot{a}\dot{b}}(u)=K_{a\dot{d}}^{0}K_{b\dot{c}}^{0}R_{\dot{a}\dot{b}}^{\dot{c}\dot{d}}(u,-u)
\end{equation}
where $K^{0}\in\mathrm{End}(\mathbb{C}^{4})$ is related to the choice
of the basis and $R$ is the $\mathfrak{su}(2\vert2)_{c}$ invariant
$R$-matrix. The two-site state built from this $K$-matrix satisfies
the achiral integrability condition and the residual symmetry is the
diagonal $\mathfrak{su}(2\vert2)_{c}$. 

The other type of solutions are the factorizing ones
\begin{equation}
K_{ab,\dot{a}\dot{b}}(z)=K_{0}(z)K_{ab}(z)K_{\dot{a}\dot{b}}(z)\label{eq:Ksu22}
\end{equation}
where $K_{ab}(z)$ solves the $\mathfrak{su}(2\vert2)_{c}$ KYBE
\begin{equation}
K_{34}(z_{2})K_{12}(z_{1})S_{14}(z_{1},-z_{2})S_{13}(z_{1},z_{2})=K_{12}(z_{1})K_{34}(z_{2})S_{32}(z_{2},-z_{1})S_{42}(-z_{2},-z_{1})
\end{equation}
here we used the rapidity variable instead of the spectral parameter,
but they can be used interchangeably. In the following we solve this
equation and classify all solutions. Motivated by the results for
$\mathfrak{su}(2\vert2)$ without central extension we search the
solutions in the bosonic form 
\begin{equation}
K(z)=K_{0}(z)\left(\begin{array}{cccc}
k_{1}(z) & k_{2}(z) & 0 & 0\\
k_{3}(z) & k_{4}(z) & 0 & 0\\
0 & 0 & h_{1}(z) & h_{2}(z)\\
0 & 0 & h_{3}(z) & h_{4}(z)
\end{array}\right)
\end{equation}

\subsubsection{Leading order solution}

Plugging back the ansatz into the KYBE and making a small $g$ expansion
one finds two classes of solutions, differing in which 2 by 2 block
is nontrivial: 
\begin{equation}
K(z)=\left(\begin{array}{cccc}
0 & 1 & 0 & 0\\
-1 & 0 & 0 & 0\\
0 & 0 & e^{i\frac{p(z)}{2}}h_{1} & e^{i\frac{p(z)}{2}}h_{2}\\
0 & 0 & e^{i\frac{p(z)}{2}}h_{3} & e^{i\frac{p(z)}{2}}h_{4}
\end{array}\right)\quad;\qquad K(z)=\left(\begin{array}{cccc}
k_{1} & k_{2} & 0 & 0\\
k_{3} & k_{4} & 0 & 0\\
0 & 0 & 0 & e^{i\frac{p(z)}{2}}\\
0 & 0 & -e^{i\frac{p(z)}{2}} & 0
\end{array}\right)
\end{equation}
where $h_{1},h_{2}=h_{3},h_{4}$ or $k_{1},k_{2}=k_{3},k_{4}$ are
arbitrary constants. Let us analyze the symmetries of these solutions. 

\subsubsection{Leading order symmetry}

In the weak coupling, $g\to0$, limit the AdS/CFT S-matrix is gauge-equivalent
to the rational $\mathfrak{su}(2|2)$ S-matrix :
\begin{equation}
S(u)=uI_{g}+P\label{eq:LOS}
\end{equation}
where $I_{g}$ and $P$ are the graded unity and permutation operators
and the rapidity is $u=x^{+}+\frac{1}{x^{+}}-\frac{i}{2g}$. The KYBE
is equivalent to the twisted BYBE (or representation changing reflection
equation). Our solutions are consistent with the solutions classified
in \cite{Arnaudon:2004sd}, which consist of constant matrices of
the form 
\begin{equation}
K=\left(\begin{array}{cc}
V_{a} & 0\\
0 & V_{s}
\end{array}\right)\qquad\text{or}\qquad K=\left(\begin{array}{cc}
V_{s} & 0\\
0 & V_{a}
\end{array}\right)
\end{equation}
where $V_{s}$ and $V_{a}$ are arbitrary symmetric and anti-symmetric
2 by 2 matrices. In order to determine the symmetry of the K-matrix
we assume that $V_{s}$ and $V_{a}$ are invertible and focus on the
first case. The symmetry transformations should commute with the scattering
matrix and annihilate the K-matrix 
\begin{equation}
[\Delta(M),S(u)]=0\quad;\quad\Delta(M)K(u)=0\quad;\qquad\Delta(M)=\left(M\otimes I+I\otimes M\right)\label{eq:Ksym}
\end{equation}
where $\otimes$ denotes the graded tensor product for which $\left(a\otimes b\right)\left(c\otimes d\right)=\left(-1\right)^{[b][c]}ac\otimes bd.$
Transformations commuting with the S-matrix (\ref{eq:LOS}) form the
$\mathfrak{gl}(2|2)$ algebra, while those which leave the $K$-matrix
invariant form a super Lie sub-algebra $\mathfrak{h}.$ In order to
identify its defining relations we elaborate (\ref{eq:Ksym})
\begin{align}
\left[\left(M\otimes I+I\otimes M\right)K\right]_{ij} & =\sum_{k}M_{ik}K_{kj}+\sum_{l}(-1)^{[i][j]+[i][l]}M_{jl}K_{il}=\\
 & =\sum_{k}M_{ik}K_{kj}+\sum_{l}(-1)^{[l][j]+[l]}M_{jl}K_{il}=\left[MK+KM^{st}\right]_{ij}\nonumber 
\end{align}
where we used that $K$ is bosonic as a matrix and the super transpose
reads as $\left[M^{st}\right]_{ij}=(-1)^{[i][j]+[i]}M_{ji}.$ Therefore
the symmetry algebra is the $\mathfrak{osp}(2|2)$ algebra, which
is usually defined by the relation
\begin{equation}
M+KM^{st}K^{-1}=0\quad;\qquad M=\left(\begin{array}{cc}
R & S\\
Q & L
\end{array}\right)\label{eq:ospdefrel}
\end{equation}
and, in the particular parametrization, looks like 
\begin{align}
R+V_{s}R^{t}V_{s}^{-1}=0\quad;\qquad L+V_{a}L^{t}V_{a}^{-1}=0\quad;\qquad Q-V_{a}S^{t}V_{s}^{-1} & =0\label{eq:Qtildedef}
\end{align}
From these relations it is obvious that the two bosonic sub-algebras
of $\mathfrak{osp}(2|2)$ are $\mathfrak{so}(2)$ and $\mathfrak{sp}(2)\cong\mathfrak{su}(2)$.

\subsubsection{All loop solutions}

In order to find the all loop solutions one starts with the one-loop
form and promotes the constants to non-trivial functions at higher
orders in $g$. We no longer demand that $k_{2}(z)=k_{3}(z)$ at higher
orders. Equations having particles only with bosonic labels fixes
the ratio of $k_{1}(z)$ and $k_{4}(z)$ to be a constant. One then
selects simple looking equations including $k_{1}(z_{1}),k_{1}(z_{2})$
and $k_{3}(z_{1}),k_{3}(z_{2})$. These equations can be evaluated
at any $z_{2}$ leading to equations for $k_{1}(z_{1})$ and $k_{3}(z_{1})$.
Particularly simple choice is $z_{2}=0$, although one should be careful
as both $x^{\pm}(z)$ go to $0$ for $z\to0$, with the ratio being
$1$. Similar equation can be derived for $k_{2}(z_{1})$ including
$k_{3}(z_{1})$ and $k_{1}(z_{1})$, which finally can be solved leading
to the most general 3 parameter family of solutions
\begin{equation}
k_{1,4}(z)=k_{1,4}\frac{x^{+}(z)(1+x^{-}(z)x^{+}(z))}{x^{-}(z)(A+x^{+}(z)^{2})}\quad;\qquad A=k_{1}k_{4}-k_{2}^{2}
\end{equation}

\[
k_{2}(z)=\frac{x^{+}(z)(k_{2}-x^{+}(z)+x^{-}(z)(A+k_{2}x^{+}(z))}{x^{-}(z)(A+x^{+}(z)^{2})}\quad;\qquad k_{3}(z)=\frac{x^{+}(z)(k_{2}+x^{+}(z)+x^{-}(z)(-A+k_{2}x^{+}(z))}{x^{-}(z)(A+x^{+}(z)^{2})}
\]
We can use the bosonic $\mathfrak{sl}(2)$ symmetries of the $S$-matrix
to bring the solution into some canonical form. The three parameter
family of such symmetries transform the solutions as $k_{1}\to e^{a}k_{1};k_{4}\to e^{-a}k_{4}$,
or as $k_{1}\to a(2k_{2}+ak_{4});k_{2}\to k_{2}+ak_{4}$ or finally
as $k_{2}\to k_{2}+ak_{1};k_{4}\to k_{4}+a(2k_{2}+ak_{1})$ and leaves
$A$ invariant. These transformations can be used to arrange $k_{1}=k_{4}=0$
or $k_{2}=0;k_{1}=k_{4}$. We can thus observe that basically we have
only one free parameter in the solution. 

In order to completely fix this solution one has to fix the overall
scalar factor, which, from the unitarity and crossing unitarity equations,
satisfies the equations 
\begin{equation}
K_{0}(z+\frac{\omega_{2}}{2})K_{0}(-z+\frac{\omega_{2}}{2})=e^{ip(z+\frac{\omega_{2}}{2})}\quad;\qquad K_{0}(z)=S_{0}(z,-z)K_{0}(-z)e^{-2ip(z)}\left(\frac{A+x^{+}(z)^{2}}{A+x^{-}(z)^{2}}\right)^{2}
\end{equation}
 Shifting variables in the first equation we might use $K_{0}(z+\omega_{2})K_{0}(-z)=1$
instead. For the moment we cannot see how these equations could be
easily solved for generic $A$. 

The other type of solution takes the form 
\begin{equation}
h_{1,4}(z)=h_{1,4}\frac{x^{-}(z)(1+x^{-}(z)x^{+}(z))}{x^{+}(z)(A+x^{+}(z)^{2})}\quad;\qquad A=h_{1}h_{4}-h_{2}^{2}
\end{equation}

\[
h_{2}(z)=\frac{x^{-}(z)(h_{2}+x^{-}(z)+x^{+}(z)(-A+h_{2}x^{-}(z))}{x^{+}(z)(A+x^{+}(z)^{2})}\quad;\qquad h_{3}(z)=\frac{x^{-}(z)(h_{2}-x^{-}(z)+x^{+}(z)(A+h_{2}x^{-}(z))}{x^{+}(z)(A+x^{+}(z)^{2})}
\]

Let us now identify the symmetries of these solutions. 

\subsubsection{All loop symmetry}

In the following we show that the solutions above can be obtained
from a centrally extended $\mathfrak{osp}(2\vert2)_{c}$ symmetry.
In doing so we generalize the $\mathfrak{osp}(2|2)$ embedding to
the centrally extended version of $\mathfrak{su}(2|2)$. See the Appendix
for the details of the defining relation of the centrally extended
$\mathfrak{su}(2\vert2)_{c}$ algebra. Following (\ref{eq:Qtildedef})
we define the fermionic generators of $\mathfrak{osp}(2|2)_{c}$ as
\begin{equation}
\tilde{\mathbb{Q}}_{\alpha}^{\:\:a}=\mathbb{Q}_{\alpha}^{\:\:a}+\epsilon_{\alpha\beta}s^{ab}\mathbb{Q}_{b}^{\dagger\:\beta}
\end{equation}
where $s^{ab}$ is any symmetric invertible matrix. These generators
have the following anti-commutation relations 
\begin{multline}
\left\{ \tilde{\mathbb{Q}}_{\alpha}^{\:\:a},\tilde{\mathbb{Q}}_{\beta}^{\:\:b}\right\} =\left\{ \mathbb{Q}_{\alpha}^{\:\:a}+\epsilon_{\alpha\gamma}s^{ac}\mathbb{Q}_{c}^{\dagger\:\gamma},\mathbb{Q}_{\beta}^{\:\:b}+\epsilon_{\beta\delta}s^{bd}\mathbb{Q}_{d}^{\dagger\:\delta}\right\} =\\
\epsilon_{\alpha\beta}\left(s^{ac}\mathbb{R}_{c}^{\:\:b}-s^{bd}\mathbb{R}_{d}^{\:\:a}\right)+s^{ab}\left(\epsilon_{\beta\delta}\mathbb{L}_{\alpha}^{\:\:\delta}+\epsilon_{\alpha\gamma}\mathbb{L}_{\beta}^{\:\:\gamma}\right)+\epsilon_{\alpha\beta}\epsilon^{ab}\mathbb{C}+\epsilon_{\alpha\beta}s^{ac}s^{bd}\epsilon_{cd}\mathbb{C}^{\dagger}=\\
=\epsilon_{\alpha\beta}\epsilon^{ab}\tilde{\mathbb{R}}+s^{ab}\left(\epsilon_{\beta\delta}\mathbb{L}_{\alpha}^{\:\:\delta}+\epsilon_{\alpha\gamma}\mathbb{L}_{\beta}^{\:\:\gamma}\right)+\epsilon_{\alpha\beta}\epsilon^{ab}\tilde{\mathbb{C}}.
\end{multline}
where 
\begin{equation}
\tilde{\mathbb{R}}=s^{ab}\mathbb{R}_{b}^{\:\:c}\epsilon_{ac}\quad;\qquad\tilde{\mathbb{C}}=\mathbb{C}+\text{det}(s^{ab})\mathbb{C}^{\dagger}
\end{equation}
Therefore the generators $\tilde{\mathbb{R}},\mathbb{L}_{\alpha}^{\:\:\beta},\tilde{\mathbb{Q}}_{\alpha}^{\:\:a},\tilde{\mathbb{C}}$
form a centrally extended $\mathfrak{osp}(2|2)_{c}$ algebra: 
\begin{align}
\left[\tilde{\mathbb{R}},\tilde{\mathbb{Q}}_{\alpha}^{\:\:a}\right]=-s^{ab}\epsilon_{bc}\tilde{\mathbb{Q}}_{\alpha}^{\:\:c}\quad;\qquad\left[\mathbb{L}_{\alpha}^{\:\:\beta},\mathbb{J}_{\gamma}\right] & =\delta_{\gamma}^{\beta}\mathbb{J}_{\alpha}-\frac{1}{2}\delta_{\alpha}^{\beta}\mathbb{J}_{\gamma}\quad;\qquad\left[\mathbb{L}_{\alpha}^{\:\:\beta},\mathbb{J}^{\gamma}\right]=-\delta_{\alpha}^{\gamma}\mathbb{J}^{\beta}+\frac{1}{2}\delta_{\alpha}^{\beta}\mathbb{J}^{\gamma},\\
\left\{ \tilde{\mathbb{Q}}_{\alpha}^{\:\:a},\tilde{\mathbb{Q}}_{\beta}^{\:\:b}\right\}  & =\epsilon_{\alpha\beta}\epsilon^{ab}\tilde{\mathbb{R}}+s^{ab}\left(\epsilon_{\beta\delta}\mathbb{L}_{\alpha}^{\:\:\delta}+\epsilon_{\alpha\gamma}\mathbb{L}_{\beta}^{\:\:\gamma}\right)+\epsilon_{\alpha\beta}\epsilon^{ab}\tilde{\mathbb{C}}.
\end{align}
Notice that $\mathcal{V}(p)\otimes\mathcal{V}(-p)$ is a representation
of the non centrally extended $\mathfrak{osp}(2|2)$ since $\tilde{\mathbb{C}}\cdot\mathcal{V}(p)\otimes\mathcal{V}(-p)=0$.

The fundamental S-matrix $S(p_{1},p_{2}):\mathcal{V}(p_{1})\otimes\mathcal{V}(p_{2})\to\mathcal{V}(p_{2})\otimes\mathcal{V}(p_{1})$
commutes with the conserved charges
\begin{equation}
\Delta(\mathbb{J})S(p_{1},p_{2})=S(p_{1},p_{2})\Delta^{op}(\mathbb{J})
\end{equation}
for all $\mathbb{J}\in\mathfrak{su}(2|2)_{c}$. Let us assume that
the K-matrix $K(p)\in\mathcal{V}(p)\otimes\mathcal{V}(-p)$ has $\mathfrak{osp}(2|2)$
symmetry i.e.
\begin{equation}
K(p)\Delta^{op}(\mathbb{J})=0.\label{eq:symK}
\end{equation}

For simplicity, we fix the embedding as
\begin{equation}
s^{ab}=\left(\begin{array}{ll}
0 & s\\
s & 0
\end{array}\right).
\end{equation}
By using bosonic generators the equation (\ref{eq:symK}) fixes the
tensor structure of $K(p)$ as
\begin{equation}
K(p)=e(p)\bigl\langle e_{1}\bigr|\otimes\bigl\langle e_{2}\bigr|+f(p)\bigl\langle e_{2}\bigr|\otimes\bigl\langle e_{1}\bigr|+\bigl\langle e_{3}\bigr|\otimes\bigl\langle e_{4}\bigr|-\bigl\langle e_{4}\bigr|\otimes\bigl\langle e_{3}\bigr|.
\end{equation}
The fermionic generators completely fix $e(p)$ and $f(p)$ as follows:
By applying $\tilde{\mathbb{Q}}_{3}^{\:\:1}=\mathbb{Q}_{3}^{\:\:1}+s\mathbb{Q}_{2}^{\dagger\:4}$
we obtain
\begin{align}
K(p)\Delta^{op}\left(\mathbb{Q}_{3}^{\:\:1}\right) & =e^{-ip/4}\left[\left(e(p)b(-p)-a(p)\right)\bigl\langle e_{1}\bigr|\otimes\bigl\langle e_{4}\bigr|+\left(f(p)b(p)-a(-p)\right)\bigl\langle e_{4}\bigr|\otimes\bigl\langle e_{1}\bigr|\right]\\
K(p)\Delta^{op}\left(\mathbb{Q}_{2}^{\dagger\:4}\right) & =e^{ip/4}\left[\left(e(p)d(-p)-c(p)\right)\bigl\langle e_{1}\bigr|\otimes\bigl\langle e_{4}\bigr|+\left(f(p)d(p)-c(-p)\right)\bigl\langle e_{4}\bigr|\otimes\bigl\langle e_{1}\bigr|\right]
\end{align}
where we used that it is a graded tensor product in moving the fermionic
generators through $\vert e_{3}\rangle$ and $\vert e_{4}\rangle$.
Therefore
\begin{align}
e(p) & =\frac{s^{-1}a(p)e^{-ip/4}+c(p)e^{ip/4}}{-s^{-1}b(p)e^{-ip/4}+d(p)e^{ip/4}}\quad,\qquad f(p)=\frac{s^{-1}a(p)e^{-ip/4}-c(p)e^{ip/4}}{s^{-1}b(p)e^{-ip/4}+d(p)e^{ip/4}}
\end{align}
where we further used that $a(-p)=a(p)=d(p),b(-p)=-b(p)=-c(p).$ We
obtain the same relations for the other fermionic generators. Using
the explicit forms of $a(p)$ and $b(p)$ we obtain 
\begin{align}
e(p) & =e^{ip/2}\frac{s^{-1}x^{-}-1}{x^{+}+s^{-1}}\quad,\qquad f(p)=e^{ip/2}\frac{1+s^{-1}x^{-}}{x^{+}-s^{-1}}.\label{eq:ef}
\end{align}
These agree with the solution of the KYBE, once $k_{1}=k_{4}=0$ and
$k_{2}=s^{-1}$ is chosen.

In summarizing, we found that the factorizing bosonic solutions of
the KYBE must have $\mathfrak{osp}(2\vert2)_{c}$ symmetry. The 3
parameters in the solutions are related how the boundary $\mathfrak{osp}(2\vert2)_{c}$
symmetry is embedded into the centrally extended $\mathfrak{su}(2\vert2)_{c}$
bulk symmetry. 

\section{Asymptotic overlaps and nesting for $K$-matrices}

In this section we demonstrate how K-matrices can be defined for various
levels of the nesting and how this ideas can be used to calculate
factorizing overlaps. 

In calculating the spectrum of a spin chain with a higher rank symmetry
we typically use the nesting method. This means that we start with
an $R$-matrix with symmetry $\mathfrak{g}$ and some representation
where one site states can be labeled as $i=1,\dots,N$. (For simplicity
we can assume an $\mathfrak{su}(N)$ spin chain). In diagonalizing
the transfer matrix we first choose a pseudo vacuum, typically picking
one of the indices say $1$, assuming that the state $\vert0\rangle=\vert1\rangle^{\otimes L}$
is an eigenstate of the transfer matrix. $R_{11}^{11}$ describes
the diagonal scattering of these excitations. Then we introduce $L_{1}$
excitations with labels $j=2,\dots,N$ over this pseudovacuum with
rapidity $u^{(1)}$ 
\begin{equation}
\bigl|u_{1}^{(1)},\dots,u_{L_{1}}^{(1)}\bigr\rangle_{j_{1}\dots j_{N}}\quad;\qquad j_{i}=2,\dots,N
\end{equation}
These excitations propagate over the pseudo vacuum through a diagonal
scattering, $R_{1j}^{1j}$, but scatter on themselves non-trivially
with a reduced $R^{(1)}$-matrix, which we calculate from the exchange
relation
\begin{equation}
\bigl|u_{1}^{(1)},u_{2}^{(1)}\bigr\rangle_{ab}=R_{\,\,ab}^{(1)dc}(u_{1}^{(1)}-u_{2}^{(1)})\bigl|u_{2}^{(1)},u_{1}^{(1)}\bigr\rangle_{cd}
\end{equation}
 In the next step we repeat the procedure for this $R^{(1)}$-matrix
having symmetry $\mathfrak{g}^{(1)}\subset\mathfrak{g}$, by choosing
a second level pseudo vacuum, say $\vert2\rangle^{\otimes L_{1}}$
and diagonal scattering $R_{22}^{(1)22}$ among themselves and second
level excitations $k=3,\dots,N$ with diagonal propagation $R_{2j}^{(2)2j}$
and non-diagonal scatterings $R^{(3)}$. We then carry on this procedure
until it terminates. As a result the spectrum of the transfer matrix
is described in terms of particles with rapidities $u_{j}^{(a)}$
of various nesting levels $a=1,\dots,N-1$, which scatter on each
other diagonally as obtained at each step of the nesting.

The aim of this section is to develop a similar procedure for $K$-matrices
and overlaps. Our procedure is a recursive one which can determine
not only the nested $K$-matrices, but also the generic overlaps if
they are factorizing. In doing so we assume that the square of the
overlap\footnote{By the abuse of terminology sometimes we call the squares as overlaps.}
has the following ``factorizing'' form

\begin{equation}
\frac{\vert\langle\Psi\vert\mathbf{u}^{(a)}\rangle\vert^{2}}{\langle\mathbf{u}^{(a)}\vert\mathbf{u}^{(a)}\rangle}=\prod_{a,i}h^{(a)}(u_{i}^{(a)})\frac{G^{+}}{G^{-}}
\end{equation}
where the norm of the state is given by the Gaudin determinant, $\langle\mathbf{u}^{(a)}\vert\mathbf{u}^{(a)}\rangle=G$
which, for states with a pair structure, can be written into a factorized
form $G=G^{+}G^{-}$ and the overlap function of the nested excitations
at level $a$ are denoted by $h^{(a)}$. Let us note that in the thermodynamic
limit (number of sites goes to infinity) the ratio of determinant
cancels and we can calculate systematically the overlaps in this limit.
As a first step we normalize the $K$-matrix appearing in the boundary
state $\langle\Psi\vert$ for the pseudo vacuum by dividing with $K_{11}$
in order to ensure a normalized boundary state $\langle\Psi^{(1)}\vert0^{(1)}\rangle=1$,
i.e. a normalized overlap with the first level pseudo vacuum $\vert0\rangle\equiv\vert0^{(1)}\rangle$.
Clearly we have to choose the pseudo vacuum, such that it has a nonzero
overlap with the boundary state. We will comment on the importance
of the choice of the pseudo vacuum later. Then the idea is to extract
the nested level $K$-matrix in the limit as
\begin{equation}
K_{ab}^{(1)}(u^{(1)})=\lim_{L\to\infty}\frac{\langle\Psi^{(1)}\vert\bigl|u^{(1)},-u^{(1)}\bigr\rangle_{ab}}{\sqrt{_{ab}\langle u^{(1)},-u^{(1)}\bigl|u^{(1)},-u^{(1)}\bigr\rangle_{ab}}}
\end{equation}
This $K$-matrix, by construction, satisfies the 
\begin{equation}
K_{ab}^{(1)}(u^{(1)})=R_{\,\,ab}^{(1)cd}K_{dc}^{(1)}(-u^{(1)})
\end{equation}
crossing equation. The one-particle overlap of the first level magnon
is simply 
\begin{equation}
h^{(1)}(u^{(1)})=k^{(1)}(u^{(1)})k^{(1)}(u^{(1)})^{*}\quad;\qquad k^{(1)}(u^{(1)})=K_{22}^{(1)}(u^{(1)})
\end{equation}
By dividing $K^{(1)}(u^{(1)})$ with $k^{(1)}(u^{(1)})$ we can build
up $\langle\Psi^{(2)}\vert$ which has a normalized overlap with the
second level pseudo vacuum $\langle\Psi^{(2)}\vert0^{(2)}\rangle=1$.
We then can proceed with the nesting at the second level. This procedure
ends up with the nested $K$-matrices and overlaps $h^{(a)}$. We
demonstrate this method in the following on various models. We use
coordinate space Bethe vectors to calculate the overlap, while explicit
formulas and technical details are relegated to Appendix C.

In the following we verify this method by reconstructing previously
known results for XXX, $\mathfrak{su}(3)$ and $\mathfrak{so}(6)$
spin chains \cite{Pozsgay:2018ixm,deLeeuw:2016umh,deLeeuw:2018mkd}.
After the verification, we propose a new overlap formula for the $\mathfrak{su}(2|2)_{c}$
spin chain using our nesting method.

\subsection{Overlaps in the XXX spin chain}

Let us start with the general integrable two-site state of the XXX
spin chain of size $L$. The elements of the normalized boundary state,
which solves the $\mathfrak{su}(2)$ KYBE, are:

\begin{equation}
K_{11}=1\quad;\quad K_{12}=-ie^{-\theta}(\cosh\beta+2\alpha\sinh\beta)\quad;\quad K_{21}=-ie^{-\theta}(\cosh\beta-2\alpha\sinh\beta)\quad;\quad K_{22}=-e^{-2\theta}.\label{eq:XXXK}
\end{equation}
where $\alpha,\beta$ and $\theta$ are arbitrary parameters. We are
interested in the overlap of the integrable state $\langle\Psi^{(1)}\vert$
with a two magnon state, built over the pseudo-vacuum $\vert1\rangle^{\otimes L}$
in the large $L$ limit. As we have only one nesting level we denote
$u^{(1)}$ by $u$ and the scalar $R^{(1)}$ by $S$. In coordinate
space Bethe ansatz the two magnon state is a plane wave of the form
\begin{equation}
\left|u,-u\right\rangle =\sum_{n_{1}=1}^{L}\sum_{n_{2}=n_{1}+1}^{L}\left(e^{ip(n_{1}-n_{2})}+e^{-ip(n_{1}-n_{2})}S(2u)\right)\left|n_{1}n_{2}\right\rangle 
\end{equation}
where
\begin{align}
p & =-i\log\frac{u-i/2}{u+i/2}\quad,\qquad S(u)=\frac{u+i}{u-i}.
\end{align}
and $\vert n_{1}n_{2}\rangle$ represents a state, in which sites
$n_{1}$ and $n_{2}$ are in state $2$. This state is not symmetric
in the rapidities, it satisfies $\vert u,-u\rangle=S(2u)\vert-u,u\rangle$.
It is also not normalized, the norm is proportional to $L$ in the
large $L$ limit. In calculating the overlap we have to analyze carefully
the parity of $n_{1}$ and $n_{2}$ and their relations. The result
from Appendix C is 

\begin{equation}
\langle\Psi^{(1)}\left|u,-u\right\rangle =\left(\Sigma(p)+\Sigma(-p)S(2u)\right)\left(K_{12}^{2}+K_{21}^{2}+\left(e^{ip}+e^{-ip}\right)K_{12}K_{21}\right)+\frac{L}{2}\left(e^{-ip}+e^{+ip}S(2u)\right)K_{22}.\label{eq:overXXX-1}
\end{equation}
where in the asymptotic limit ($L\to\infty$), after proper regularization,
$\Sigma(p)$ can be written as
\begin{equation}
\Sigma(p)=\frac{L}{2}\frac{1}{e^{2ip}-1}.
\end{equation}
By substituting into (\ref{eq:overXXX-1}) and dividing by the norm
of the state in the $L\to\infty$ limit we obtain that
\begin{equation}
K^{(1)}(u)=k(u)=\frac{1}{L}\left\langle \Psi\right.\left|u,-u\right\rangle =e^{-2\theta}\sinh(\beta)^{2}\frac{u^{2}+\alpha^{2}}{u(u-i/2)}.\label{eq:kfunc-1}
\end{equation}
We can calculate the normalized overlap square leading to 
\begin{equation}
h(u)=k(u)k^{*}(u)=e^{-4\theta}\sinh(\beta)^{4}\frac{(u^{2}+\alpha^{2})^{2}}{u^{2}(u^{2}+\frac{1}{4})}\label{eq:XXXoverlap}
\end{equation}
which is the known one particle overlap function of the XXX spin chain
\cite{Pozsgay:2018ixm}. Thus we provided an alternative calculation
of those results. In the following we check how nesting works. For
this we first analyze an $\mathfrak{su}(3)$ spin chain with $\mathfrak{so}(3)$
symmetry.

\subsection{Overlaps in $\mathfrak{su}(3)$ spin chains with $\mathfrak{so}(3)$
symmetry}

We analyze a two site state and a matrix product state for this model. 

\subsubsection{Two-site state}

We take the integrable two-site state to be 
\begin{equation}
\left\langle \Psi\right|=\left(\left\langle 1\right|\otimes\left\langle 1\right|+\left\langle 2\right|\otimes\left\langle 2\right|+\left\langle 3\right|\otimes\left\langle 3\right|\right)^{\otimes L/2}
\end{equation}
We choose the pseudo-vacuum as $\left|1\right\rangle ^{\otimes L}$
and introduce excitations with labels $2$ and $3$. We would like
to calculate the two-site $K$-matrix, $K^{(1)}(u^{(1)})$ of these
$\mathfrak{su}(2)$ excitations. We read off the $K^{(1)}$-matrix
from an overlap with a two magnon state in Appendix C
\begin{equation}
K_{ab}^{(1)}(u^{(1)}):=\frac{1}{L}\langle\Psi\left|u^{(1)},-u^{(1)}\right\rangle =\frac{1}{2}\left(e^{-ip}\delta_{ab}+e^{ip}R_{ab}^{(1)cc}(2u^{(1)})\right)=\frac{u^{(1)}}{u^{(1)}-i/2}\delta_{ab}
\end{equation}
By defining 
\begin{equation}
k^{(1)}(u):=K_{11}^{(1)}(u)=\frac{u}{u-i/2}
\end{equation}
we can see that it is related as $h^{(1)}(u)=k^{(1)}(u)k^{(1)}(u^{*})^{*}=\frac{u^{2}}{u^{2}+1/4}$
to the one particle overlap function of the $u^{(1)}$ magnons which
agrees with \cite{Piroli:2018ksf,Piroli:2018don}. By normalizing
with this factor the second level state 
\begin{align}
\psi_{ab}^{(2)} & :=\frac{K_{ab}^{(1)}(u)}{k^{(1)}(u)}=\delta_{ab}
\end{align}
is an $SU(2)$ integrable initial state with $\alpha=0$ and $\beta=\theta=i\pi/2$
from which the k-function turns out to be $k^{(2)}(u^{(2)})=\frac{u^{(2)}}{(u^{(2)}-i/2)}$
(see (\ref{eq:XXXoverlap})). Notice that $h^{(2)}(u)=k^{(2)}(u)k^{(2)}(u^{*})^{*}=\frac{u^{2}}{u^{2}+1/4}$
is the one particle overlap function of the $u^{(2)}$ magnons, see
\cite{Piroli:2018ksf,Piroli:2018don}.

\subsubsection{Matrix product state with Pauli matrices}

Here we show that similar ideas can be used also for boundaries with
inner degrees of freedom. Let the MPS be
\begin{equation}
^{\alpha,\beta}\left\langle \mathrm{MPS}\right|=\left[\left(\left\langle 1\right|\sigma_{1}+\left\langle 2\right|\sigma_{2}+\left\langle 3\right|\sigma_{3}\right)^{\otimes L}\right]^{\alpha,\beta}
\end{equation}
where $\alpha,\beta=1,2$ are the ``inner'' indexes of the Pauli
matrices. The pseudo vacuum is $\left|1\right\rangle ^{\otimes L}$
and we calculate the overlap $^{\alpha,\beta}\left\langle \mathrm{MPS}\right|u,-u\rangle_{ab}$.
From the results of the Appendix C we can see that the overlap is
diagonal in $\alpha$ and $\beta$: 
\begin{align}
^{1,1}\left\langle \mathrm{MPS}\right.\left|u^{(1)},-u^{(1)}\right\rangle _{a,b} & =K_{ab}^{(1)+}(u^{(1)})\quad;\qquad{}^{2,2}\left\langle \mathrm{MPS}\right.\left|u^{(1)},-u^{(1)}\right\rangle _{a,b}=K_{ab}^{(1)-}(u^{(1)})
\end{align}
with
\begin{equation}
K^{(1)\pm}(u)=\frac{1}{u}\left(\begin{array}{cc}
u+\frac{i}{2} & \pm\frac{1}{2}\\
\mp\frac{1}{2} & u+\frac{i}{2}
\end{array}\right)\quad;\qquad k^{(1)}(u)=K_{11}^{(1)\pm}(u).
\end{equation}
Notice again that $h^{(1)}(u)=\frac{u^{2}+1/4}{u^{2}}$ and $\psi_{ab}^{(2)\pm}$s
are integrable $SU(2)$ states for inhomogeneous spin chains. Taking
the homogeneous limit, we obtain an integrable two site state with
the parameters
\begin{align}
\theta^{\pm} & =i\frac{\pi}{2}, & \beta^{\pm} & =i\frac{\pi}{2}, & \alpha^{\pm} & =\pm\frac{1}{2}
\end{align}
therefore $h^{(2)}(u)=\frac{u^{2}+1/4}{u^{2}}$. We can see that $h^{(1)}(u)$
and $h^{(2)}(u)$ agree with \cite{deLeeuw:2016umh}.

\subsection{Overlaps in $\mathfrak{so}(6)$ spin chains with $\mathfrak{so}(3)\oplus\mathfrak{so}(3)$
symmetry}

This model is relevant for the weak coupling limit of AdS/dCFT. The
six dimensional one site Hilbert space is parametrized by $\phi_{i}$,
$i=1,\dots,6$ and we introduce the notation 
\begin{align}
Z & =\frac{1}{\sqrt{2}}\left(\phi_{5}+i\phi_{6}\right)\quad,\qquad\bar{Z}=\frac{1}{\sqrt{2}}\left(\phi_{5}-i\phi_{6}\right)
\end{align}
We are going to analyze a two site state and a matrix product state
and point out the importance of the right choice for the pseudovacuum. 

\subsubsection{Two-site state\label{subsec:Two-site-state}}

Let the two-site state be
\begin{equation}
\left\langle \Psi\right|=\left(Z\otimes Z+\bar{Z}\otimes\bar{Z}+\phi_{1}\otimes\phi_{1}-\phi_{2}\otimes\phi_{2}+\phi_{3}\otimes\phi_{3}-\phi_{4}\otimes\phi_{4}\right)^{\otimes L/2}
\end{equation}
We choose the pseudo vacuum as $Z^{\otimes L}$ and excitations are
labeled with $a,b=1,2,3,4$. The excitations have an $\mathfrak{so}(4)=\mathfrak{su}(2)\oplus\mathfrak{su}(2)$
symmetry. The overlap with a two magnon state $\left|u^{(1)},-u^{(1)}\right\rangle _{ab}$
is calculated in the Appendix C to be 
\begin{equation}
K_{ab}^{(1)}(u^{(1)})=\frac{1}{L}\left\langle \Psi\right.\left|u^{(1)},-u^{(1)}\right\rangle _{ab}=\frac{u^{(1)}}{u^{(1)}-i/2}F_{ab}\quad;\qquad F=\mathrm{diag}(1,-1,1,-1)
\end{equation}
This K-matrix satisfies the KYBE and has a factorized form in the
spinor basis 
\begin{equation}
K^{(1)}(u)\cong\frac{u}{u-i/2}\left(\begin{array}{cc}
1 & 0\\
0 & 1
\end{array}\right)\otimes\left(\begin{array}{cc}
1 & 0\\
0 & 1
\end{array}\right)=k^{(1)}(u)\psi^{(L)}\otimes\psi^{(R)}.
\end{equation}
Notice that $h^{(1)}(u)=\frac{u^{2}}{u^{2}+1/4}$ is the one particle
overlap function of the $u^{(1)}$ magnons and $\psi^{(L/R)}$ are
$\mathfrak{su}(2)$ integrable initial states for which the k-functions
are $k^{(L/R)}(u)=\frac{u}{(u-i/2)}$ (see (\ref{eq:XXXoverlap})).
Clearly $h^{(L/R)}(u)=k^{(L/R)}(u)k^{(L/R)}(u^{*})^{*}=\frac{u^{2}}{u^{2}+1/4}$
is the one particle overlap function of the $u^{(L/R)}$ magnons.
We can see that $h^{(1)}(u)$ and $h^{(2)}(u)$ agree with \cite{deLeeuw:2019ebw}.

Let us emphasize that the choice of the direction of the pseudo vacuum
is crucial in obtaining and integrable $K$-matrix in the nesting.
Indeed, the boundary state has the symmetry $\mathfrak{so}(3)\oplus\mathfrak{so}(3)$,
which is broken to $\mathfrak{so}(2)\oplus\mathfrak{so}(2)$ by choosing
the pseudo vacuum as $Z^{\otimes L}$ and $\mathfrak{so}(4)$ excitations
over it. Since $\mathfrak{so}(2)\oplus\mathfrak{so}(2)$ is an integrable
residual symmetry of the $\mathfrak{so}(4)$ model the excitation
$K$-matrix satisfies the KYBE. Rotating the pseudovacuum or equivalently
the two site state as
\[
\phi_{3}\to\cos(\alpha)\phi_{3}+\sin(\alpha)\phi_{5}\quad;\qquad\phi_{5}\to-\sin(\alpha)\phi_{3}+\cos(\alpha)\phi_{5}
\]
the full symmetry of the boundary state $\mathfrak{so}(3)\oplus\mathfrak{so}(3)$
is unchanged, however the excitation symmetry becomes only $\mathfrak{so}(2)$
which is no longer an integrable residual symmetry of the $\mathfrak{so}(4)$
model. The excitation $K$-matrix also has an $\mathfrak{so}(2)$
symmetry only, therefore it cannot be a solution of the KYBE, which
can be checked by an explicit calculation.

Thus the excitation $K$-matrix is meaningful only with the proper
pseudo vacuum. This is not at all surprising from the boundary nested
BA point of view, since the correct choice of pseudo vacuum was important
when the boundary was in space. For reflections in space, the excitation
reflection matrices can be defined only when the labels of the pseudo
vacuum reflects to themselves on the boundary. In \cite{Gombor:2017qsy}
it was shown that the nesting of the $\mathfrak{so}(3)\oplus\mathfrak{so}(3)$
symmetric reflection matrix was related to following symmetry breaking
\[
\left(\mathfrak{so}(6),\mathfrak{so}(3)\oplus\mathfrak{so}(3)\right)\longrightarrow\left(\mathfrak{so}(4),\mathfrak{so}(3)\right).
\]
These two integrable symmetries are related to special residual symmetries.
In contrast, we have just seen above that when the boundary is in
time then the nesting of the residual symmetries is
\[
\left(\mathfrak{so}(6),\mathfrak{so}(3)\oplus\mathfrak{so}(3)\right)\longrightarrow\left(\mathfrak{so}(4),\mathfrak{so}(2)\oplus\mathfrak{so}(2)\right)
\]
which are both related to chiral overlaps. In this example, for space
boundaries and reflections nesting preserves the regular/special residual
symmetries, while for time boundaries and overlaps the chiral/achiral
nature of the pairings. This indicates that the nesting of the K-matrix
can be different when the boundaries are in time or in space.

\subsubsection{Matrix product state with Pauli matrices}

Let the MPS be
\begin{equation}
^{\alpha,\beta}\left\langle \mathrm{MPS}\right|=\left[\left(\sqrt{2}\phi_{1}\sigma_{1}+\sqrt{2}\phi_{3}\sigma_{2}+\sqrt{2}\phi_{5}\sigma_{3}\right)^{\otimes L}\right]^{\alpha,\beta}=\left[\left(\sqrt{2}\phi_{1}\sigma_{1}+\sqrt{2}\phi_{3}\sigma_{2}+(Z+\bar{Z})\sigma_{3}\right)^{\otimes L}\right]^{\alpha,\beta}.
\end{equation}
We take the pseudo vacuum and the excitations as before. The overlaps,
calculated in the Appendix C, turns out to be diagonal in $\alpha$
and $\beta$ with non-vanishing components
\begin{align}
^{1,1}\left\langle \mathrm{MPS}\right.\left|n_{1}n_{2}\right\rangle _{a,b} & =K_{ab}^{(1)+}(u)\quad;\qquad{}^{2,2}\left\langle \mathrm{MPS}\right.\left|n_{1}n_{2}\right\rangle _{a,b}=K_{ab}^{(1)-}(u)
\end{align}
where the K-matrices can be written as
\begin{equation}
K^{(1)\pm}(u)=\left(\begin{array}{cccc}
\frac{u^{2}+iu-1/2}{u(u+i/2)} & 0 & \mp\frac{1}{u} & 0\\
0 & -\frac{u+i}{u+i/2} & 0 & 0\\
\pm\frac{1}{u} & 0 & \frac{u^{2}+iu-1/2}{u(u+i/2)} & 0\\
0 & 0 & 0 & -\frac{u+i}{u+i/2}
\end{array}\right)
\end{equation}
It is factorized in the spinor basis:
\begin{equation}
K^{(1)\pm}(u)=\frac{u+i/2}{u}\left(\begin{array}{cc}
1 & \pm\frac{1}{2u+i}\\
\mp\frac{1}{2u+i} & 1
\end{array}\right)\otimes\left(\begin{array}{cc}
1 & \pm\frac{1}{2u+i}\\
\mp\frac{1}{2u+i} & 1
\end{array}\right).\label{eq:so6mps}
\end{equation}
showing that it is an integrable K-matrix. We can define the k-function
and the $\mathfrak{su}(2)$ two-site states as
\begin{align}
k^{(1)}(u)=\frac{u+i/2}{u}\quad;\quad\psi_{ab}^{(L/R)\pm} & =\left.\left(\begin{array}{cc}
1 & \pm\frac{1}{2u+i}\\
\mp\frac{1}{2u+i} & 1
\end{array}\right)\right|_{u=0}=\left(\begin{array}{cc}
1 & \mp i\\
\pm i & 1
\end{array}\right)\label{eq:so6mps0}
\end{align}
Notice that $h^{(1)}(u)=\frac{u^{2}+1/4}{u^{2}}$ is the one particle
overlap function for the $u^{(1)}$ magnons and $\psi_{ab}^{(2)\pm}$s
are integrable initial states for which the k-function is (see (\ref{eq:XXXoverlap}))
$k^{(2)}(u)=k^{(2)\pm}(u)=\frac{(u+i/2)}{u}.$ Clearly $h^{(2)}(u)=\frac{u^{2}+1/4}{u^{2}}$
is the one particle overlap function for the $u^{(2)}$ magnons. We
can see that $h^{(1)}(u)$ and $h^{(2)}(u)$ agree with \cite{deLeeuw:2018mkd}.

\subsection{Overlaps in $\mathfrak{su}(2\vert2)_{c}$ spin chains}

We would like to analyze an all loop two-site integrable state which
in the weakly coupled limit reproduces the result in the previous
section (\ref{eq:so6mps0}). This can be obtained from 
\begin{equation}
K(p)=\left(\begin{array}{cccc}
k_{1}(p) & k_{2}(p) & 0 & 0\\
k_{3}(p) & k_{4}(p) & 0 & 0\\
0 & 0 & 0 & e^{i\frac{p}{2}}\\
0 & 0 & -e^{i\frac{p}{2}} & 0
\end{array}\right)
\end{equation}
by choosing $k_{1}=k_{4}=s(g)=g^{-1}+\dots$ and $k_{2}=0$. One can
indeed check that the weak coupling expansion of $K(p)$ reproduces
$\psi_{ab}^{(L/R)}(u)$ in the upper 2 by 2 block and gives zero elsewhere.
The all coupling integrable boundary state for $2L$ sites then takes
the form 
\begin{equation}
\left\langle \Psi\right|=\left\langle K(p_{1})\right|\otimes\dots\otimes\left\langle K(p_{L})\right|\quad;\qquad\left\langle K(p)\right|=K_{ij}(p)\left\langle i\right|\otimes\left\langle j\right|\mathbb{I}_{g}.
\end{equation}
The pseudo-vacuum is going to be $\vert1\rangle^{\otimes2L}$ and
excitations are labeled by $3,4$ and $2$. States of the nested Bethe
ansatz are labeled by $\vert\mathbf{y},\mathbf{w}\rangle$ or $\vert\mathbf{p},\mathbf{y},\mathbf{w}\rangle$
if we want to emphasize the dependence on the inhomogeneities $\mathbf{p}$.
The normalized overlap square is assumed to factorize as 
\begin{equation}
\frac{\left|\left\langle \Psi\right|\left.\mathbf{p},\mathbf{y},\mathbf{w}\right\rangle \right|^{2}}{\left\langle \mathbf{p},\mathbf{y},\mathbf{w}\right.\left|\mathbf{p},\mathbf{y},\mathbf{w}\right\rangle }=\prod_{i=1}^{L}h^{p}(p_{i})\prod_{i=1}^{N/2}h^{y}(v_{i})\prod_{i=1}^{M/2}h^{w}(w_{i})\frac{G^{+}}{G^{-}}.
\end{equation}
As a first step we renormalize the boundary state by pulling out the
overlap of the pseudovacuum:
\begin{equation}
\frac{K(p)}{K_{11}(p)}\quad;\qquad h^{p}(p)=\left|K_{1,1}(p)\right|^{2}
\end{equation}
The corresponding state is denoted by $\langle\Psi^{(1)}\vert$. In
the following let us use special inhomogeneities $p_{2k-1}=p$ and
$p_{2k}=p$ for $k=1,\dots L$ as the boundary overlaps does not depend
on the inhomogeneities. These inhomogeneities merely influence how
the second level excitations are propagating and do not effect the
overlaps. Using these special momenta we can calculate the two particle
overlap in the asymptotic limit. Using the explicit form of the two-particle
coordinate space Bethe vectors \cite{deLeeuw:2007akd} we obtain in
Appendix C that 
\begin{equation}
K_{\alpha\beta}^{(1)}(y)=\frac{\left|\langle\Psi^{(1)}\left|y,-y\right\rangle _{\alpha,\beta}\right|^{2}}{_{\alpha\beta}\left\langle y,-y\right.\left|y,-y\right\rangle _{\alpha,\beta}}=\frac{4g^{2}}{s^{2}}\frac{(y^{2}+s^{2})^{2}}{y^{2}+4g^{2}(y^{2}+1)^{2}}\epsilon_{\alpha\beta}
\end{equation}
Observe that we obtained a Dimer state, $\theta=0,\beta=\frac{i\pi}{2},\alpha\to\infty$,
for the inhomogeneous $\mathfrak{su}(2)$ model at the second level
of the nesting. For this state the overlap can be written as (the
number of magnons has to equal to the half length of the spin chain
i.e. $N=2M$) \cite{Bajnok:2020xoz}
\begin{equation}
\frac{\left|\left\langle Dimer\right.\left|\mathbf{y},\mathbf{w}\right\rangle \right|^{2}}{\left\langle \mathbf{y},\mathbf{w}\right.\left|\mathbf{y},\mathbf{w}\right\rangle }=\left(-1\right)^{N/2}\prod_{i=1}^{N/2}(v_{i}^{2}+\frac{1}{4g^{2}})\prod_{i=1}^{\left\lfloor N/4\right\rfloor }\frac{1}{w_{i}^{2}\left(w_{i}^{2}+\frac{1}{4g^{2}}\right)}\frac{\hat{G}^{+}}{\hat{G}^{-}}\quad,\quad v_{i}=y_{i}+y_{i}^{-1}
\end{equation}
with the determinant corresponding to the subchains. Using the results
above, the proposed overlap formula is
\begin{equation}
\frac{\left|\left\langle \Psi\right|\left.\mathbf{p},\mathbf{y},\mathbf{w}\right\rangle \right|^{2}}{\left\langle \mathbf{p},\mathbf{y},\mathbf{w}\right.\left|\mathbf{p},\mathbf{y},\mathbf{w}\right\rangle }=\prod_{i=1}^{L}\left|K_{1,1}(p_{i})\right|^{2}\prod_{i=1}^{N/2}\frac{-(y_{i}^{2}+s^{2})^{2}}{s^{2}y_{i}^{2}}\prod_{i=1}^{\left\lfloor N/4\right\rfloor }\frac{1}{w_{i}^{2}\left(w_{i}^{2}+\frac{1}{4g^{2}}\right)}\frac{G^{+}}{G^{-}}.\label{eq:AdSoverlap}
\end{equation}
where the determinants $G^{\pm}$ can be written in terms of the roots
as follows. Let us parametrize the Bethe Ansatz equations as
\begin{align}
e^{i\phi_{v_{j}}} & =1=\prod_{i=1}^{2L}e^{-ip_{i}/2}\frac{y_{j}-x_{i}^{+}}{y_{j}-x_{i}^{-}}\prod_{k=1}^{M}\frac{v_{j}-w_{k}+\frac{i}{2g}}{v_{j}-w_{k}-\frac{i}{2g}},\quad\text{for all}\quad j=1,\dots,N,\label{eq:su22BA}\\
e^{i\phi_{w_{j}}} & =1=\prod_{i=1}^{N}\frac{w_{j}-v_{i}+\frac{i}{2g}}{w_{j}-v_{i}-\frac{i}{2g}}\prod_{\begin{array}{c}
l=1\\
l\neq j
\end{array}}^{M}\frac{w_{j}-w_{l}-\frac{i}{g}}{w_{j}-w_{l}+\frac{i}{g}},\quad\text{for all}\quad j=1,\dots,M,
\end{align}
where $v_{i}=y_{i}+y_{i}^{-1}$. Using the rescaled Bethe roots $\hat{v}_{i}=gv_{i}$
and $\hat{w}_{i}=gw_{i}$, one can define the Gaudin determinant as
\begin{equation}
G=\left|\begin{array}{cc}
\left(\partial_{\hat{v}_{i}}\phi_{v_{j}}\right)_{N\times N} & \left(\partial_{\hat{v}_{i}}\phi_{w_{j}}\right)_{N\times M}\\
\left(\partial_{\hat{w}_{i}}\phi_{v_{j}}\right)_{M\times N} & \left(\partial_{\hat{w}_{i}}\phi_{w_{j}}\right)_{M\times M}
\end{array}\right|=G_{+}G_{-}.
\end{equation}
For the definition of $G_{+}$ and $G_{-}$, we have to separate two
cases.

\paragraph{M is even}

For even M, the pair structure is
\begin{align}
\mathbf{p} & =\left\{ \mathbf{p}^{+},\mathbf{p}^{-}\right\} , & \mathbf{v} & =\left\{ \mathbf{v}^{+},\mathbf{v}^{-}\right\} , & \mathbf{w} & =\left\{ \mathbf{w}^{+},\mathbf{w}^{-}\right\} ,
\end{align}
where $p_{i}^{+}=-p_{i}^{-}$, $v_{i}^{+}=-v_{i}^{-}$ and $w_{i}^{+}=-w_{i}^{-}$.
The Gaudin-like determinants can be written as
\begin{equation}
G_{\pm}=\left|\begin{array}{cc}
\left(\partial_{\hat{v}_{i}^{+}}\phi_{v_{j}^{+}}\pm\partial_{\hat{v}_{i}^{+}}\phi_{v_{j}^{-}}\right)_{N/2\times N/2} & \left(\partial_{\hat{v}_{i}^{+}}\phi_{w_{j}^{+}}\pm\partial_{\hat{v}_{i}^{+}}\phi_{w_{j}^{-}}\right)_{N/2\times M/2}\\
\left(\partial_{\hat{w}_{i}^{+}}\phi_{v_{j}^{+}}\pm\partial_{\hat{w}_{i}^{+}}\phi_{v_{j}^{-}}\right)_{M/2\times N/2} & \left(\partial_{\hat{w}_{i}^{+}}\phi_{w_{j}^{+}}\pm\partial_{\hat{w}_{i}^{+}}\phi_{w_{j}^{-}}\right)_{M/2\times M/2}
\end{array}\right|.
\end{equation}
where $\partial_{v}=\frac{y^{2}}{y^{2}-1}\partial_{y}$. The Gaudin
matrix is factorized as $G=G_{+}G_{-}.$

\paragraph{M is odd}

For odd M, the pair structure is
\begin{align}
\mathbf{p} & =\left\{ \mathbf{p}^{+},\mathbf{p}^{-}\right\} , & \mathbf{v} & =\left\{ \mathbf{v}^{+},\mathbf{v}^{-}\right\} , & \mathbf{w} & =\left\{ \mathbf{w}^{+},\mathbf{w}^{-},w^{0}\right\} ,
\end{align}
where $p_{i}^{+}=-p_{i}^{-}$, $v_{i}^{+}=-v_{i}^{-}$, $w_{i}^{+}=-w_{i}^{-}$
and $w^{0}=0$. The Gaudin-like determinants can be written in this
case as
\begin{equation}
G_{+}=\left|\begin{array}{ccc}
\left(\partial_{\hat{v}_{i}^{+}}\phi_{v_{j}^{+}}+\partial_{\hat{v}_{i}^{+}}\phi_{v_{j}^{-}}\right)_{N/2\times N/2} & \left(\partial_{\hat{v}_{i}^{+}}\phi_{w_{j}^{+}}+\partial_{\hat{v}_{i}^{+}}\phi_{w_{j}^{-}}\right)_{N/2\times\left\lfloor M/2\right\rfloor } & 2\left(\partial_{\hat{v}_{i}^{+}}\phi_{w^{0}}\right)_{N/2\times1}\\
\left(\partial_{\hat{w}_{i}^{+}}\phi_{v_{j}^{+}}+\partial_{\hat{w}_{i}^{+}}\phi_{v_{j}^{-}}\right)_{\left\lfloor M/2\right\rfloor \times N/2} & \left(\partial_{\hat{w}_{i}^{+}}\phi_{w_{j}^{+}}+\partial_{\hat{w}_{i}^{+}}\phi_{w_{j}^{-}}\right)_{\left\lfloor M/2\right\rfloor \times\left\lfloor M/2\right\rfloor } & 2\left(\partial_{\hat{w}_{i}^{+}}\phi_{w^{0}}\right)_{\left\lfloor M/2\right\rfloor \times1}\\
\left(\partial_{\hat{v}_{i}^{+}}\phi_{w^{0}}\right)_{1\times N/2} & \left(\partial_{\hat{w}_{i}^{+}}\phi_{w^{0}}\right)_{1\times\left\lfloor M/2\right\rfloor } & \partial_{\hat{w}^{0}}\phi_{w^{0}}
\end{array}\right|
\end{equation}
and $G_{-}$ is the same as above. The Gaudin determinant is again
factorized as $G=G_{+}G_{-}.$

We extensively tested these formulas for various sizes $2L=4,6,8$
and $M=2N=2,4$ , numerically, by specifying $\mathbf{p}$ and keeping
$s$ generic. In all the cases we found perfect agreement.

\subsection{Summary}

So far we have conjectured $K$-matrices for nested excitation from
two particle states. The question is how we can generalize the ideas
for more excitations and whether the obtained $K$-matrices are integrable.
We elaborate on this in the following. 

Let us assume we start the nesting with the top level excited Bethe
states and denote them as 
\begin{equation}
\bigl|u_{1},\dots,u_{N}\bigr\rangle_{a_{1}\dots a_{N}}.
\end{equation}
This state satisfies the following exchange relation
\begin{equation}
\bigl|\dots,u_{i},u_{i+1},\dots\bigr\rangle_{\dots a_{i}a_{i+1}\dots}=S_{a_{i}a_{i+1}}^{bc}(u_{1}-u_{2})\bigl|\dots,u_{i+1},u_{i},\dots\bigr\rangle_{\dots cb\dots}\label{eq:perm}
\end{equation}
where $S(u)$ is the scattering matrix of the excitations which satisfy
the YBE and the unitarity relation and derives from the $R$-matrix
of the spin chain. We can define the $N$-particle $K$-matrix as
the matrix element
\begin{equation}
K_{a_{1}b_{1}\dots a_{N/2}b_{N/2}}(u_{1},\dots,u_{N/2})=\bigl\langle\Psi\bigl|u_{1},-u_{1},\dots,u_{N/2},-u_{N/2}\bigr\rangle_{a_{1}b_{1}\dots a_{N/2}b_{N/2}}.
\end{equation}
From the exchange relation (\ref{eq:perm}) we can derive that the
four-particle $K$-matrix automatically satisfy the following equation
\begin{equation}
S_{13}(u_{1}-u_{2})S_{14}(u_{1}+u_{2})K_{3412}(u_{2},u_{1})=S_{42}(u_{1}-u_{2})S_{32}(u_{1}+u_{2})K_{1234}(u_{1},u_{2}).\label{eq:4particleKYBE}
\end{equation}
In order to connect to the previous investigations we take the $L\to\infty$
limit. From equation (\ref{eq:4particleKYBE}), we can see that if
the $N$-particle $K$-matrix factorizes into the product of two-particle
$K$-matrices in the $L\to\infty$ limit as
\begin{equation}
K_{a_{1}b_{1}\dots a_{N/2}b_{N/2}}(u_{1},\dots,u_{N/2})=K_{a_{1}b_{1}}(u_{1})\dots K_{a_{N/2}b_{N/2}}(u_{N/2})
\end{equation}
then the two-particle $K$-matrix automatically satisfies the KYBE.
From the integrability point of view, naively one could think that
the question is whether the four-particle $K$-matrices satisfy the
KYBE or not. However, we can see that this comes from the construction
of the Bethe states and the real question is whether the $N$-particle
$K$-matrix factorizes into two-particle $K$-matrices in the $L\to\infty$
limit or not. This question is not easy to decide and it is not obvious
at all how it is connected to the integrability of the state $\bigl\langle\Psi\bigl|$.

We have already seen that the proper direction of the pseudo vacuum
in the nesting is relevant, which can be supported by symmetry arguments.
Thus we can formulate necessary requirements for the existence of
integrable nested $K$-matrices.

Recall that an integrable state can be labeled by a symmetric pair
$(\mathfrak{g},\mathfrak{h})$ where $\mathfrak{g}$ is the symmetry
algebra of the spin chain and $\mathfrak{h}$ is the residual symmetry
which annihilates the state. For the top-level excitations, these
symmetries are reduced to $(\mathfrak{g}^{(1)},\mathfrak{h}^{(1)})$.
If the excitation $K$-matrix satisfies the KYBE (i.e. it is factorized,
then $(\mathfrak{g}^{(1)},\mathfrak{h}^{(1)})$ has to be a symmetric
pair, too. The consistency of the pair structures requires that both
symmetries $(\mathfrak{g},\mathfrak{h})$ and $(\mathfrak{g}^{(1)},\mathfrak{h}^{(1)})$
have to belong to the same chiral or achiral pair structures. Since
the pair $(\mathfrak{g}^{(1)},\mathfrak{h}^{(1)})$ depends on the
choice of the pseudo-vacuum the factorizability depends on the choice
of the pseudo vacuum, too. We can repeat the analysis for each next
nesting level. 

\section{Application to one-point functions in AdS/CFT}
\begin{onehalfspace}
\begin{flushleft}
\begin{table}
\begin{onehalfspace}
\begin{centering}
\begin{tabular}{|c|c|c|c|c|c||c|c|c|c|c||c|}
\hline 
 & $x^{0}$ & $x^{1}$ & $x^{2}$ & $x^{3}$ & $x^{4}$ & $x^{5}$ & $x^{6}$ & $x^{7}$ & $x^{8}$ & \multirow{1}{*}{$x^{9}$} & symmetries\tabularnewline
\hline 
\hline 
D3 & $\newmoon$ & $\newmoon$ & $\newmoon$ & $\newmoon$ &  &  &  &  &  &  & \tabularnewline
\hline 
\hline 
D5 & $\newmoon$ & $\newmoon$ & $\newmoon$ &  & $\newmoon$ & $\newmoon$ & $\newmoon$ &  &  &  & $\mathfrak{osp}(4\vert4)$\tabularnewline
\hline 
D7 & $\newmoon$ & $\newmoon$ & $\newmoon$ &  & $\newmoon$ & $\newmoon$ & $\newmoon$ & $\newmoon$ & $\newmoon$ &  & $\mathfrak{so}(2,3)\times\mathfrak{so}(5)$\tabularnewline
\hline 
\end{tabular}
\par\end{centering}
\end{onehalfspace}
\caption{Various brane structures on $AdS_{5}\times S^{5}$ and their symmetries.}
\label{tab:branes}
\end{table}
In culminating all the investigations in the previous sections we
investigate the defect $\mathcal{N}=4$ SYM with the aim of providing
all loop asymptotic one-point functions for local gauge invariant
operators. There are two types of defects which are integrable for
the scalar sector at tree level. Their gravity duals belong the following
$D$-brane configurations: 
\par\end{flushleft}
\end{onehalfspace}
\begin{enumerate}
\item D5-brane which wraps around $AdS_{4}$ and $S_{2}$ (see table \ref{tab:branes}).
\item D7-brane which wraps around $AdS_{4}$ and $S_{4}$ (see table \ref{tab:branes}).
\end{enumerate}
The symmetry of the $D5$ brane is $\mathfrak{osp}(4|4)$ while that
of the $D7$ brane is $\mathfrak{so}(2,3)\times\mathfrak{so}(5)$.
Since in the integrability argumentations the Lorentzian signature
is irrelevant we do not write it out explicitly, i.e. we write $\mathfrak{su}(2,2)\equiv\mathfrak{su}(4)$
and $\mathfrak{so}(2,3)\equiv\mathfrak{so}(5)$. At week coupling
the spectrum of single trace operators can be described by an $\mathfrak{gl}(4|4)$
spin chain. The tree level one-point functions can be obtained from
overlaps between one-loop Bethe states and boundary states. Since
integrable states belong to symmetric pairs we can easily decide which
configurations can be integrable. The pair $\left(\mathfrak{gl}(4|4),\mathfrak{osp}(4|4)\right)$
is a symmetric pair but $\left(\mathfrak{gl}(4|4),\mathfrak{so}(5)\oplus\mathfrak{so}(5)\right)$
is not, therefore this argument suggests that the D7 configuration
may be not integrable for the full spectrum. In the following we investigate
only the D5 defect configuration.

Since the D5 defect corresponds to the symmetric pair $\left(\mathfrak{gl}(4|4),\mathfrak{osp}(4|4)\right)$
it must have chiral pair structures. As a consequence the excitation
$K$-matrix must have the same chiral pair structure, too. For the
$AdS_{5}/CFT_{4}$ $S$-matrix, there are factorisable and non-factorisable
$K$-matrices which have chiral and achiral pair structures, respectively.
Thus we have to use the factorisable $K$-matrix which have residual
symmetry $\mathfrak{osp}(2|2)\oplus\mathfrak{osp}(2|2)$. This is
exactly what we determined in section 5. This argument is used the
integrability requirement only. Let us continue now with the explicit
symmetries of the top level excitations.

The bosonic symmetries of the D5 defect are $\mathfrak{so}(5)\oplus\mathfrak{so}(3)\oplus\mathfrak{so}(3)$,
where $\mathfrak{so}(5)$ comes form the conformal while $\mathfrak{so}(3)\oplus\mathfrak{so}(3)$
from the $R$-symmetry. Let us continue with the symmetries of the
excitations over the pseudo-vacuum $\mathrm{Tr}Z^{L}.$ This pseudo-vacuum
breaks the conformal symmetry to Lorentz symmetry $\mathfrak{so}(1,3)\equiv\mathfrak{so(4)}$
and the $R$-symmetry to $\mathfrak{so}(4)$. The D5 defect breaks
the Lorentz symmetry to $\mathfrak{so}(1,2)\equiv\mathfrak{su}(2)$
which is the diagonal algebra of the Lorentz algebra $\mathfrak{so}(1,3)\equiv\mathfrak{su}(2)\oplus\mathfrak{su}(2)$.
The residual $R$-symmetry depends on the orientation $D$-brane.
It can be $\mathfrak{so}(2)\oplus\mathfrak{so}(2)$, $\mathfrak{so}(3)$
or $\mathfrak{so}(2)$. Clearly only the residual $R$-symmetry $\mathfrak{so}(2)\oplus\mathfrak{so}(2)$
is consistent with the symmetry algebra $\mathfrak{osp}(2|2)\oplus\mathfrak{osp}(2|2)$
which is obtained from the integrability argument. We can see however
that the residual Lorentz symmetry, the diagonal $\mathfrak{su}(2)$,
is not factorized as it is required. This problem is similar to choosing
a pseudo vacuum with a non-proper direction (see subsection \ref{subsec:Two-site-state}).
The problem can be cured by global conformal transformations. From
symmetry argumentations we saw that if and integrable $K$-matrix
exists then the defect must respect the full Lorentz symmetry, therefore
we have to use a symmetry transformation which transforms the plane
defect to a spherical one. This can be done by using special conformal
transformations.

Now let us focus on the tree level overlaps. In the SYM theory with
defects some of the scalar fields require nonzero vacuum expectation
values. In the simplest case they are \emph{
\begin{align}
\phi_{i}(x) & =\frac{\sigma_{i}}{x_{4}}\quad;\quad i=1,2,3 & \phi_{i}(x) & =0\quad;\quad i=4,5,6.
\end{align}
}At tree level the excitation $K$-matrix in the $\mathfrak{so}(2,4)$
sector is zero and the $K$-matrix of the $\mathfrak{so}(6)$ sector
is given in (\ref{eq:so6mps}). In the inner boundary space in which
the Pauli matrices act it is diagonal, thus a direct sum of two scalar
$K$-matrices. In order to get these scalar $K$-matrices we have
to put together the $\mathfrak{so}(2,4)$ and $\mathfrak{so}(6)$
subsectors. This means we have to find rational scalar $\mathfrak{su}(2|2)$
$K$-matrices with bosonic symmetry $\mathfrak{su}(2)\oplus\mathfrak{u}(1)$.
This symmetry constrains the matrix structure as
\begin{equation}
\left(\begin{array}{cccc}
1 & B & 0 & 0\\
-B & 1 & 0 & 0\\
0 & 0 & 0 & A\\
0 & 0 & -A & 0
\end{array}\right)\quad;\qquad\begin{array}{c}
A=0\\
B=\frac{c}{i-2u}
\end{array}.
\end{equation}
Since the tree level $K$-matrix is zero at the $\mathfrak{so}(2,4)$
sector, we have to choose $A=0$. Substituting to the KYBE, we obtain
that $B(u)=\frac{c}{i+2u}$ is the most general solution. From comparing
to (\ref{eq:so6mps}) we have to choose $c=\pm1$, where the signs
are related to the inner degrees of freedom. In extending this tree
level $K$-matrix for AdS/dCFT at all loops we have to choose a solution
of the KYBE equation of the form: 
\begin{equation}
K(p)=\left(\begin{array}{cccc}
k_{1}(p) & k_{2}(p,s) & 0 & 0\\
k_{2}(p) & k_{4}(p,s) & 0 & 0\\
0 & 0 & 0 & e^{i\frac{p}{2}}\\
0 & 0 & -e^{i\frac{p}{2}} & 0
\end{array}\right)\quad;\qquad\begin{array}{c}
k_{1}=k_{4}=\frac{sx^{+}(1+x^{+}x^{-})}{x^{-}(s^{2}+(x^{+})^{2})}\\
k_{2}=\frac{x^{+}(s^{2}x^{-}-x^{+})}{x^{-}(s^{2}+(x^{+})^{2})}\\
k_{3}=\frac{x^{+}(x^{+}-s^{2}x^{-})}{x^{-}(s^{2}+(x^{+})^{2})}
\end{array}\label{eq:K1pt}
\end{equation}
Indeed, choosing $s=\pm\frac{1}{2}g^{-1}+O(1)$ and expanding at weak
coupling we reproduce the tree level result. Unfortunately we cannot
see at the moment how the function $s(g)$ could be fixed from some
symmetry argumentations. One possible way is to compare the all loop
asymptotic overlaps with explicit string theory or higher loop YM
calculations.

The full $\mathfrak{su}(2|2)_{c}\oplus\mathfrak{su}(2|2)_{c}$ scalar
$K$-matrix can be written as a tensor product of two identical copies
of (\ref{eq:K1pt}) as written in (\ref{eq:Ksu22}). The full transfer
matrix also has a factorized form (\ref{eq:tAdS}) and each factor
can be diagonalized independently. The eigenvalues can be written
in terms of Bethe roots (\ref{eq:rootsu22}). We denote the Bethe
roots of the left wing by $\mathbf{y}^{(1)},\mathbf{w}^{(1)}$, while
that of the right wing by $\mathbf{y}^{(2)},\mathbf{w}^{(2)}$. Within
each wing they all satisfy their own Bethe ansatz equations (\ref{eq:su22BA}).
The full transfer matrix eigenvalue depends additionally on the physical
momenta which connect the Bethe roots on the two sides by the middle
node BA equation 
\begin{equation}
e^{i\phi_{p_{j}}}=1=e^{ip_{j}(L-N+\frac{M_{1}+M_{2}}{2}+1)}\prod_{i=1;i\neq j}^{N}e^{ip_{i}}\frac{x_{j}^{+}-x_{i}^{-}}{x_{j}^{-}-x_{i}^{+}}\frac{1-\frac{1}{x_{j}^{+}x_{i}^{-}}}{1-\frac{1}{x_{j}^{-}x_{i}^{+}}}\sigma(p_{j,}p_{i})^{2}\prod_{\nu=1}^{2}\prod_{k=1}^{M_{1}}\frac{x_{j}^{-}-y_{k}^{(\nu)}}{x_{j}^{+}-y_{k}^{(\nu)}},\quad\text{for all}\quad j=1,\dots,N.
\end{equation}
where the number of $p$, $y^{(\alpha)},w^{(\alpha)}$ variables are
$N,M_{\alpha},K_{\alpha}$, respectively \cite{Martins:2007hb}. Let
us denote the eigenvector of the full transfer as $\vert\mathbf{p},\mathbf{y}^{(\alpha)},\mathbf{w}^{(\alpha)}\rangle.$
Based on the overlap formula we obtained in the previous section (\ref{eq:AdSoverlap})
we conjecture the overlap for the full spectrum to take the form 
\begin{align}
\frac{\left|\left\langle \Psi\right|\left.\mathbf{p},\mathbf{y}^{(\alpha)},\mathbf{w}^{(\alpha)}\right\rangle \right|^{2}}{\left\langle \mathbf{p},\mathbf{y}^{(\alpha)},\mathbf{w}^{(\alpha)}\right.\left|\mathbf{p},\mathbf{y}^{(\alpha)},\mathbf{w}^{(\alpha)}\right\rangle }=\prod_{i=1}^{N/2}\left|K_{0}(p_{1})\right|^{2}\left|k_{1}(p_{i})\right|^{4} & \prod_{\nu=1}^{2}\prod_{i=1}^{M_{\nu}/2}\frac{-((y_{i}^{(\nu)})^{2}+s^{2})^{2}}{s^{2}(y_{i}^{(\nu)})^{2}}\prod_{i=1}^{\left\lfloor M_{\nu}/4\right\rfloor }\frac{1}{(w_{i}^{(\nu)})^{2}\left((w_{i}^{(\nu)})^{2}+\frac{1}{4g^{2}}\right)}\frac{G_{+}}{G_{-}}
\end{align}
where the new determinants involve differentiation wrt. the momenta,
too: 
\begin{equation}
G_{\pm}=\left(\partial_{\hat{U}_{i}^{+}}\phi_{U_{j}^{+}}\pm\partial_{\hat{U}_{i}^{+}}\phi_{U_{j}^{-}}\right)
\end{equation}
Here $\mathbf{U}^{+}=\left\{ \mathbf{p}^{+},\mathbf{v}^{(\alpha)+},\mathbf{w}^{(\alpha)+}\right\} $
collects all the variables and $\mathbf{\hat{U}}=\left\{ \mathbf{u}^{+},\hat{\mathbf{w}}^{(\alpha)+},\hat{\mathbf{w}}^{(\alpha)+}\right\} $
is the collection of the properly normalized rapidities. In particular
$u_{i}$ is the rapidity parameter belonging to $p_{i}$. For simplicity
we assumed that $K_{1},K_{2}$ are even.

This is the contribution of the scalar $K$-matrices, out of which
we have two in the simplest case. If their $s$-parameter is the opposite
of each other their contributions just double.

\section{Conclusions}

In this paper we analyzed integrable boundary states, overlaps, nesting
and their applications in bootstrapping the simplest asymptotic all
loop 1-point functions in AdS/dCFT. We started by formulating the
YBE for the boundary state (KYBE) which can annihilate pairs of particles
corresponding to different representations. We called this boundary
state twisted in order to distinguish from the one which annihilates
particles in the same representation, which is called untwisted. This
twisting is related to a symmetry of the Dynkin diagram of the full
symmetry, which can be charge conjugation (exchanging representation
with contragradient representation) or some other involutions of the
algebra. We then showed that for each solution of the KYBE one can
associate a reflection matrix in the mirror theory, which solves the
boundary YBE. Since the crossing involves a charge conjugation the
BYBE is untwisted only if the conjugated twisted transformation is
a trivial one, otherwise it is twisted.

We then turned to the investigation of the overlap of the finite size
boundary state with a periodic multiparticle state, which is an eigenstate
of the transfer matrix. Eigenvalues and eigenvectors can be formulated
in terms of Bethe roots, which satisfy the Bethe equations, following
from the regularity of the transfer matrix. The usual definition of
the integrable finite size boundary states demands that the difference
of the transfer matrix and the parity transformed transfer matrix
annihilates the boundary state. We used this definition to derive
a pair structure between Bethe roots. In doing so we observed that
the definition does not fix uniquely the type of the pair structure.
For theories with extra symmetries, corresponding to symmetries of
the Dynkin diagram, two different pair structures are allowed. If
the Bethe roots are paired within each type we called the overlap
chiral. If however, different Bethe roots, related by the Dynkin symmetry
were paired, we called the overlap achiral. 

We then focused on boundary states which are built up from $K$-matrices,
solutions of the KYBE. We could relate the chirality of the overlap
to the twisted nature of the KYBE, i.e. twisted KYBE leads to achiral
overlaps, while untwisted equations to chiral ones. The chirality
property of the overlap can be related to the unbroken symmetries.
These symmetries are the same for $K$-matrices and the corresponding
mirror reflections, and has been classified for spin chains. It was
found that the residual symmetry of the BYBE together with the symmetry
of the bulk theory must form a symmetric pair. These symmetric pairs
are classified and they all are related to involutions of the algebra.
It was found that for inner involutions the BYBE equations are not
twisted, while for outer involutions they are twisted and the twist
is related to a nontrivial symmetry of the Dynkin diagram. We then
analyzed in detail how the twisted nature of the BYBE, can be related
to the twisted nature of the KYBE. As a result we could relate the
chirality of the overlap to the type of the unbroken symmetry. This
was all crucial for formulating the nesting program for overlaps.
This analysis was for spin chains, but we wanted also to see how it
extends to AdS/CFT. For this reason we determined the most general
solution of the KYBE for the centrally extended $\mathfrak{su}(2\vert2)_{c}$
scattering matrix. We found that it has three parameters, out of which
two can be transformed away. The corresponding solution had an $\mathfrak{os}(2\vert2)_{c}$
symmetry, which together with $\mathfrak{su}(2\vert2)_{c}$ formed
a symmetric pair, the only one of this kind.

The next step was the calculation of the overlap formulas. In doing
so we suggested a completely new and original way how nesting could
be used for overlaps and $K$-matrices. The framework was the nested
Bethe ansatz, in which at each step a pseudo vacuum is chosen and
excitation with smaller symmetry and reduced scattering matrix is
identified. We followed the same idea and defined the nested $K$-matrices
by the overlap of the infinite volume two particle state with the
boundary state. This two particle state was a coordinate space BA
eigenstate of the nested excitations. In choosing the right pseudo
vacuum and excitations the residual symmetries played a crucial role.
Indeed, the boundary state determines the symmetries of the problem,
which fixes the chirality of the overlaps. At each step we have to
choose such a pseudo vacuum whose symmetry is in the same chirality
class. We tested these ideas for various spin chains relevant also
for 1-point functions for $D5$ branes in AdS/dCFT. We then carried
out this program for the newly calculated $\mathfrak{su}(2\vert2)_{c}$
$K$-matrix. 

As an application of our results we investigated the symmetries of
various D-branes in the AdS/dCFT setting. We found that the D5 brane
have the chance to be integrable at leading order in the coupling
for the whole theory, not only for a subsector. We identified the
leading order $K$-matrix, in case of the simplest defect with matrix
product states of Pauli matrices. The symmetry investigations suggested
that in order the whole theory be integrable at finite couplings the
symmetry related to the Lorentz transformations has to be enhanced.
We speculated about a mechanism, how this can happen. If it happens
our proposal describes the asymptotic 1-point functions for any couplings
and any sector of the theory. We thus spelled out our conjectured
overlaps corresponding to this $K$-matrix in terms of the Bethe roots
of the $\mathfrak{su}(2\vert2)_{c}$ algebra together with the ratio
of Gaudin determinants. We tested this formula for various sizes and
Bethe roots and found convincing evidence of its correctness.

Our solution for the $K$-matrix and overlaps is just the first step
in solving completely the 1-point functions in AdS/dCFT. First of
all we found a free parameter in the solution which should be related
to some physical data. Also we did not calculate the prefactor of
the solution, which should be fixed from unitarity of the mirror reflection
factor and crossing symmetry of the $K$-matrix. In order to perform
these tasks many new data are needed for AdS/dCFT both from the weak
and strong coupling type. They should extend the available overlaps
both in the coupling and also for larger sectors of the theory.

The situation we analyzed was the simplest possible boundary state
as far as the boundary degrees of freedom was concerned. It would
be interesting to understand how it could be extended for higher dimensional
representations and boundary spaces. Particularly interesting problem
is the calculation of the overlaps from the CFT side. Especially the
D7 brane, which seems to be integrable at 1 loop for the subsector
considered so far, but does not seem to be integrable from our point
of view.

Our analysis provides an asymptotic overlap, which does not include
finite size wrapping effects. These finite size effects are due to
virtual particles and can be dealt with using the thermodynamic Bethe
ansatz \cite{Jiang:2019xdz,Jiang:2019zig}. Recent works \cite{Kostov:2018dmi,Kostov:2019fvw,Kostov:2019sgu,Caetano:2020dyp}
on excited state g-functions can help to include such corrections
also in our case.

Our developments are relevant also for spin chains. The new nesting
procedure we initiated was tested only in a few examples. It would
be very nice to extend systematically the calculations for any algebras
and for all boundary conditions. 

\subsection*{Note added}

While our draft was finalized the paper \cite{Komatsu:2020sup} appeared
on the arxiv, with partially overlapping results. The paper \cite{Komatsu:2020sup}
has two parts: The first calculates the 1-point functions of BPS operators
for the D5 system using supersymmetric localization. The second deals
with non-BPS operators in the bootstrap setting. By identifying the
unbroken symmetries the authors calculate the solution of the KYBE,
which is relevant for the defect problem and use crossing symmetry
and unitarity to fix the scalar factor. In order to apply for AdS/dCFT
and to fix the CDD ambiguity they analyze the asymptotic overlaps
in the $\mathfrak{su}(2)$ sector. By comparing with 1-loop calculations
they fix the scalar factor and use excited boundary states to describe
$\mathfrak{su}(2)$ overlaps for a large class of defects.

Our paper analyzed AdS/dCFT and spin chains in the same time, and
formulated a nesting program how the generic overlaps can be calculated.
We found two types of solutions of the KYBE, and calculated the generic
asymptotic overlaps for all sectors, which is relevant for the simplest
AdS/dCFT setting.

We are overlapping with \cite{Komatsu:2020sup} with the calculation
of the K-matrix from symmetries. Indeed our K-matrix (\ref{eq:ef})
is equivalent to their (4.30), once we change the sign of the momentum
originating from a different definition of the S-matrix and take into
account that their S-matrix uses a different phase factor for bosons.

All other results we obtained are independent. Actually their results
nicely extends ours by determining the scalar factor and by fixing
our free parameter in the solution. By combining the two results the
asymptotic all loop 1-point functions for all sectors are available
now.

\subsection*{Acknowledgments}

We thank Balázs Pozsgay and Shota Komatsu for discussions and the
NKFIH research Grant K116505 for supports. We thank the Hungarian
Academy of Sciences for providing us their infrastructure for free.

\appendix

\section{Notations and conventions for $\mathfrak{su}(2\vert2)_{c}$}

In this Appendix we summarize in a selfcontained way the notations
and convention we used for the $AdS_{5}/CFT_{4}$ integrable model
thorough the paper. This model has an $\mathfrak{su}(2\vert2)_{c}\oplus\mathfrak{su}(2\vert2)_{c}$
symmetry, where the centrally extended $\mathfrak{su}(2|2)_{c}$ algebra
is defined by the relations

\begin{align}
\left[\mathbb{R}_{a}^{\:\:b},\mathbb{J}_{c}\right] & =\delta_{c}^{b}\mathbb{J}_{a}-\frac{1}{2}\delta_{a}^{b}\mathbb{J}_{c}, & \left[\mathbb{L}_{\alpha}^{\:\:\beta},\mathbb{J}_{\gamma}\right] & =\delta_{\gamma}^{\beta}\mathbb{J}_{\alpha}-\frac{1}{2}\delta_{\alpha}^{\beta}\mathbb{J}_{\gamma},\nonumber \\
\left[\mathbb{R}_{a}^{\:\:b},\mathbb{J}^{c}\right] & =-\delta_{a}^{c}\mathbb{J}^{b}+\frac{1}{2}\delta_{a}^{b}\mathbb{J}^{c}, & \left[\mathbb{L}_{\alpha}^{\:\:\beta},\mathbb{J}^{\gamma}\right] & =-\delta_{\alpha}^{\gamma}\mathbb{J}^{\beta}+\frac{1}{2}\delta_{\alpha}^{\beta}\mathbb{J}^{\gamma},\nonumber \\
\left\{ \mathbb{Q}_{\alpha}^{\:\:a},\mathbb{Q}_{\beta}^{\:\:b}\right\}  & =\epsilon_{\alpha\beta}\epsilon^{ab}\mathbb{C}, & \left\{ \mathbb{Q}_{a}^{\dagger\:\alpha},\mathbb{Q}_{b}^{\dagger\:\beta}\right\}  & =\epsilon_{ab}\epsilon^{\alpha\beta}\mathbb{C}^{\dagger},\nonumber \\
\left\{ \mathbb{Q}_{\alpha}^{\:\:b},\mathbb{Q}_{a}^{\dagger\:\beta}\right\}  & =\delta_{\alpha}^{\beta}\mathbb{R}_{a}^{\:\:b}+\delta_{a}^{b}\mathbb{L}_{\alpha}^{\:\:\beta}+\frac{1}{2}\delta_{\alpha}^{\beta}\delta_{a}^{b}\mathbb{H}
\end{align}
and the central charges are related to the worldsheet momenta as $\mathbb{C}=\mathbb{C}^{\dagger}=ig\left(e^{-i\mathbb{P}/2}-e^{i\mathbb{P}/2}\right)$.
The algebra form a bialgebra with the following co-product:
\begin{align}
\Delta(\mathbb{J}) & =\mathbb{J}\otimes\mathbb{U}^{-[\mathbb{J}]}+\mathbb{U}^{[\mathbb{J}]}\otimes\mathbb{J}
\end{align}
where $\mathbb{U}=e^{i\mathbb{P}/4}$ and $\left[\mathbb{R}_{a}^{\:\:b}\right]=\left[\mathbb{L}_{\alpha}^{\:\:\beta}\right]=\left[\mathbb{H}\right]=0$,
$\left[\mathbb{Q}_{\alpha}^{\:\:a}\right]=1$, $\left[\mathbb{Q}_{a}^{\dagger\:\beta}\right]=-1$,
$\left[\mathbb{C}\right]=2$, $\left[\mathbb{C}^{\dagger}\right]=-2$.
Let us choose the parameterization of the defining representation
$\mathcal{V}(p)$ as

\begin{align}
\mathbb{Q}_{\alpha}^{\:\:a}\left|e_{b}\right\rangle  & =a(p)\delta_{b}^{a}\left|e_{\alpha}\right\rangle , & \mathbb{Q}_{a}^{\dagger\:\alpha}\left|e_{b}\right\rangle  & =c(p)\epsilon_{ab}\epsilon^{\alpha\beta}\left|e_{\beta}\right\rangle ,\nonumber \\
\mathbb{Q}_{\alpha}^{\:\:a}\left|e_{\beta}\right\rangle  & =b(p)\epsilon_{\alpha\beta}\epsilon^{ab}\left|e_{b}\right\rangle , & \mathbb{Q}_{a}^{\dagger\:\alpha}\left|e_{\beta}\right\rangle  & =d(p)\delta_{\beta}^{\alpha}\left|e_{a}\right\rangle 
\end{align}
where bosonic labels are $a,b=1,2$, fermionic ones are $\alpha,\beta=3,4$
and we take the symmetric choice
\begin{align}
a(p) & =d(p)=\eta(p)=\sqrt{ig(x^{-}(p)-x^{+}(p))}, & b(p) & =-\frac{1}{x^{-}(p)}e^{-ip/2}\eta(p)=c(p)=-\frac{1}{x^{+}(p)}e^{ip/2}\eta(p)
\end{align}
with 
\begin{equation}
x^{\pm}(p)=\text{\ensuremath{e^{\pm i\frac{p}{2}}\frac{1+\sqrt{1+16g^{2}\sin^{2}\frac{p}{2}}}{4g\sin\frac{p}{2}}} }
\end{equation}
and $g$ is related to the t' Hooft coupling of the $\mathcal{N}=4$
SYM theory as $g=\frac{\sqrt{\lambda}}{4\pi}.$ The dispersion relation
follows from the algebra 
\begin{equation}
e^{ip}=\frac{x^{+}(p)}{x^{-}(p)}\quad;\quad E(p)=ig\left(x^{-}(p)-\frac{1}{x^{-}(p)}-x^{+}(p)+\frac{1}{x^{+}(p)}\right)=\sqrt{1+16g^{2}\sin^{2}\frac{p}{2}}
\end{equation}
By demanding that the fundamental S-matrix $S(p_{1},p_{2}):\mathcal{V}(p_{1})\otimes\mathcal{V}(p_{2})\to\mathcal{V}(p_{2})\otimes\mathcal{V}(p_{1})$
commutes with the conserved charges
\begin{equation}
\Delta(\mathbb{J})S(p_{1},p_{2})=S(p_{1},p_{2})\Delta^{op}(\mathbb{J})
\end{equation}
for all $\mathbb{J}\in\mathfrak{su}(2|2)_{c}$ we can obtain \cite{Arutyunov:2007tc}:
\begin{align}
S_{aa}^{aa} & =\frac{x_{2}^{-}-x_{1}^{+}}{x_{2}^{+}-x_{1}^{-}}\frac{\eta_{1}\eta_{2}}{\tilde{\eta}_{1}\tilde{\eta}_{2}}\quad;\qquad S_{ab}^{ba}=-\frac{(x_{1}^{-}-x_{1}^{+})(x_{2}^{-}-x_{2}^{+})(x_{2}^{-}+x_{1}^{+})}{(x_{1}^{-}-x_{2}^{+})(x_{1}^{-}x_{2}^{-}-x_{1}^{+}x_{2}^{+})}\frac{\eta_{1}\eta_{2}}{\tilde{\eta}_{1}\tilde{\eta}_{2}}\quad;\qquad S_{ab}^{ab}=S_{aa}^{aa}-S_{ab}^{ba}\nonumber \\
S_{\alpha\alpha}^{\alpha\alpha} & =-1\quad;\qquad S_{\alpha\beta}^{\beta\alpha}=-\frac{(x_{1}^{-}-x_{1}^{+})(x_{2}^{-}-x_{2}^{+})(x_{1}^{-}+x_{2}^{+})}{(x_{1}^{-}-x_{2}^{+})(x_{1}^{-}x_{2}^{-}-x_{1}^{+}x_{2}^{+})}\quad;\qquad S_{\alpha\beta}^{\alpha\beta}=S_{\alpha\alpha}^{\alpha\alpha}-S_{\alpha\beta}^{\beta\alpha}\\
S_{a\alpha}^{a\alpha} & =\frac{x_{2}^{-}-x_{1}^{-}}{x_{2}^{+}-x_{1}^{-}}\frac{\eta_{1}}{\tilde{\eta}_{1}}\quad;\quad S_{a\alpha}^{\alpha a}=\frac{x_{2}^{-}-x_{2}^{+}}{x_{2}^{+}-x_{1}^{-}}\frac{\eta_{1}}{\tilde{\eta}_{2}}\quad;\quad S_{\alpha a}^{a\alpha}=\frac{x_{1}^{-}-x_{1}^{+}}{x_{2}^{+}-x_{1}^{-}}\frac{\eta_{2}}{\tilde{\eta}_{1}}\quad;\quad S_{\alpha a}^{\alpha a}=\frac{x_{2}^{+}-x_{1}^{+}}{x_{2}^{+}-x_{1}^{-}}\frac{\eta_{2}}{\tilde{\eta}_{2}}\nonumber \\
S_{ab}^{\alpha\beta} & =-\epsilon_{ab}\epsilon^{\alpha\beta}\frac{ix_{1}^{-}x_{2}^{-}(x_{2}^{+}-x_{1}^{+})\eta_{1}\eta_{2}}{x_{1}^{+}x_{2}^{+}(x_{2}^{+}-x_{1}^{-})(1-x_{1}^{-}x_{2}^{-})}\quad;\qquad S_{\alpha\beta}^{ab}=-\epsilon_{\alpha\beta}\epsilon^{ab}\frac{i(x_{1}^{+}-x_{1}^{-})(x_{2}^{-}-x_{2}^{+})(x_{1}^{+}-x_{2}^{+})}{(x_{2}^{+}-x_{1}^{-})(1-x_{1}^{-}x_{2}^{-})\tilde{\eta}_{1}\tilde{\eta}_{2}}\nonumber 
\end{align}
where $a,b=1,2$ ; $a\neq b$, while $\alpha,\beta=3,4$ ; $\alpha\neq\beta$
and we have streamlined the notations as $S_{aa}^{aa}\equiv S_{aa}^{aa}(p_{1},p_{2})$,
$x_{i}^{\pm}=x^{\pm}(p_{i})$ and $\eta_{1}=e^{ip_{1}/4}e^{ip_{2}/2}\eta(p_{1})$,
$\eta_{2}=e^{ip_{2}/4}\eta(p_{2})$, $\tilde{\eta}_{1}=e^{ip_{1}/4}\eta(p_{1})$
and $\tilde{\eta}_{2}=e^{ip_{1}/2}e^{ip_{2}/4}\eta(p_{2})$.

The energy and momentum can be parametrized in terms of the torus
variable $z$ using Jacobi elliptic functions: 
\begin{equation}
p(z)=2\text{am}(z,k)\quad;\qquad E(z)=\text{dn}(z,k)\quad;\qquad k=-16g^{2}
\end{equation}
where the rapidity torus has two periods $2\omega_{1}=4K(k)$ and
$2\omega_{2}=4iK(1-k)-4K(k)$, with $K(k)$ being the elliptic K function.
Crossing transformation and reflections are easy to implement on the
torus:
\begin{equation}
x^{\pm}(z+\omega_{2})=\frac{1}{x^{\pm}(z)}\quad;\qquad x^{\pm}(-z)=-x^{\mp}(z)
\end{equation}
where explicitly 
\begin{equation}
x^{\pm}(z)=\frac{1}{2g}(\frac{\text{cn}z}{\text{sn}z}\pm i)(1+\text{dn}z)
\end{equation}
In the weak coupling limit the spectral parameter is also useful.
It is defined as 
\begin{equation}
u=x^{+}+\frac{1}{x^{+}}-\frac{i}{2g}=x^{-}+\frac{1}{x^{-}}+\frac{i}{2g}
\end{equation}
Thus we can define $x(u)$ such that 
\begin{equation}
x^{\pm}(u)=x(u\pm\frac{i}{2g})
\end{equation}
All of the three: the momentum, $p$, the spectral parameter $u$
or the torus rapidity $z$ can be used to label the particle's representation.

\section{KYBE, symmetries and selection rules for spin chains\label{sec:KYBE,-symmetries-and}}

In this Appendix we focus on rational spin chains of type $\mathfrak{su}(N)$
and $\mathfrak{so}(2N)$ and investigate the relations between the
type of the residual symmetries, YBE and chirality of the overlaps.
The $R$-matrix of the $\mathfrak{su}(N)$ model involves the identity
and the permutation (\ref{eq:Rsun}), while that of the $\mathfrak{so}(2N)$
contains additionally the trace operator (\ref{eq:Rso6}). There are
two types of BYBEs: the untwisted one
\begin{equation}
R_{12}(u-v)K_{1}(u)R_{12}(u+v)K_{2}(v)=K_{2}(v)R_{12}(u+v)K_{1}(u)R_{12}(u-v)
\end{equation}
which exist both for $\mathfrak{su}(N)$ and $\mathfrak{so}(2N)$
and the twisted one
\begin{equation}
R_{12}(u-v)K_{1}(u)\bar{R}_{12}(u+v)K_{2}(v)=K_{2}(v)\bar{R}_{12}(u+v)K_{1}(u)R_{12}(u-v)
\end{equation}
which exist only for the $\mathfrak{su}(N)$ R-matrix. Here $\bar{R}_{12}(u)=R_{\bar{1}2}(u)=R_{1\bar{2}}(u)$
describes the scattering of the fundamental representation and the
antifundamental and $R_{12}(u)=R_{\bar{1}\bar{2}}(u)$. The classification
of the solutions of these equations together with the type of the
residual symmetry is shown in Table \ref{Rclass}. 

\begin{table}[H]
\begin{centering}
\begin{tabular}{|c|c|c||c|c|c|}
\hline 
\multicolumn{1}{|c}{} & \multicolumn{1}{c}{$\mathfrak{su}(N)$} &  & \multicolumn{1}{c}{} & \multicolumn{1}{c}{$\mathfrak{so}(2N)$} & \tabularnewline
\hline 
residual symmetry & type  & reflection  & residual symmetry & type & reflection\tabularnewline
\hline 
\hline 
$\mathfrak{su}(M)\oplus\mathfrak{su}(N-M)\oplus\mathfrak{u}(1)$ & regular & untwisted  & $\mathfrak{so}(2m)\oplus\mathfrak{so}(2N-2m)$ & regular & untwisted\tabularnewline
\hline 
$\mathfrak{so}(N)$ & special & twisted & $\mathfrak{u}(N)$ & regular & untwisted\tabularnewline
\hline 
$\mathfrak{sp}(N)$ & special & twisted & $\mathfrak{so}(2m+1)\oplus\mathfrak{so}(2N-2m-1)$ & special & twisted\tabularnewline
\hline 
\end{tabular}
\par\end{centering}
\caption{Classification of the solution of the BYBEs together with their symmetries. }

\label{Rclass}
\end{table}

In the following we investigate the boundary states belonging to these
BYBEs. 

\subsection{Connection between the reflection and KYBE for rational spin chains}

\begin{figure}
\begin{centering}
\includegraphics[width=0.8\textwidth]{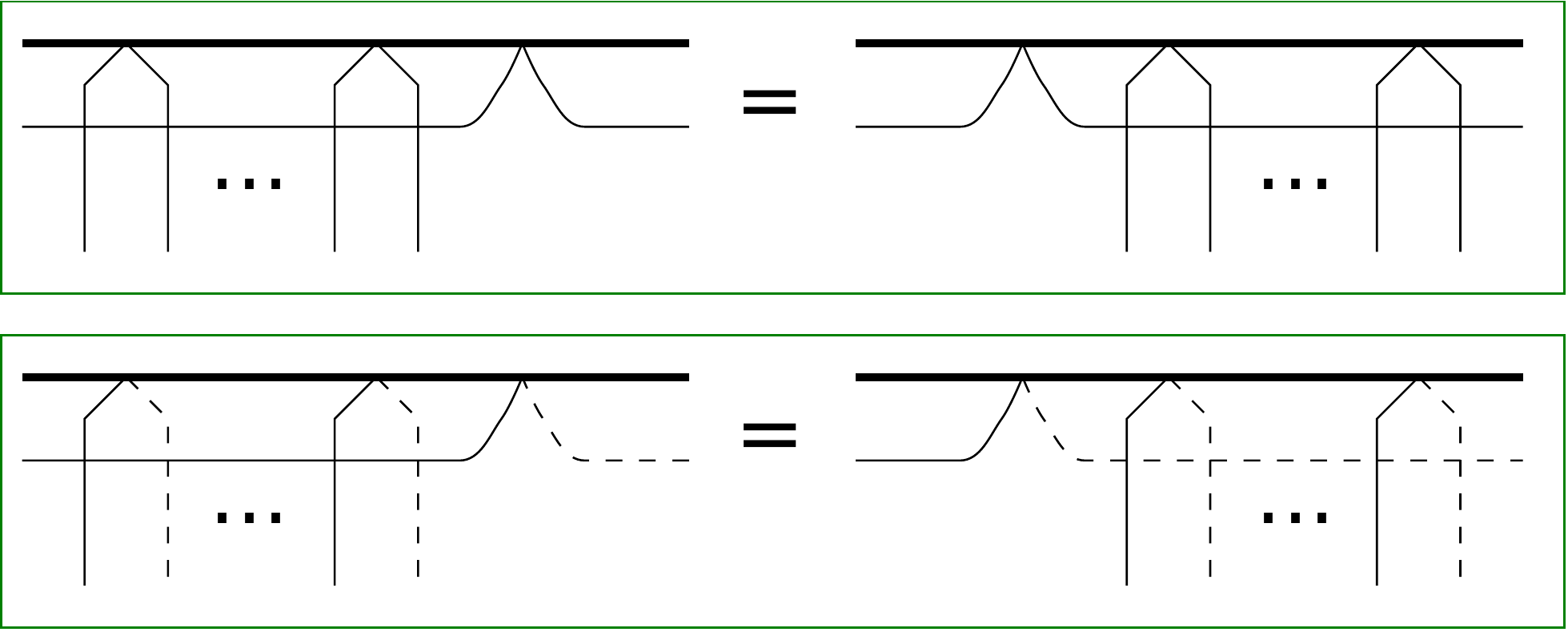}
\par\end{centering}
\caption{The proof of (\ref{eq:intcond-1}).}

\label{fig:intcond}
\end{figure}

The $K$-matrices can be used to define integrable two-site states:
\begin{equation}
\left\langle \Psi\right|=\left\langle \psi\right|^{\otimes L/2}\quad;\qquad\left\langle \psi\right|=\left\langle a\right|\otimes\left\langle b\right|K^{ab}(0).
\end{equation}
If the $K$- matrix is not twisted the KYBE ensures that the boundary
state satisfies the following requirement\emph{
\begin{equation}
\left\langle \Psi\right|K_{0}(u)T_{0}(u)=\left\langle \Psi\right|\Pi T_{0}(u)\Pi K_{0}(u)\quad;\qquad T_{0}(u)=R_{0L}(u)R_{0,L-1}(u)\dots R_{02}(u)R_{01}(u)\label{eq:intcond-1}
\end{equation}
}If however, the $K$-matrix is a twisted one we have to introduce
a spin chain with alternating inhomogeneities. In this spin chain
we can introduce two monodromy matrices 
\begin{align}
T_{0}(u) & =\bar{R}_{0L}(u)R_{0,L-1}(u)\dots\bar{R}_{02}(u)R_{01}(u),\\
\bar{T}_{0}(u) & =R_{0L}(u)\bar{R}_{0,L-1}(u)\dots R_{02}(u)\bar{R}_{01}(u),
\end{align}
such that the boundary state satisfies the relation
\begin{equation}
\left\langle \Psi\right|K_{0}(u)T_{0}(u)=\left\langle \Psi\right|\Pi\bar{T}_{0}(u)\Pi K_{0}(u)\label{eq:intcond2}
\end{equation}
The proof of (\ref{eq:intcond-1}) and (\ref{eq:intcond2}) is shown
in figure \ref{fig:intcond}. From (\ref{eq:intcond-1}) and (\ref{eq:intcond2}),
the following integrability conditions follow
\[
\text{untwisted:}\left\langle \Psi\right|t(u)=\left\langle \Psi\right|\Pi t(u)\Pi\quad;\qquad\text{twisted: }\left\langle \Psi\right|t(u)=\left\langle \Psi\right|\Pi\bar{t}(u)\Pi
\]

The untwisted case is simpler. It is a natural assumption that the
space reflection exchanges the sign of the Bethe roots $\Pi\left|\mathbf{u}^{(a)}\right\rangle =\left|-\mathbf{u}^{(a)}\right\rangle $.
This assumption was shown in (\ref{subsec:Pair-structure-insuN})
from the explicit form of the transfer matrices' eigenvalues. The
non-vanishing overlap implies that the eigenvalues satisfy

\begin{equation}
\Lambda\left(u|\mathbf{u}^{(a)}\right)=\Lambda\left(u|-\mathbf{u}^{(a)}\right)\quad;\qquad t(u)\left|\mathbf{u}^{(a)}\right\rangle =\Lambda\left(u|\mathbf{u}^{(a)}\right)\left|\mathbf{u}^{(a)}\right\rangle .
\end{equation}
The eigenvalue is symmetric in each type of Bethe roots therefore
this equation implies
\begin{equation}
\mathbf{u}^{(a)}=\left\{ u_{1}^{(a)},-u_{1}^{(a)},\dots,u_{N_{a}/2}^{(a)},-u_{N_{a}/2}^{(a)}\right\} \label{eq:pairstr}
\end{equation}
the chirality of the overlaps. 
\begin{figure}
\begin{centering}
\includegraphics[width=0.7\textwidth]{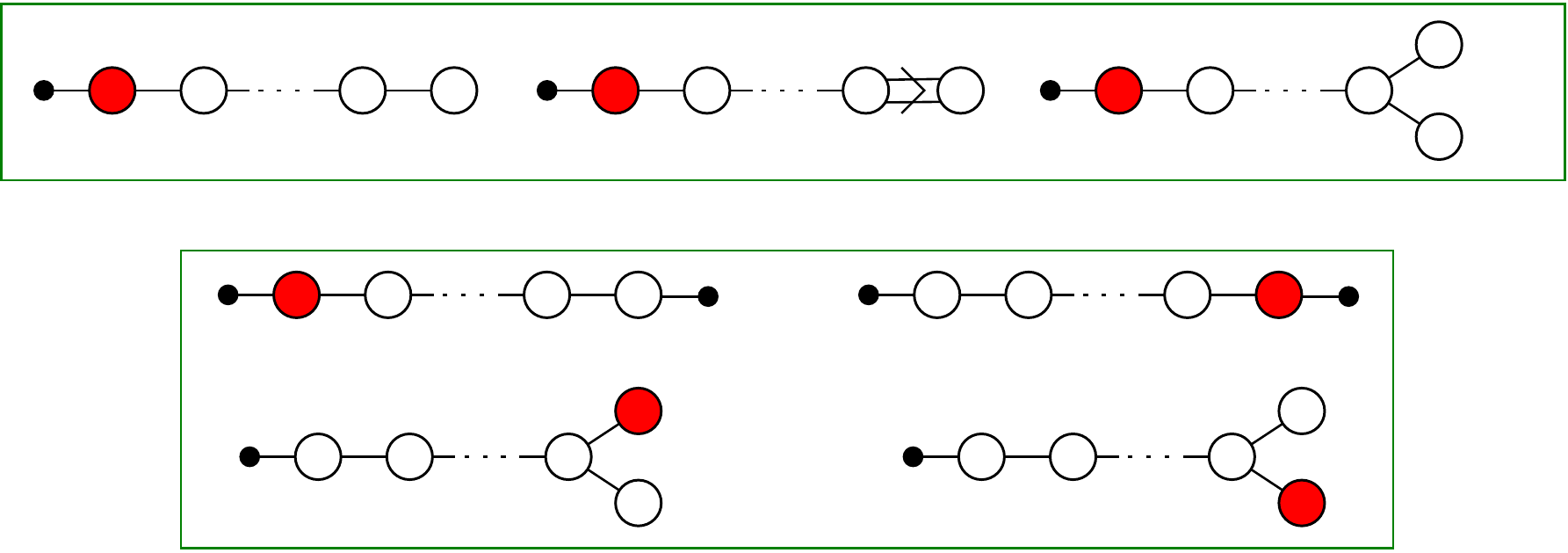}
\par\end{centering}
\caption{Decorated Dynkin diagrams.}

\label{fig:Dynkin}
\end{figure}

There can be other symmetries of the transfer matrix eigenvalue beyond
the permutation of the Bethe roots with the same type. These are the
symmetries of the Dynkin diagram which leave invariant the representations
of the quantum and auxiliary spaces. It can be illustrated with the
Dynkin diagram which is decorated by the used representations. We
use only the fundamental representations which is identified to one
of the Dynkin node. The red note indicates the auxiliary representation.
The additional black nodes\textbf{ }are connected to the Dynkin nodes
of the quantum spaces. The first box in figure \ref{fig:Dynkin} shows
the decorated Dynkin diagrams of the untwisted cases. We can see that
the $\mathfrak{su}(N)$ and the $\mathfrak{so}(2n+1)$ cases has no
additional symmetry therefore the symmetric pairs

\begin{equation}
(\mathfrak{su}(N),\mathfrak{so}(N))\quad;\qquad(\mathfrak{su}(2n),\mathfrak{sp}(2n))\quad;\qquad(\mathfrak{so}(2n+1),\mathfrak{so}(M)\oplus\mathfrak{so}(2n+1-M))
\end{equation}
has pair structure (\ref{eq:pairstr}) i.e. these are chiral symmetric
pairs. 

In the case of the decorated $\mathfrak{so}(2n)$ diagram there is
a symmetry which interchanges the roots $\mathbf{u}^{+}$ and $\mathbf{u}^{-}$,
therefore for the $\mathfrak{so}(2n)$ spin chain there can be two
types of pair structures
\begin{itemize}
\item chiral pair structure for which 
\begin{align}
\mathbf{u}^{(a)} & =\left\{ u_{1}^{(a)},-u_{1}^{(a)},\dots,u_{m_{a}/2}^{(a)},-u_{m_{a}/2}^{(a)}\right\} ,\quad\text{for}\:a=1,\dots n-2\quad;\qquad\mathbf{u}^{(\pm)}=\left\{ u_{1}^{(\pm)},-u_{1}^{(\pm)},\dots,u_{m_{\pm}/2}^{(\pm)},-u_{m_{\pm}/2}^{(\pm)}\right\} .
\end{align}
\item achiral pair structure for which 
\begin{align}
\mathbf{u}^{(a)} & =\left\{ u_{1}^{(a)},-u_{1}^{(a)},\dots,u_{m_{a}/2}^{(a)},-u_{m_{a}/2}^{(a)}\right\} ,\qquad\text{for}\:a=1,\dots n-2\quad;\qquad\mathbf{u}^{(+)}=-\mathbf{u}^{(-)}
\end{align}
\end{itemize}
Before we decide which pair structures belong to the concrete examples,
let us continue with the twisted case for the $\mathfrak{su}(N)$
model, which belongs to the symmetric pair $(\mathfrak{su}(N),\mathfrak{su}(M)\oplus\mathfrak{su}(N-M)\oplus\mathfrak{u}(1))$.
In this case the quantum space is an alternate tensor product of the
particles and antiparticles and there are two transfer matrices $t(u)$
and $\bar{t}(u)$ where the auxiliary spaces are particle and antiparticle,
respectively. They are diagonalizable simultaneously and let $\left|\mathbf{u}^{(a)}\right\rangle $
be a Bethe state for which
\begin{align}
t(u)\left|\mathbf{u}^{(a)}\right\rangle  & =\Lambda\left(u|\mathbf{u}^{(a)}\right)\left|\mathbf{u}^{(a)}\right\rangle \quad;\qquad\bar{t}(u)\left|\mathbf{u}^{(a)}\right\rangle =\bar{\Lambda}\left(u|\mathbf{u}^{(a)}\right)\left|\mathbf{u}^{(a)}\right\rangle .
\end{align}
The eigenvalue of the transfer matrices can be written as \cite{Arnaudon_2005}
\begin{align}
\Lambda(u) & =\sum_{k=1}^{N}\frac{Q_{k-1}^{[k+1]}\left(u\right)}{Q_{k-1}^{[k-1]}\left(u\right)}\frac{Q_{k}^{[k-2]}\left(u\right)}{Q_{k}^{[k]}\left(u\right)}\quad;\qquad\bar{\Lambda}(u)=\sum_{k=1}^{N}\frac{Q_{k-1}^{[N-k-1]}\left(u\right)}{Q_{k-1}^{[N-k+1]}\left(u\right)}\frac{Q_{k}^{[N-k+2]}\left(u\right)}{Q_{k}^{[N-k]}\left(u\right)},
\end{align}
where
\begin{equation}
Q_{0}(u)=Q_{N}(u)=u^{L/2}.
\end{equation}
Using this explicit form we can show that
\begin{equation}
t(u)\Pi\left|\mathbf{u}^{(a)}\right\rangle =\Pi\bar{t}(-u-N/2)\left|\mathbf{u}^{(a)}\right\rangle =\bar{\Lambda}\left(-u-N/2|\mathbf{u}^{(a)}\right)\Pi\left|\mathbf{u}^{(a)}\right\rangle =\Lambda\left(u|-\mathbf{u}^{(a)}\right)\Pi\left|\mathbf{u}^{(a)}\right\rangle 
\end{equation}
therefore the parity transformation acts on the Bethe roots as $\Pi\left|\mathbf{u}^{(a)}\right\rangle =\left|-\mathbf{u}^{(a)}\right\rangle $.
The non-vanishing overlap implies that 
\begin{equation}
\Lambda\left(u|\mathbf{u}^{(a)}\right)=\bar{\Lambda}\left(u|-\mathbf{u}^{(a)}\right).
\end{equation}
The extended Dynkin diagrams of these eigenvalues is in the first
row of the second box of figure \ref{fig:Dynkin}. Now there are two
additional black nodes since there are particles and antiparticles
in the quantum space. We can see that these Diagram is connected by
the original Dynkin diagram isomorphism which transforms the Bethe
roots as $\tilde{\mathbf{u}}^{(a)}=\mathbf{u}^{(N-a)}$ therefore
it implies that
\begin{equation}
\bar{\Lambda}\left(u|\mathbf{u}^{(a)}\right)=\Lambda\left(u|\tilde{\mathbf{u}}^{(a)}\right)\quad\longrightarrow\qquad\Lambda\left(u|\mathbf{u}^{(a)}\right)=\Lambda\left(u|-\tilde{\mathbf{u}}^{(a)}\right).
\end{equation}
Since there is no symmetry of the extended diagram of $\Lambda$ the
sets $\mathbf{u}^{(a)}$, $-\tilde{\mathbf{u}}^{(a)}$ have to be
the same which means that the symmetric pair $(\mathfrak{su}(N),\mathfrak{su}(M)\oplus\mathfrak{su}(N-M)\oplus\mathfrak{u}(1))$
belongs to an achiral pair structure i.e.
\begin{align}
\mathbf{u}^{(a)} & =\left\{ +u_{1}^{(a)},+u_{2}^{(a)},\dots,+u_{m_{a}-1}^{(a)},+u_{m_{a}}^{(a)}\right\} =-\mathbf{u}^{(N-a)},\qquad\text{for}\:a=1,\dots,\left\lfloor \frac{N-1}{2}\right\rfloor 
\end{align}
and
\begin{equation}
\mathbf{u}^{(N/2)}=\left\{ u_{1}^{(N/2)},-u_{1}^{(N/2)},\dots,u_{m_{N/2}/2}^{(N/2)},-u_{m_{N/2}/2}^{(N/2)}\right\} .
\end{equation}

We can see that in the $\mathfrak{su}(N)$ case we could decide which
symmetric pair is chiral or achiral because the auxiliary space representation
breaks the symmetry of the original Dynkin diagram. We can do the
same $\mathfrak{so}(2n)$ by choosing another auxiliary space, such
as the spinor representations, as they are not invariant under the
Dynkin diagram symmetry. The corresponding transfer matrices are denoted
by $t^{(+)}(u)$ and $t^{(-)}(u)$. Using these matrices, we can define
two integrability conditions 
\begin{align}
\left\langle \Psi\right|t^{(+)}(u) & =\left\langle \Psi\right|\Pi t^{(+)}(u)\Pi,\label{eq:chiralsoN}\\
\left\langle \Psi\right|t^{(+)}(u) & =\left\langle \Psi\right|\Pi t^{(-)}(u)\Pi.\label{eq:achiralsoN}
\end{align}
From the previous argument, we can see that the conditions (\ref{eq:chiralsoN})
and (\ref{eq:achiralsoN}) lead to chiral and achiral pair structures,
respectively.

\begin{figure}
\begin{centering}
\includegraphics[width=0.8\textwidth]{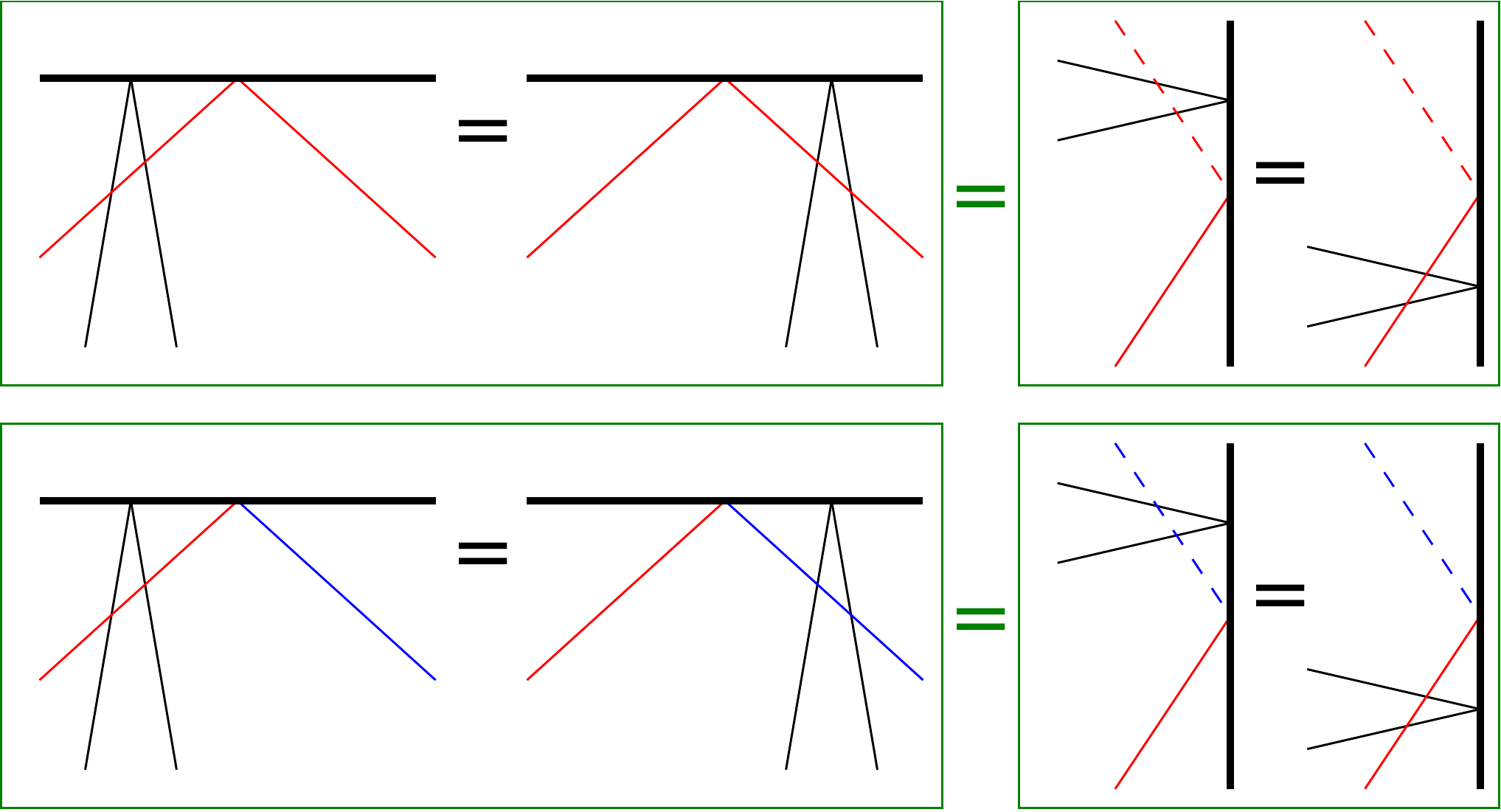}
\par\end{centering}
\caption{Reflection equations for $\mathfrak{so}(2n)$ model. Black line: vector
rep. Red and blue line: spinor reps. Dashed lines: contragradient
reps.}
\label{fig:KYBESON}
\end{figure}

In figure \ref{fig:KYBESON} we can see the KYBEs for the $K$-matrices
which solves the integrable conditions (\ref{eq:chiralsoN}) and (\ref{eq:achiralsoN}).
The red and blue lines belongs two the spinor representations. The
first line and second lines belong to (\ref{eq:chiralsoN}) and (\ref{eq:achiralsoN}),
respectively. We recall that switching from reflection to $K$-matrices
we have to use a conjugation, which changes the representations to
their contragradients. We know that 
\begin{itemize}
\item For $\mathfrak{so}(4n)$, the spinor representations are pseudo-reals
therefore the dashed red line is equivalent to the simple red line.
Which means that the first and the second lines describe non representation
changing and representation changing reflections, i.e. untwisted and
twisted reflections, respectively.
\item For $\mathfrak{so}(4n+2)$, the spinor representations are contragradients
of each other therefore the dashed blue line is equivalent to the
simple red line. Which means that the first and the second lines describe
representation changing and non representation changing reflections,
i.e. twisted and untwisted reflections, respectively.
\end{itemize}
The table \ref{tab:class} summarize the results of this section and
naturally extends them for $\mathfrak{gl}(N\vert M)$.

\section{Calculation of two particle overlaps }

In this appendix we elaborate on the calculation of the overlaps in
the $L\to\infty$ limit using two particle coordinate Bethe ansatz
state of Section 6. Bethe ansatz states represent a plane wave containing
two particles with momentum $p$ and $-p$. We are not going to impose
periodicity for the wave functions, thus our states are of-shell Bethe
states. Actually in the $L\to\infty$ limit off-shell and on-shell
states become equivalent.

\subsection{XXX spin chain}

Let us start with the general integrable two-site state of the XXX
spin chain of size $L$. We take the boundary state with $\langle\Psi\vert=\langle\psi_{11},\psi_{12},\psi_{21},\psi_{22}\vert$
and calculate the overlap of the integrable state $\langle\Psi\vert$
with a two magnon state, built over the pseudo-vacuum $\vert1\rangle^{\otimes L}$
in the large $L$ limit. In coordinate space Bethe ansatz the two
magnon state is a plane wave of the form 
\begin{equation}
\left|u,-u\right\rangle =\sum_{n_{1}=1}^{L}\sum_{n_{2}=n_{1}+1}^{L}\left(e^{ip(n_{1}-n_{2})}+e^{-ip(n_{1}-n_{2})}S(2u)\right)\left|n_{1}n_{2}\right\rangle 
\end{equation}
where$p=-i\log\frac{u-i/2}{u+i/2}$ and $S(u)=\frac{u+i}{u-i}$. The
state $\vert n_{1}n_{2}\rangle$ has excitation $2$ at sites $n_{1}$
and $n_{2}$. To obtain the overlap $\left\langle \Psi\right.\left|u,-u\right\rangle $
we have to use the following elementary overlaps
\begin{equation}
\left\langle \Psi\right.\left|n_{1}n_{2}\right\rangle =\begin{cases}
\psi_{12}^{2} & \text{if }n_{1},n_{2}\text{ are even}\\
\psi_{21}^{2} & \text{if }n_{1},n_{2}\text{ are odd}\\
\psi_{12}\psi_{21} & \text{if }n_{1}\text{ is even and }n_{2}\text{ is odd}\\
\psi_{21}\psi_{12} & \text{if }n_{1}\text{ is odd and }n_{2}\text{ is even and }n_{2}-n_{1}>1\\
\psi_{22} & \text{if }n_{1}\text{ is odd and }n_{2}\text{ is even and }n_{2}-n_{1}=1
\end{cases}
\end{equation}
By plugging back these into the overlap $\left\langle \Psi\right.\left|u,-u\right\rangle $
we obtain
\begin{align}
\left\langle \Psi\right.\left|u,-u\right\rangle  & =\sum_{m_{1}=1}^{L/2}\sum_{m_{2}=m_{1}+1}^{L/2}\left(e^{ip(2m_{1}-2m_{2})}+e^{-ip(2m_{1}-2m_{2})}S(2u)\right)\psi_{12}^{2}+\nonumber \\
 & +\sum_{m_{1}=1}^{L/2}\sum_{m_{2}=m_{1}+1}^{L/2}\left(e^{ip(2m_{1}-2m_{2})}+e^{-ip(2m_{1}-2m_{2})}S(2u)\right)\psi_{21}^{2}+\nonumber \\
 & +\sum_{m_{1}=1}^{L/2}\sum_{m_{2}=m_{1}+1}^{L/2}\left(e^{ip(2m_{1}-2m_{2}+1)}+e^{-ip(2m_{1}-2m_{2}+1)}S(2u)\right)\psi_{12}\psi_{21}+\nonumber \\
 & +\sum_{m_{1}=1}^{L/2}\sum_{m_{2}=m_{1}+1}^{L/2}\left(e^{ip(2m_{1}-2m_{2}-1)}+e^{-ip(2m_{1}-2m_{2}-1)}S(2u)\right)\psi_{21}\psi_{12}+\nonumber \\
 & +\sum_{m=1}^{L/2}\left(e^{-ip}+e^{+ip}S(2u)\right)\psi_{22}.\label{eq:overlapXXX}
\end{align}
It is convenient to introduce the following quantity
\begin{equation}
\Sigma(p)=\sum_{m_{1}=1}^{L/2}\sum_{m_{2}=m_{1}+1}^{L/2}e^{ip(2m_{1}-2m_{2})}.
\end{equation}
with which the overlap (\ref{eq:overlapXXX}) can be written as
\begin{equation}
\left\langle \Psi\right.\left|u,-u\right\rangle =\left(\Sigma(p)+\Sigma(-p)S(2u)\right)\left(\psi_{12}^{2}+\psi_{21}^{2}+\left(e^{ip}+e^{-ip}\right)\psi_{12}\psi_{21}\right)+\frac{L}{2}\left(e^{-ip}+e^{+ip}S(2u)\right)\psi_{22}.\label{eq:overXXX}
\end{equation}

\subsection{$SU(3)$ spin chain with $SO(3)$ symmetry}

We analyze a two site state and a matrix product state for these models. 

\subsubsection{Two-site state}

Let the two-site state be
\begin{equation}
\left\langle \Psi\right|=\left(\left\langle 1\right|\otimes\left\langle 1\right|+\left\langle 2\right|\otimes\left\langle 2\right|+\left\langle 3\right|\otimes\left\langle 3\right|\right)^{\otimes L/2}
\end{equation}
We choose the pseudo-vacuum as $\left|3\right\rangle ^{\otimes L}$
and introduce excitations with labels $1$ and $2$. The two magnon
state can be written as
\begin{equation}
\left|u,-u\right\rangle _{ab}=\sum_{n_{1}=1}^{L}\sum_{n_{2}=n_{1}+1}^{L}\left(e^{ip(n_{1}-n_{2})}\left|n_{1}n_{2}\right\rangle _{ab}+e^{-ip(n_{1}-n_{2})}S_{ab}^{cd}(2u)\left|n_{1}n_{2}\right\rangle _{cd}\right)
\end{equation}
where $\vert n_{1}n_{2}\rangle_{ab}$ represent state $a$ at site
$n_{1}$ and $b$ at site $n_{2}$, while $a,b,c,d=1,2$. The scattering
matrix of the top level Bethe ansatz excitations is 
\begin{align}
S(u) & =\frac{i}{u-i}\mathbf{1}+\frac{u}{u-i}\mathbf{P}
\end{align}
with normalization $S_{11}^{11}=1$. We have to calculate the following
scalar product
\begin{equation}
\left\langle \Psi\right.\left|n_{1}n_{2}\right\rangle _{ab}=\begin{cases}
\delta_{ab} & \text{if }n_{1}\text{ is odd and }n_{2}\text{ is even and }n_{2}-n_{1}=1\\
0 & \text{otherwise}
\end{cases}
\end{equation}
By dividing with the asymptotic norm of the states we obtain the K-matrix
of the top level excitations: 
\begin{equation}
K_{ab}^{(1)}(u):=\frac{1}{L}\left\langle \Psi\right.\left|u,-u\right\rangle _{ab}=\frac{1}{2}\left(e^{-ip}\delta_{ab}+e^{ip}S_{ab}^{cc}(2u)\right)=\frac{u}{u-i/2}\delta_{ab}
\end{equation}

\subsubsection{Matrix product state with Pauli matrices}

Let the MPS be
\begin{equation}
^{\alpha,\beta}\left\langle \mathrm{MPS}\right|=\left[\left(\left\langle 1\right|\sigma_{1}+\left\langle 2\right|\sigma_{2}+\left\langle 3\right|\sigma_{3}\right)^{\otimes L}\right]^{\alpha,\beta}
\end{equation}
where $\alpha,\beta=1,2$ are the ``inner'' indexes of the Pauli
matrices. The pseudo vacuum is $\left|3\right\rangle ^{\otimes L}$
and we calculate the overlap $^{\alpha,\beta}\left\langle \mathrm{MPS}\right|u,-u\rangle_{ab}$.
The elementary overlaps with the states $\left|n_{1}n_{2}\right\rangle _{ab}$
can be written as
\begin{equation}
^{\alpha,\beta}\left\langle \mathrm{MPS}\right.\left|n_{1}n_{2}\right\rangle _{a,b}=\begin{cases}
\left(\sigma_{3}\sigma_{a}\sigma_{3}\sigma_{b}\right)^{\alpha\beta}=-\left(\sigma_{a}\sigma_{b}\right)^{\alpha\beta} & \text{if }n_{1},n_{2}\text{ are even}\\
\left(\sigma_{a}\sigma_{3}\sigma_{b}\sigma_{3}\right)^{\alpha\beta}=-\left(\sigma_{a}\sigma_{b}\right)^{\alpha\beta} & \text{if }n_{1},n_{2}\text{ are odd}\\
\left(\sigma_{3}\sigma_{a}\sigma_{b}\sigma_{3}\right)^{\alpha\beta}=\left(\sigma_{a}\sigma_{b}\right)^{\alpha\beta} & \text{if }n_{1}\text{ is even and }n_{2}\text{ is odd}\\
\left(\sigma_{a}\sigma_{b}\sigma_{3}\sigma_{3}\right)^{\alpha\beta}=(\sigma_{a}\sigma_{b})^{\alpha\beta} & \text{if }n_{1}\text{ is odd and }n_{2}\text{ is even}
\end{cases}\label{eq:sig}
\end{equation}
Let us introduce the following notation $F_{ab}^{\alpha\beta}=\left(\sigma_{a}\sigma_{b}\right)^{\alpha\beta}$.
The full overlap thus can be written as
\begin{multline}
^{\alpha,\beta}\left\langle \mathrm{MPS}\right.\left|u,-u\right\rangle _{a,b}=\left(\Sigma(p)F_{ab}^{\alpha\beta}+\Sigma(-p)S_{ab}^{cd}(2u)F_{cd}^{\alpha\beta}\right)\left(e^{ip}+e^{-ip}-2\right)+\\
+\frac{L}{2}\left(e^{-ip}F_{ab}^{\alpha\beta}+e^{+ip}S_{ab}^{cd}(2u)F_{cd}^{\alpha\beta}\right)
\end{multline}
This overlap is diagonal in $\alpha$ and $\beta$. The remaining
components can be written after normalizing with $L^{-1}$ as 
\begin{align}
^{1,1}\left\langle \mathrm{MPS}\right.\left|u,-u\right\rangle _{a,b} & =K_{ab}^{(1)+}(u)\quad;\qquad{}^{2,2}\left\langle \mathrm{MPS}\right.\left|u,-u\right\rangle _{a,b}=K_{ab}^{(1)-}(u)
\end{align}
with
\begin{equation}
K^{(1)\pm}(u)=\frac{1}{u}\left(\begin{array}{cc}
u+\frac{i}{2} & \pm\frac{1}{2}\\
\mp\frac{1}{2} & u+\frac{i}{2}
\end{array}\right)=k^{(1)}(u)\psi^{(2)\pm}\quad;\qquad k^{(1)}(u)=K_{11}^{(1)\pm}(u).
\end{equation}

\subsection{$\mathfrak{so}(6)$ spin chains with $\mathfrak{so}(3)\times\mathfrak{so}(3)$
symmetry}

The six dimensional one site Hilbert space is parametrized by $\phi_{i}$,
$i=1,\dots,6$ and we introduce the notation 
\begin{align}
Z & =\frac{1}{\sqrt{2}}\left(\phi_{5}+i\phi_{6}\right)\quad,\qquad\bar{Z}=\frac{1}{\sqrt{2}}\left(\phi_{5}-i\phi_{6}\right)
\end{align}
We are going to analyze a two site state and a matrix product state. 

\subsubsection{Two-site state}

Let the two-site state be
\begin{equation}
\left\langle \Psi\right|=\left(Z\otimes Z+\bar{Z}\otimes\bar{Z}+\phi_{1}\otimes\phi_{1}-\phi_{2}\otimes\phi_{2}+\phi_{3}\otimes\phi_{3}-\phi_{4}\otimes\phi_{4}\right)^{\otimes L/2}
\end{equation}
We choose the pseudo vacuum as $Z^{\otimes L}$ and then the two magnon
state can be written as
\begin{equation}
\left|u,-u\right\rangle _{ab}=\sum_{n_{1}=1}^{L}\sum_{n_{2}=n_{1}+1}^{L}\left(e^{ip(n_{1}-n_{2})}\left|n_{1}n_{2}\right\rangle _{ab}+e^{-ip(n_{1}-n_{2})}S_{ab}^{cd}(2u)\left|n_{1}n_{2}\right\rangle _{cd}\right)+\delta_{ab}e(2u)\sum_{n=1}^{L}\left|n\right\rangle 
\end{equation}
where the excitations are labeled with $a,b,c,d=1,2,3,4$ and
\begin{align}
S(u) & =\frac{i}{u-i}\mathbf{1}+\frac{u}{u-i}\mathbf{P}-i\frac{u}{(u-i)(u+i)}\mathbf{K},\quad e(u)=-\frac{u}{u+i},
\end{align}
while $\left|n\right\rangle =\left|Z\dots Z\bar{Z}Z\dots Z\right\rangle .$
We have to calculate the following elementary scalar products
\begin{equation}
\left\langle \Psi\right.\left|n_{1}n_{2}\right\rangle _{ab}=\begin{cases}
F_{ab} & \text{if }n_{1}\text{ is odd and }n_{2}\text{ is even and }n_{2}-n_{1}=1\\
0 & \text{otherwise}
\end{cases}
\end{equation}
where
\begin{equation}
F=\mathrm{diag}(1,-1,1,-1)
\end{equation}
Since $\left\langle \Psi\right.\left|n\right\rangle =0$ the K-matrix
of the top level $\mathfrak{so}(4)$ excitations is 
\begin{equation}
K_{ab}^{(1)}(u)=\frac{1}{L}\left\langle \Psi\right.\left|u,-u\right\rangle _{ab}=\frac{1}{2}\left(e^{-ip}F_{ab}+e^{ip}S_{ab}^{cd}(2u)F_{cd}\right)=\frac{u}{u-i/2}F_{ab}
\end{equation}

\subsubsection{Matrix product state with Pauli matrices}

Let the MPS be
\begin{equation}
^{\alpha,\beta}\left\langle \mathrm{MPS}\right|=\left[\left(\sqrt{2}\phi_{1}\sigma_{1}+\sqrt{2}\phi_{3}\sigma_{2}+\sqrt{2}\phi_{5}\sigma_{3}\right)^{\otimes L}\right]^{\alpha,\beta}=\left[\left(\sqrt{2}\phi_{1}\sigma_{1}+\sqrt{2}\phi_{3}\sigma_{2}+(Z+\bar{Z})\sigma_{3}\right)^{\otimes L}\right]^{\alpha,\beta}.
\end{equation}
We take the pseudo vacuum and the excitations as before. The overlaps
with the states $\left|n_{1}n_{2}\right\rangle _{ab}$ can be written
similarly as (\ref{eq:sig}) and for $\left|n\right\rangle $ we obtain
\begin{equation}
^{\alpha,\beta}\left\langle \mathrm{MPS}\right.\left|n\right\rangle =\delta^{\alpha\beta}
\end{equation}
The full overlap can be written as
\begin{multline}
^{\alpha,\beta}\left\langle \mathrm{MPS}\right.\left|u,-u\right\rangle _{a,b}=\left(\Sigma(p)F_{ab}^{\alpha\beta}+\Sigma(-p)S_{ab}^{cd}(2u)F_{cd}^{\alpha\beta}\right)\left(e^{ip}+e^{-ip}-2\right)+\\
+\frac{L}{2}\left(e^{-ip}F_{ab}^{\alpha\beta}+e^{+ip}S_{ab}^{cd}(2u)F_{cd}^{\alpha\beta}\right)+L\delta_{ab}\delta^{\alpha\beta}e(2u).
\end{multline}
This overlap is diagonal in $\alpha$ and $\beta$. The remaining
components can be written again as
\begin{align}
^{1,1}\left\langle \mathrm{MPS}\right.\left|n_{1}n_{2}\right\rangle _{a,b} & =K_{ab}^{(1)+}(u)\quad;\qquad{}^{2,2}\left\langle \mathrm{MPS}\right.\left|n_{1}n_{2}\right\rangle _{a,b}=K_{ab}^{(1)-}(u)
\end{align}
where the K-matrices can be written as
\begin{equation}
K^{(1)\pm}(u)=\left(\begin{array}{cccc}
\frac{u^{2}+iu-1/2}{u(u+i/2)} & 0 & \mp\frac{1}{u} & 0\\
0 & -\frac{u+i}{u+i/2} & 0 & 0\\
\pm\frac{1}{u} & 0 & \frac{u^{2}+iu-1/2}{u(u+i/2)} & 0\\
0 & 0 & 0 & -\frac{u+i}{u+i/2}
\end{array}\right)
\end{equation}

\subsection{Two-site state in $\mathfrak{su}(2\vert2)_{c}$ with $\mathfrak{osp}(2\vert2)_{c}$
symmetry}

Let the parameters of the K-matrix be $k_{1}=k_{4}=s$, $k_{2}=0$
and the boundary state take the form 
\begin{equation}
\left\langle B\right|=\left\langle K(p_{1})\right|\otimes\dots\otimes\left\langle K(p_{L})\right|\quad;\qquad\left\langle K(p)\right|=K_{i,j}(p)\left\langle i\right|\otimes\left\langle j\right|\mathbb{I}_{g}.
\end{equation}
 The pseudovacuum is $\vert1\rangle^{\otimes2L}$ and excitations
are labeled with $3,4$ and $2$. The two-particle Bethe state in
coordinate space can be written as \cite{deLeeuw:2007akd}
\begin{align}
\left|y_{1},y_{2}\right\rangle _{\alpha,\beta}= & \sum_{1\leq n_{1}<n_{2}\leq2L}\Psi_{n_{1}}(y_{1})\Psi_{n_{2}}(y_{2})\left|n_{1},n_{2}\right\rangle _{\alpha,\beta}-\Psi_{n_{1}}(y_{2})\Psi_{n_{2}}(y_{1})R_{\alpha\beta}^{\gamma\delta}(y_{1},y_{2})\left|n_{1},n_{2}\right\rangle _{\gamma,\delta}\nonumber \\
 & +\epsilon_{\alpha\beta}\sum_{1\leq n\leq2L}\Psi_{n}(y_{1})\Psi_{n}(y_{2})g_{n}(y_{1},y_{2})\left|n\right\rangle 
\end{align}
where $\alpha,\beta$=3,4, and $R$ is the R-matrix of the XXX model
\begin{equation}
R(y_{1},y_{2})=\frac{-i/g}{v_{1}-v_{2}-i/g}\mathbf{1}+\frac{v_{1}-v_{2}}{v_{1}-v_{2}-i/g}\mathbf{P}\quad;\qquad v_{i}=y_{i}+1/y_{i}
\end{equation}
The basis vectors are 
\begin{equation}
\left|n_{1},n_{2}\right\rangle _{\alpha,\beta}=\left|1,\dots,1,\alpha,1,\dots1,\beta,1\dots,1\right\rangle ,\quad;\qquad\left|n\right\rangle =\left|1,\dots,1,2,1,\dots1\right\rangle 
\end{equation}
while the wave functions and S-matrices read as
\begin{align}
\Psi_{n}(y) & =\psi_{n}(y)\prod_{k=1}^{n-1}S(y,x_{k})\quad;\qquad S(y,x_{k})=\frac{y-x_{k}^{+}}{y-x_{k}^{-}}e^{-ip_{k}/2},\\
g_{n}(y_{1},y_{2}) & =e^{-ip_{n}/2}\frac{y_{1}y_{2}-x_{n}^{+}x_{n}^{-}}{y_{1}y_{2}x_{n}^{-}}\frac{y_{1}-y_{2}}{v_{1}-v_{2}-i/g}\quad;\qquad\psi_{n}(y)=e^{-ip_{n}/4}\frac{y\sqrt{i(x_{n}^{-}-x_{n}^{+})}}{y-x_{n}^{-}}.
\end{align}
where $\mathbb{I}_{g}$ is the graded identity: the product of the
permutation and the graded permutation. Graded permutation picks up
a minus sign, whenever two fermions are interchanged. We normalize
the $K$ function as 
\begin{equation}
\left\langle K(p)\right|\left.K(p)\right\rangle =1.
\end{equation}
Let us assume that the overlap with a general Bethe state looks in
the $L\to\infty$ limit looks like as\footnote{One should actually put the ratio of Gaudin type determinant at each
step of the nesting. In this appendix we do not write them out as
they are irrelevant for nested K-matrices.}
\begin{equation}
\frac{\left|\left\langle B\right|\left.\mathbf{p},\mathbf{y},\mathbf{w}\right\rangle \right|^{2}}{\left\langle \mathbf{p},\mathbf{y},\mathbf{w}\right.\left|\mathbf{p},\mathbf{y},\mathbf{w}\right\rangle }=\prod_{i=1}^{L}h^{p}(p_{i})\prod_{i=1}^{N/2}h^{y}(v_{i})\prod_{i=1}^{M/2}h^{w}(w_{i})
\end{equation}
First we define the renormalized boundary state $\left\langle \bar{B}\right|$
for which the K-matrix is
\begin{equation}
\bar{K}(p)=\frac{K(p)}{K_{1,1}(p)}.
\end{equation}
For this boundary state the overlap with pseudo-vacuum is 1 therefore
\begin{equation}
h^{p}(p)=\left|K_{1,1}(p)\right|^{2}
\end{equation}
and the remaining overlap is 
\begin{equation}
\frac{\left|\left\langle \bar{K}\right|\left.\mathbf{p},\mathbf{y},\mathbf{w}\right\rangle \right|^{2}}{\left\langle \mathbf{p},\mathbf{y},\mathbf{w}\right.\left|\mathbf{p},\mathbf{y},\mathbf{w}\right\rangle }=\prod_{i=1}^{N/2}h^{y}(v_{i})\prod_{i=1}^{M/2}h^{w}(w_{i}).
\end{equation}
This is basically the nesting for the boundary states and overlaps.
In the following let us use special inhomogeneities 
\begin{equation}
p_{2k-1}=p,\qquad p_{2k}=-p,\qquad\text{for}\;k=1,\dots,L.
\end{equation}
as the boundary overlaps does not depend on the inhomogeneities. Due
to the special form of the boundary state nonzero contributions comes
only from states $n_{1}=2m-1$ and $n_{2}=2m$ and $n$ arbitrarily:
\begin{align}
\langle\bar{B}\left|y,-y\right\rangle _{\alpha,\beta}= & -\epsilon_{\alpha\beta}LK_{34}(p)\left[\psi_{2m-1}(y)\psi_{2m}(-y)S(-y,x_{2m-1})+\frac{2v+\frac{i}{g}}{2v-\frac{i}{g}}\psi_{2m-1}(-y)\psi_{2m}(y)S(y,x_{2m-1})\right]+\\
 & \epsilon_{\alpha\beta}LK_{12}(p)\left[\psi_{2m-1}(y)\psi_{2m-1}(-y)g_{2m-1}(y,-y)-\psi_{2m}(-y)\psi_{2m}(y)S(-y,x_{2m-1})S(y,x_{2m-1})g_{2m}(y,-y)\right]\nonumber 
\end{align}
here we used that 
\begin{equation}
S(y,x_{2k-1})S(y,x_{2k})S(-y,x_{2k-1})S(-y,x_{2k})=1
\end{equation}
Dividing by the leading order norm of the state leads to 
\begin{equation}
\frac{\left|\langle\bar{B}\left|y,-y\right\rangle \right|^{2}}{\left\langle y_{1},y_{2}\right.\left|y,-y\right\rangle _{\alpha,\beta}}=\frac{4g^{2}}{s^{2}}\frac{(y^{2}+s^{2})^{2}}{y^{2}+4g^{2}(y^{2}+1)^{2}}\epsilon_{\alpha\beta}
\end{equation}
and the boundary state at the nested level is the SU(2) dimer state,
what we already know.

\bibliographystyle{elsarticle-num}
\bibliography{refs}

\end{document}